\input harvmac
\noblackbox
\newcount\figno
\figno=0
\def\fig#1#2#3{
\par\begingroup\parindent=0pt\leftskip=1cm\rightskip=1cm\parindent=0pt
\baselineskip=11pt \global\advance\figno by 1 \midinsert
\epsfxsize=#3 \centerline{\epsfbox{#2}} \vskip 12pt
\centerline{{\bf Figure \the\figno:} #1}\par
\endinsert\endgroup\par}
\def\figlabel#1{\xdef#1{\the\figno}}

\def\np#1#2#3{Nucl. Phys. {\bf B#1} (#2) #3}

\def\IR{\relax{\rm I\kern-.18em R}}


\font\cmss=cmss10 \font\cmsss=cmss10 at 7pt
\def\rlx{\relax\leavevmode}
\def\inbar{\vrule height1.5ex width.4pt depth0pt}
\def\IC{\relax\,\hbox{$\inbar\kern-.3em{\rm C}$}}
\def\IN{\relax{\rm I\kern-.18em N}}
\def\IP{\relax{\rm I\kern-.18em P}}
\def\ZZ{\rlx\leavevmode\ifmmode\mathchoice{\hbox{\cmss Z\kern-.4em Z}}
 {\hbox{\cmss Z\kern-.4em Z}}{\lower.9pt\hbox{\cmsss Z\kern-.36em Z}}
 {\lower1.2pt\hbox{\cmsss Z\kern-.36em Z}}\else{\cmss Z\kern-.4em
 Z}\fi}
\def\IZ{\relax\ifmmode\mathchoice
{\hbox{\cmss Z\kern-.4em Z}}{\hbox{\cmss Z\kern-.4em Z}}
{\lower.9pt\hbox{\cmsss Z\kern-.4em Z}} {\lower1.2pt\hbox{\cmsss
Z\kern-.4em Z}}\else{\cmss Z\kern-.4em Z}\fi}

\def\narrowplus{\kern -.04truein + \kern -.03truein}
\def\narrowminus{- \kern -.04truein}
\def\narrowminussub{\kern -.02truein - \kern -.01truein}

\def\ker{{\rm Ker}}

\def\O{{\cal O}}

\def\a{{\alpha}}
\def\g{{\gamma}}

\def\frac#1#2{{#1\over #2}}

\def\ol{{\rm lim}}

\def\IZ{\relax\ifmmode\mathchoice
{\hbox{\cmss Z\kern-.4em Z}}{\hbox{\cmss Z\kern-.4em Z}}
{\lower.9pt\hbox{\cmsss Z\kern-.4em Z}} {\lower1.2pt\hbox{\cmsss
Z\kern-.4em Z}}\else{\cmss Z\kern-.4em Z}\fi}
\def\IB{\relax{\rm I\kern-.18em B}}
\def\IC{{\relax\hbox{$\inbar\kern-.3em{\rm C}$}}}
\def\ID{\relax{\rm I\kern-.18em D}}
\def\IE{\relax{\rm I\kern-.18em E}}
\def\IF{\relax{\rm I\kern-.18em F}}
\def\IG{\relax\hbox{$\inbar\kern-.3em{\rm G}$}}
\def\IGa{\relax\hbox{${\rm I}\kern-.18em\Gamma$}}
\def\IH{\relax{\rm I\kern-.18em H}}
\def\II{\relax{\rm I\kern-.18em I}}
\def\IK{\relax{\rm I\kern-.18em K}}
\def\IP{\relax{\rm I\kern-.18em P}}

\font\cmss=cmss10 \font\cmsss=cmss10 at 7pt
\def\IR{\relax{\rm I\kern-.18em R}}

\def\ker{{\rm ker\ }}

\def\1{{\bf 1}}
\def\3{{\bf 3}}
\def\7{{\bf 7}}
\def\2{{\bf 2}}
\def\8{{\bf 8}}

\def\hat{\widehat}
\def\quabla{{\sqcap}\!\!\!\!{\sqcup}}

\def\sp{superpotential}
%

%
%
\def\eqnn#1{\xdef #1{(\secsym\the\meqno)}\writedef{#1\leftbracket#1}%
\global\advance\meqno by1\wrlabeL#1}
\def\eqna#1{\xdef #1##1{\hbox{$(\secsym\the\meqno##1)$}}
\writedef{#1\numbersign1\leftbracket#1{\numbersign1}}%
\global\advance\meqno by1\wrlabeL{#1$\{\}$}}
\def\eqn#1#2{\xdef #1{(\secsym\the\meqno)}\writedef{#1\leftbracket#1}%
\global\advance\meqno by1$$#2\eqno#1\eqlabeL#1$$}


\lref\DuffWD{
M.~J.~Duff, J.~T.~Liu and R.~Minasian, {\it
``Eleven-Dimensional Origin of String/String Duality: A One-Loop Test''},
Nucl.\ Phys. {\bf B452} (1995) 261, hep-th/9506126. }

\lref\rBB{
K.~Becker and M.~Becker, {\it ``${\cal M}$-Theory on Eight-Manifolds,''},
Nucl.\ Phys.\ {\bf B477} (1996) 155, hep-th/9605053.}

\lref\DasguptaSS{ K.~Dasgupta, G.~Rajesh and S.~Sethi, {\it ``M
theory, Orientifolds and $G$-flux''}, JHEP {\bf 9908} (1999) 023,
hep-th/9908088. }

\lref\beckerD{
K.~Becker and K.~Dasgupta,
{\it ``Heterotic Strings with Torsion,''}
hep-th/0209077.}

\lref\banks{T. Banks, ``Cosmological Breaking of Supersymmetry? or little
Lambda Goes Back to the Future'', hep-th/0007146.}

\lref\rBHO{E. Bergshoeff, C. Hull and T. Ortin,
{\it ``Duality in the Type II Superstring Effective Action''},
\np{451} {1995}{547}, hep-th/9504081. }
 \lref\rsenorien{A. Sen, {\it ``F-theory and Orientifolds''},
 Nucl. Phys. {\bf B475} (1996) 562,
hep-th/9605150.}

\lref\rstrom{A. ~Strominger, {\it ``Superstrings With Torsion''},
Nucl.\ Phys.\ {\bf B274} (1986) 253.}

\lref\witteno{E.~Witten,
{\it ``Some Properties Of O(32) Superstrings,''}
Phys.\ Lett.\ B {\bf 149}, 351 (1984).}

\lref\sencount{A.~Sen, {\it ``Local Gauge and Lorentz Invariance
of the Heterotic String Theory,''} Phys.\ Lett.\ B {\bf 166}, 300
(1986)}

\lref\xenwit{M.~Dine, N.~Seiberg, X.~G.~Wen and E.~Witten, {\it
``Nonperturbative Effects on the String World Sheet,''} Nucl.\
Phys.\ B {\bf 278}, 769 (1986); {\it ``Nonperturbative Effects on
the String World Sheet. 2,''} Nucl.\ Phys.\ B {\bf 289}, 319
(1987).}

\lref\louisL{S.~Gurrieri, J.~Louis, A.~Micu and D.~Waldram, {\it
``Mirror Symmetry in Generalized Calabi-Yau Compactifications,''}
hep-th/0211102}

\lref\HULL{C.~M.~Hull, {\it ``Superstring Compactifications with
Torsion and Space-Time Supersymmetry,''} In Turin 1985,
Proceedings, Superunification and Extra Dimensions, 347-375, 29p;
{\it ``Sigma Model Beta Functions and String Compactifications,''}
Nucl.\ Phys.\ B {\bf 267}, 266 (1986); {\it ``Compactifications of
the Heterotic Superstring,''} Phys.\ Lett.\ B {\bf 178}, 357
(1986); {\it ``Lectures on Nonlinear Sigma Models and Strings,''}
Lectures given at Super Field Theories Workshop, Vancouver,
Canada, Jul 25 - Aug 6, 1986.}

\lref\hetcit{D.~J.~Gross, J.~A.~Harvey, E.~J.~Martinec and R.~Rohm,
{\it ``The Heterotic String,''}
Phys.\ Rev.\ Lett.\  {\bf 54}, 502 (1985);
{\it ``Heterotic String Theory. 1. The Free Heterotic String,''}
Nucl.\ Phys.\ B {\bf 256}, 253 (1985);
{\it ``Heterotic String Theory. 2. The Interacting Heterotic String,''}
Nucl.\ Phys.\ B {\bf 267}, 75 (1986).}

\lref\rDJM{A.~Sen, {\it ``Strong Coupling Dynamics of Branes from
M-theory,''} JHEP {\bf 9710}, 002 (1997), 9708002; K. Dasgupta,
D. P. Jatkar and S. Mukhi, {\it ``Gravitational Couplings and
$Z_2$ Orientifolds''}, Nucl. Phys. {\bf B523} (1998) 465,
hep-th/9707224; J.~F.~Morales, C.~A.~Scrucca and M.~Serone, {\it
``Anomalous Couplings for D-branes and O-planes,''} Nucl.\ Phys.\
B {\bf 552}, 291 (1999), hep-th/9812071; B.~J.~Stefanski, {\it
``Gravitational Couplings of D-branes and O-planes,''} Nucl.\
Phys.\ B {\bf 548}, 275 (1999), hep-th/9812088. }

\lref\nemanja{N.~Kaloper and R.~C.~Myers, {\it ``The O(dd) Story
of Massive Supergravity,''} JHEP {\bf 9905}, 010 (1999),
hep-th/9901045; G.~Curio, A.~Klemm, B.~Kors and D.~Lust, {\it
``Fluxes in Heterotic and Type II String Compactifications,''}
Nucl.\ Phys.\ B {\bf 620}, 237 (2002), hep-th/0106155; J.~Louis
and A.~Micu, {\it ``Heterotic String Theory with Background
Fluxes,''} Nucl.\ Phys.\ B {\bf 626}, 26 (2002), hep-th/0110187.}

\lref\kapulov{V.~Kaplunovsky, J.~Louis and S.~Theisen, {\it
``Aspects of Duality in N=2 String Vacua,''} Phys.\ Lett.\ B {\bf
357}, 71 (1995); hep-th/9506110.}

\lref\nati{N.~Seiberg, {\it ``Observations on the Moduli Space of
Superconformal Field Theories,''} Nucl.\ Phys.\ B {\bf 303}, 286
(1988); A.~Ceresole, R.~D'Auria, S.~Ferrara and A.~Van Proeyen,
{\it ``Duality Transformations in Supersymmetric Yang-Mills
Theories Coupled to Supergravity,''} Nucl.\ Phys.\ B {\bf 444}, 92
(1995), hep-th/9502072.}

\lref\narainJ{K.~S.~Narain, {\it ``New Heterotic String Theories
in Uncompactified Dimensions $< 10$,''} Phys.\ Lett.\ B {\bf 169},
41 (1986); K.~S.~Narain, M.~H.~Sarmadi and E.~Witten, {\it ``A
Note on Toroidal Compactification of Heterotic String Theory,''}
Nucl.\ Phys.\ B {\bf 279}, 369 (1987); J.~Maharana and
J.~H.~Schwarz, {\it ``Noncompact Symmetries in String Theory,''}
Nucl.\ Phys.\ B {\bf 390}, 3 (1993), hep-th/9207016.}

\lref\senM{A.~Sen, {\it ``A Note on Enhanced Gauge Symmetries in
M- and String Theory,''} JHEP {\bf 9709}, 001 (1997),
hep-th/9707123}

\lref\rBUSH{T. Buscher, {\it ``Quantum Corrections and Extended
Supersymmetry in New Sigma Models''}, Phys. Lett. {\bf B159}
(1985) 127; {\it ``A Symmetry of the String Background Field
Equations''}, Phys. Lett. {\bf B194} (1987) 59; {\it ``Path
Integral Derivation of Quantum Duality in Nonlinear Sigma
Models''}, Phys. Lett. {\bf B201} (1988) 466.}

\lref\rKKL{E.
Kiritsis, C. Kounnas and D. Lust,
{\it ``A Large Class of New Gravitational and Axionic
Backgrounds for Four-dimensional Superstrings''},
Int. J. Mod. Phys. {\bf A9} (1994) 1361, hep-th/9308124. }

\lref\kst{S. Giddings, S. Kachru and J. Polchinski, {\it ``Hierarchies
{}From Fluxes in String Compactifications''}, hep-th/0105097;
S.~Kachru, M.~B.~Schulz and S.~Trivedi, {\it
``Moduli Stabilization from Fluxes in a Simple IIB Orientifold''},
hep-th/0201028;
A.~R.~Frey and J.~Polchinski, {\it
``N = 3 Warped Compactifications''}, Phys.\ Rev.\ {\bf D65}
(2002) 126009, hep-th/0201029. }

\lref\kstt{S.~Kachru, M.~B.~Schulz, P.~K.~Tripathy and
S.~P.~Trivedi, {\it ``New Supersymmetric String
Compactifications,''} hep-th/0211182.}

\lref\pktspt{
S.~Gurrieri and A.~Micu,
{\it ``Type IIB theory on half-flat manifolds,''} hep-th/0212278;
P.~K.~Tripathy and
S.~P.~Trivedi, {\it Compactifications with Flux on K3 and Tori,''}
hep-th/0301139.}

\lref\gates{ S.~J.~Gates, {\it ``Superspace Formulation Of New
Nonlinear Sigma Models,''} Nucl.\ Phys.\ B {\bf 238}, 349 (1984);
S.~J.~Gates, C.~M.~Hull and M.~Rocek, {\it ``Twisted Multiplets
and New Supersymmetric Nonlinear Sigma Models,''} Nucl.\ Phys.\ B
{\bf 248}, 157 (1984); S.~J.~Gates, S.~Gukov and E.~Witten, {\it
``Two-dimensional Supergravity Theories from Calabi-Yau
Four-folds,''} Nucl.\ Phys.\ B {\bf 584}, 109 (2000),
hep-th/0005120.}

\lref\rkehagias{A. Kehagias, {\it ``New Type IIB Vacua and
their F-theory Interpretation''}, Phys. Lett. {\bf B435} (1998)
337, hep-th/9805131. }

\lref\GukovYA{
S.~Gukov, C.~Vafa and E.~Witten, {\it ``CFT's from Calabi-Yau Four-folds''},
Nucl.\ Phys.\ {\bf B584} (2000) 69, hep-th/9906070. }
\lref\BeckerPM{
K.~Becker and M.~Becker, {\it ``Supersymmetry Breaking, ${\cal M}$-theory
and Fluxes''},
JHEP {\bf 0107} (2001) 038 (2001), hep-th/0107044. }
\lref\DineRZ{ M.~Dine, R.~Rohm, N.~Seiberg and E.~Witten, {\it
``Gluino Condensation in Superstring Models''}, Phys.\ Lett.\ {\bf
B156}, 55 (1985).}
\lref\KachruHE{
S.~Kachru, M.~B.~Schulz and S.~Trivedi, {\it
``Moduli Stabilization from Fluxes in a Simple IIB Orientifold''},
hep-th/0201028.}
\lref\FreyHF{ A.~R.~Frey and J.~Polchinski, {\it
``N = 3 Warped Compactifications''}, Phys.\ Rev.\ {\bf D65}
(2002) 126009, hep-th/0201029.}
\lref\CurioAE{
G.~Curio, A.~Klemm, B.~Kors and D.~Lust, {\it
``Fluxes in Heterotic and Type II String Compactifications''},
Nucl.\ Phys.\ {\bf B620} (2202) 237, hep-th/0106155.}

\lref\Vafawitten{ C.~Vafa and E.~Witten, {\it ``A One Loop Test of
String Duality''}, Nucl.\ Phys.\ {\bf B447} (1995) 261,
hep-th/9505053.}

\lref\SethiVW{ S.~Sethi, C.~Vafa and E.~Witten, {\it ``Constraints on
Low-dimensional String Compactifications''},  Nucl.\ Phys.\
{\bf B480} (1996) 213, hep-th/9606122.}

\lref\HananyK{
A.~Hanany and B.~Kol,
{\it ``On Orientifolds, Discrete Torsion, Branes and M Theory''},
JHEP {\bf 0006} (2000) 013, hep-th/0003025.}

\lref\Ganor{
O.~J.~Ganor,{\it
``Compactification of Tensionless String Theories''}, hep-th/9607092.}

\lref\DMtwo{
K.~Dasgupta and S.~Mukhi,
{\it ``A Note on Low-Dimensional String Compactifications''},
Phys.\ Lett.\ {\bf B398} (1997) 285, hep-th/9612188.}

\lref\ShiuG{
B.~R.~Greene, K.~Schalm and G.~Shiu,
{\it ``Warped Compactifications in M and F Theory''},
Nucl.\ Phys.\ {\bf B584} (2000) 480, hep-th/0004103.}

\lref\harmoni{ G.~W.~Gibbons and P.~J.~Ruback, {\it ``The Hidden
Symmetries Of Multicenter Metrics,''} Commun.\ Math.\ Phys.\ {\bf
115}, 267 (1988); N.~S.~Manton and B.~J.~Schroers, {\it ``Bundles
over Moduli Spaces and the Quantization Of BPS Monopoles,''}
Annals Phys.\  {\bf 225}, 290 (1993); A.~Sen, {\it ``Dyon -
Monopole Bound States, Selfdual Harmonic Forms on the Multi -
Monopole Moduli Space, and SL(2,Z) Invariance in String Theory,''}
Phys.\ Lett.\ B {\bf 329}, 217 (1994), hep-th/9402032.}

\lref\mesO{P.~Meessen and T.~Ortin, {\it ``An Sl(2,Z) Multiplet
of Nine-Dimensional Type II Supergravity Theories''}, Nucl.\
Phys.\ B {\bf 541} (1999) 195, hep-th/9806120; E. Bergshoeff, C.
Hull and T. Ortin, {\it ``Duality in the Type II Superstring
Effective Action''}, \np{451} {1995}{547}, hep-th/9504081;
S.~F.~Hassan, {\it ``T-duality, Space-Time Spinors and R-R fields
in Curved Backgrounds,''} Nucl.\ Phys.\ B {\bf 568}, 145 (2000),
hep-th/9907152.}

\lref\SmitD{
B.~de Wit, D.~J.~Smit and N.~D.~Hari Dass,
{\it ``Residual Supersymmetry Of Compactified D = 10 Supergravity''},
Nucl.\ Phys.\ {\bf B283} (1987) 165 (1987).}

\lref\PapaDI{ S.~Ivanov and G.~Papadopoulos, {\it ``A No-Go
Theorem for String Warped Compactifications''},
Phys.\ Lett.{\bf B497} (2001) 309, hep-th/0008232.}

\lref\DineSB{M.~Dine and N.~Seiberg, {\it ``Couplings and Scales
in Superstring Models''}, Phys.\ Rev.\ Lett.\  {\bf 55}, 366
(1985).}

\lref\olgi{O. DeWolfe and S. B. Giddings,
{\it ``Scales and Hierarchies in Warped Compactifications
and Brane Worlds''}, hep-th/0208123.}

\lref\hellermanJ{S.~Hellerman, J.~McGreevy and B.~Williams,
{\it ``Geometric Constructions of Non-Geometric String Theories''},
hep-th/0208174.}

\lref\WIP{K. Becker, M. Becker, K. Dasgupta, {\it Work in Progress}.}

\lref\tatar{K.~Dasgupta, K.~h.~Oh, J.~Park and R.~Tatar, {\it
``Geometric Transition Versus Cascading Solution,''} JHEP {\bf
0201}, 031 (2002), hep-th/0110050.}

\lref\civ{F.~Cachazo, B.~Fiol, K.~A.~Intriligator, S.~Katz and C.~Vafa,
{\it ``A geometric unification of dualities,''}
Nucl.\ Phys.\ B {\bf 628}, 3 (2002), hep-th/0110028.}

\lref\renata{K.~Dasgupta, C.~Herdeiro, S.~Hirano and R.~Kallosh,
{\it ``D3/D7 Inflationary Model and M-theory,''} Phys.\ Rev.\ D
{\bf 65}, 126002 (2002), hep-th/0203019;
K.~Dasgupta, K.~h.~Oh, J.~Park and R.~Tatar, {\it
``Geometric Transition Versus Cascading Solution,''} JHEP {\bf
0201}, 031 (2002), hep-th/0110050.}

\lref\carlos{R.~Kallosh, {\it ``N = 2 Supersymmetry and de Sitter
Space,''} hep-th/0109168; C.~Herdeiro, S.~Hirano and R.~Kallosh,
{\it ``String Theory and Hybrid Inflation / Acceleration,''} JHEP
{\bf 0112}, 027 (2001), hep-th/0110271.}

\lref\PolchinskiRR{
J.~Polchinski,
{\it ``String Theory. Vol. 2: Superstring Theory And Beyond''}.}

\lref\taylor{E.~Cremmer, S.~Ferrara, L.~Girardello and A.~Van
Proeyen, {\it ``Yang-Mills Theories with Local Supersymmetry:
Lagrangian, Transformation Laws and Superhiggs Effect,''} Nucl.\
Phys.\ B {\bf 212}, 413 (1983); T.~R.~Taylor and C.~Vafa, {\it
``RR Flux on Calabi-Yau and Partial Supersymmetry Breaking,''}
Phys.\ Lett.\ B {\bf 474}, 130 (2000), hep-th/9912152.}

\lref\guko{S.~Gukov, C.~Vafa and E.~Witten, {\it ``CFT's from
Calabi-Yau Four-folds,''} Nucl.\ Phys.\ B {\bf 584}, 69 (2000)
[Erratum-ibid.\ B {\bf 608}, 477 (2001)], hep-th/9906070.}

\lref\hullwitten{ C.~M.~Hull and E.~Witten, {\it ``Supersymmetric
Sigma Models and the Heterotic String,''} Phys.\ Lett.\ B {\bf
160}, 398 (1985);
A.~Sen,
{\it ``Local Gauge And Lorentz Invariance Of The Heterotic String Theory,''}
Phys.\ Lett.\ B {\bf 166}, 300 (1986);
{\it ``The Heterotic String In Arbitrary Background Field,''}
Phys.\ Rev.\ D {\bf 32}, 2102 (1985);
{\it ``Equations Of Motion For The Heterotic String Theory From
The Conformal Invariance Of The Sigma Model,''}
Phys.\ Rev.\ Lett.\  {\bf 55}, 1846 (1985).}

\lref\dasmukhi{ K.~Dasgupta and S.~Mukhi, {\it ``F-theory at
Constant Coupling,''} Phys.\ Lett.\ B {\bf 385}, 125 (1996),
hep-th/9606044.}

\lref\vafasen{
C.~Vafa,
{\it ``Evidence for F-Theory,''}
Nucl.\ Phys.\ B {\bf 469}, 403 (1996), hep-th/9602022;
A.~Sen,
{\it ``F-theory and Orientifolds,''}
Nucl.\ Phys.\ B {\bf 475}, 562 (1996), hep-th/9605150;
T.~Banks, M.~R.~Douglas and N.~Seiberg,
{\it ``Probing F-theory with branes,''}
Phys.\ Lett.\ B {\bf 387}, 278 (1996), hep-th/9605199.}

\lref\zwee{
M.~R.~Gaberdiel and B.~Zwiebach,
{\it ``Exceptional groups from open strings,''}
Nucl.\ Phys.\ B {\bf 518}, 151 (1998), hep-th/9709013.}

\lref\callan{
C.~G.~Callan, E.~J.~Martinec, M.~J.~Perry and D.~Friedan,
{\it ``Strings In Background Fields,''}
Nucl.\ Phys.\ B {\bf 262}, 593 (1985);
A.~Sen,
{\it ``Equations Of Motion For The Heterotic String
Theory From The Conformal Invariance Of The Sigma Model,''}
Phys.\ Rev.\ Lett.\  {\bf 55}, 1846 (1985);
{\it ``The Heterotic String In Arbitrary Background Field,''}
Phys.\ Rev.\ D {\bf 32}, 2102 (1985).}

\lref\WittenBS{
E.~Witten, {\it ``Toroidal Compactification without Vector Structure''},
JHEP {\bf 9802} (1998) 006 (1998), hep-th/9712028.}

\lref\carluest{ G.~L.~Cardoso, G.~Curio, G.~Dall'Agata, D.~Lust,
P.~Manousselis and G.~Zoupanos, {\it ``Non-K\"ahler String
Backgrounds and their Five Torsion Classes,''} hep-th/0211118.}

\lref\grifhar{
P.~ Griffiths and J.~ Harris
{\it ``Principles of Algebraic Geometry,''}
Wiley Classics Library.}

\lref\GP{ E.~Goldstein and S.~Prokushkin, {\it ``Geometric Model
for Complex non-Kaehler Manifolds with SU(3) Structure,''}
hep-th/0212307.}

\lref\toappear{ K.~Becker, M.~Becker, K.~Dasgupta, P.~S.~Green,
..., {\it ``Work in Progress.''}}

\lref\MeessenQM{
P.~Meessen and T.~Ortin,
{\it ``An Sl(2,Z) Multiplet of Nine-Dimensional
Type II Supergravity Theories''}, Nucl.\ Phys.\ B {\bf 541}
(1999) 195, hep-th/9806120.}

\lref\HUT{ C.~M.~Hull and P.~K.~Townsend, {\it ``The Two Loop
Beta Function for Sigma Models with Torsion,''} Phys.\ Lett.\ B
{\bf 191}, 115 (1987); {\it ``World Sheet Supersymmetry and
Anomaly Cancellation in the Heterotic String,''} Phys.\ Lett.\ B
{\bf 178}, 187 (1986).}

\lref\SenJS{
A.~Sen, {\it ``Dynamics of Multiple Kaluza-Klein Monopoles
in M and String Theory''}, Adv.\ Theor.\ Math.\ Phys.\
{\bf 1} (1998) 115, hep-th/9707042.
}

\lref\sav{
K.~Dasgupta, G.~Rajesh and S.~Sethi,
{\it ``M theory, orientifolds and G-flux,''}
JHEP {\bf 9908}, 023 (1999), hep-th/9908088.}

\lref\BeckerNN{
K.~Becker, M.~Becker, M.~Haack and J.~Louis, {\it
``Supersymmetry Breaking and $\alpha'$-Corrections to Flux
Induced  Potentials''}, JHEP {\bf 0206} (2002) 060,
hep-th/0204254.}

\lref\renshamit{
S.~Kachru, R.~Kallosh, A.~Linde, S.~Trivedi,
{\it ``de-Sitter Vacua in String Theory,''} hep-th/0301240.}

\lref\hullambig{C.~M.~Hull,
{\it ``Anomalies, Ambiguities And Superstrings,''}
Phys.\ Lett.\ B {\bf 167}, 51 (1986).}

\lref\becsuper{K.~Becker, M.~Becker, K.~Dasgupta and S.~Prokushkin,
{\it ``Properties Of Heterotic Vacua From Superpotentials,''}
 hep-th/0304001.}

\lref\eva{E.~Silverstein,
{\it ``{\rm (A)dS} backgrounds from asymmetric orientifolds,''} hep-th/0106209;
A.~Dabholkar and C.~Hull,
{\it ``Duality twists, orbifolds, and fluxes,''} hep-th/0210209;
S.~Hellerman, J.~McGreevy and B.~Williams,
{\it ``Geometric constructions of nongeometric string theories,''}
hep-th/0208174.}

\lref\weba{J.~Bagger and J.~Wess, {\it ``Supersymmetry and
Supergravity,''} Princeton University Press.}

\lref\becons{ M.~Becker and D.~Constantin,{\it ``A Note on Flux
Induced Superpotentials in String Theory ''}, hep-th/0210131.}

\lref\EdNew{E.~Witten, {\it ``New Issues in Manifolds of $SU(3)$
Holonomy}, Nucl.\ Phys.\ B {\bf 268} (1986) 79.}

\Title{\vbox{\hbox{hep-th/0301161} \hbox{UMD-PP-03-030,
SU-ITP-02/46}}} {\vbox{ \hbox{\centerline{Compactifications of
Heterotic Theory}}
\hbox{\centerline{on Non-K\"ahler Complex Manifolds: I}}}}

\vskip-.3in

\centerline{\bf Katrin Becker{$^1$}, Melanie Becker{$^2$},
Keshav Dasgupta{$^3$}, Paul S. Green{$^4$}}
\vskip 0.1in
\centerline
{ ${}^1$ Department of Physics,
University of Utah, Salt Lake City, UT 84112-0830}
\centerline{\tt katrin@physics.utah.edu} 
\centerline
{ ${}^2$ Department of Physics,
University of Maryland, College Park, MD
20742-4111}
\centerline{\tt melanieb@physics.umd.edu} 
\centerline
{${}^3$ Department of Physics, Varian Lab, Stanford
University, Stanford
CA 94305-4060}
\centerline{\tt keshav@itp.stanford.edu} 
\centerline
{${}^4$ Department of Mathematics,
University of Maryland, College Park, MD
20742-4111}
\centerline{\tt psg@math.umd.edu}

\vskip.1in

\centerline{\bf Abstract}

\noindent We study new compactifications of the $SO(32)$
heterotic string theory on compact complex non-K\"ahler manifolds.
These manifolds have many interesting features like fewer moduli,
torsional constraints, vanishing Euler character and vanishing
first Chern class, which make the four-dimensional theory
phenomenologically attractive. We take a particular compact
example studied earlier and determine various geometrical
properties of it. In particular we calculate the warp factor and
study the sigma model description of strings propagating on these
backgrounds. The anomaly cancellation condition and enhanced
gauge symmetry are shown to arise naturally in this framework, if
one considers the effect of singularities carefully.

\noindent We then give a detailed mathematical analysis of these
manifolds and construct a large class of them. The existence of a
holomorphic (3,0) form is important for the construction. We
clarify some of the topological properties of these manifolds and
evaluate the Betti numbers. We also determine the superpotential
and argue that the radial modulus of these manifolds can actually
be stabilized.

\Date{}

\listtoc
\writetoc

\newsec{Introduction and Summary}



Compactifications of string theory and ${\cal M}$-theory with
non-vanishing background fluxes are of great interest and an
active area of research because of their implications for particle
phenomenology. In such compactifications a potential for the
moduli fields is generated at string tree level, so that many of
the moduli fields can actually be stabilized with a mechanism,
that is simple enough to do actual computations. String theory is
then able to make definite predictions for the coupling constants
of the standard model, such as the pattern of quark and lepton
masses and the size of the gauge hierarchy. From the
phenomenological point of view, compactifications of the
heterotic string with background fluxes are particularly
interesting. However, once the fluxes are turned on, the internal
manifold is no longer K\"ahler and has torsion. This type of
string theory compactifications was first discussed in
\rstrom,\HULL\ and \SmitD\ some time ago. Yet, not much is known
about such models in the literature, neither from the mathematics
point of view nor from the physics point of view. It is the
purpose of this paper to partially fill this gap by studying
compactifications of the $SO(32)$ heterotic string on compact,
complex, non-K\"ahler manifolds both from the mathematics and
physics point of view. Very generally, the internal manifolds,
that we consider have a vanishing first Chern class and a
vanishing Euler character. The existence of such non-K\"ahler
manifolds and their realization as compact, complex three-folds
was more recently pointed out in \sav, as they originate from the
warped ${\cal M}$-theory background of \rBB\ by using a set of
$U$-duality transformations. This warping is important, as the
absence of warping will bring us back to Calabi-Yau (CY) type
compactifications, where the three-form flux vanishes. After
turning on the fluxes, these manifolds become inherently
non-K\"ahler. This is to be contrasted with the non-K\"ahler
four-folds, that we get in ${\cal M}$-theory with $G$-fluxes
\rBB. These four-folds are conformally CY, with the conformal
factor being the warp factor. The three-folds, that we get in the
heterotic theory are, in general, {\it not} conformally CY, as
one cannot extract a conformal factor to obtain a Ricci-flat CY
manifold. In fact the models presented in \sav\ and \beckerD,
consist of a four-dimensional CY base, that is warped in some
specific way by the warp factor and a fiber over this
base, that remains unwarped. However, there is a non-trivial {\it
twisting}, which mixes the fiber coordinates with the complex
coordinates of the base. This twist of the fiber and the warp
factor of the base is responsible for making the manifold non
Ricci-flat. Compactifications of $SO(32)$ heterotic theory on
these six-manifolds still give rise to a four-dimensional
Minkowski vacuum \HULL,\rstrom, \SmitD, as the warp factor is just
a function of the internal space and the three-form flux is
non-vanishing only on the internal manifold. The three-form flux, which is 
real,
is a {\it quantized} quantity, because it originates from
quantized fluxes in ${\cal M}$-theory. It acts as the torsion for
the manifold, in the presence of which, there is a preferred
connection. All curvature related quantities are therefore
measured with respect to this connection. There are also gauge bundles on the
six-manifold, which are to be embedded in some specific way. As
we will point out, the simple embedding of the spin connection
into the gauge connection is not allowed for this kind of
compactification, even though the second Chern classes satisfy
\eqn\seccher{c_2( R) = c_2 (F)} where $R$ and $F$ are the
curvatures of the spin connection and the gauge connection
respectively. In the above equation, somewhat unexpectedly, as we
will show, the choice of connection is rather irrelevant. In the
concrete example recently presented in \beckerD, one can
explicitly derive many of these facts, because the background is
written is terms of an orbifold base, as we will discuss in this
paper. However it is important to understand whether we can
extend this example to the more general case, where we consider
the base to be away from the orbifold limit. In this paper we
shall present a detailed mathematical construction of a large
class of complex non-K\"ahler manifolds, compute their Betti
numbers and show, that the model of \beckerD\ is a special case
of this general construction. From the physics point of view we
will present the sigma model description of strings propagating in
such backgrounds and study the effect of gauge fluxes, that were
not taken into account in \beckerD. In particular, we show how
the anomaly cancellation condition and enhanced gauge symmetry
naturally arise in this framework, if one considers carefully the
effect of singularities. Finally, we determine the form of the
superpotential for compactifications of the heterotic string on
non-K\"ahler manifolds and show, that the radial modulus can be
stabilized in this type of compactifications.

\vskip.2in

\noindent $\underline{\rm Organization~of~the~Paper}$

\vskip.2in

This paper is organized as follows. In section 2 we discuss the
geometrical properties of non-K\"ahler manifolds. In sub-section
2.1 we recapitulate the concrete heterotic background described in
\beckerD, where the gauge fluxes originating from the orbifold
singularities of the internal manifold had not been taken into
account. In sub-section 2.2 we find the explicit solution to the
warp factor equation, ignoring the localized fluxes coming from
the orbifold singularities. In sub-section 2.3 we describe the
torsional connection appearing in the model of \beckerD\ and show,
that this model has vanishing first Chern class, by taking the
gauge fields into account. In sub-section 2.4 we recapitulate the
sigma model description and use this description to show, that the
internal manifolds considered herein are not K\"ahler. In
sub-section 2.5 we study the effects of the gauge fields and show
how the anomaly relation of the heterotic theory originates from
the Type IIB theory. In sub-section 2.6 we show, how the full
non-abelian symmetry originates from the ${\cal M}$-theory dual
description.

In section 3 we discuss the mathematical aspects of
non-K\"ahler manifolds, in particular their topological
properties. We present various algebraic geometric properties of
these manifolds and give a generic construction of a large class
of compact, complex non-K\"ahler manifolds. In sub-section 3.1 we
first recapitulate some basic features of non-K\"ahler manifolds,
which might be lesser known to the reader. As a special case of
the Serre spectral sequence we consider the case of a torus bundle
over a base manifold $B$. If $B$ is a simply-connected
four-dimensional manifold, we can easily compute the value of the
Betti numbers and show that the Euler characteristic of these
non-K\"ahler manifolds vanishes. We use this general result to
compute the Betti numbers of the model considered in \beckerD. In
sub-section 3.2 we give a generic construction of non-K\"ahler
manifolds, that are generalizations of the Hopf surface, which is
a torus bundle over $CP^1$ and show, that the model considered in
\beckerD\ is a special case of this general construction.

In section 4 we compute the background superpotential for
compactifications of the heterotic string on a non-K\"ahler
manifold by dimensional reduction of the quadratic term appearing
in the action of ${\cal N}=1$, $D=10$ supergravity coupled to Yang-Mills
theory. We also use T-duality applied to the Type IIB
superpotential, that is well known in the literature, to get the
form of the heterotic superpotential. In sub-section 4.1 we
discuss the existence of a unique holomorphic three-form for the
particular model of Type IIB theory compactified on $T^4/{\cal
I}_4 \times T^2/\IZ_2$ and interpret the norm of the holomorphic
three-form in terms of the torsion classes presented in
\carluest,\louisL. In sub-section 4.2 we carry out the explicit
calculation of the superpotential. In sub-section 4.3 we shall
see, that many of the moduli fields for the heterotic
compactification are lifted by switching on the ${\cal H}$-fluxes.
In sub-section 4.4 we compute the potential for the radius moduli
and show, that the value of this field can actually be fixed.

Finally, in section 5 we present our conclusions and discuss
some related examples of non-K\"ahler
manifolds that have recently appeared in the literature. Some of these
examples can be constructed from the generic structure in section 3. We also
point out some directions for phenomenological applications and
elaborate some implications of our results that could be pursued in
near future.

{\bf Note added:} While the draft was being written, there
appeared a couple of papers, which have some overlap with the
contents of this paper. The paper of \carluest\ discusses a
particular kind of non-K\"ahler manifold called the Iwasawa
manifold. They also classified the torsion classes. More
discussions on this classification also appeared in \louisL.
Detailed mathematical aspects of non-K\"ahler manifolds with
$SU(3)$ structure in the heterotic theory were given in \GP.
While this paper was being written we became aware, that an
independent calculation of the superpotential for the heterotic
string on a manifold with torsion was being performed (see the
second reference in  \pktspt). We thank the authors for informing
us about their calculation prior to publication.

\newsec{Geometrical Properties of Non-K\"ahler Manifolds}

In the previous paper \beckerD\ two of us gave an explicit example
of an $SO(8)^4$ heterotic theory compactified on a non-K\"ahler
manifold with non-vanishing Ricci tensor and a background ${\cal
H}$-field. It was shown, how this background satisfies the
torsional constraints expected from supersymmetry. This analysis
was mainly motivated from the existence of a solvable warped
background \rBB\ in ${\cal M}$-theory with $G$-fluxes on the
$T^4/{\cal I}_4 \times T^4/{\cal I}_4$ manifold, where ${\cal
I}_4$ is a purely orbifold action, which reverses the sign of all
the toroidal directions. We denote the unwarped metric of the
four-fold by ${\tilde g}_{a \bar b}$. The $G$-flux, that we
consider is \eqn\gflux{{G \over 2 \pi} = A~d{\bar z}^1 dz^2
d{\bar z}^3 dz^4 + B~ dz^1 d{\bar z}^2 d{\bar z}^3 dz^4 +
\sum_{i=1}^4 ~F^i \wedge \Omega^i + {\rm c.c}.} Here $z's$
describe the coordinates on the internal manifold, $\Omega^i$
with $i= 1,..4$ are the harmonic two forms on $T^4/{\cal I}_4$,
$F^i$ are the $SO(8)^4$ gauge bundles at the singular orbifold
points of $T^4/{\cal I}_4$ with an $SO(8)$ gauge bundle at each
point and $A$,$B$ are complex constants, whose explicit value is
given later on. By construction \gflux\ is a primitive ($2,2$)
form and therefore preserves supersymmetry. In fact, if we ignore
the localized piece in \gflux, then one can easily show that in
the Type IIB theory we get three form fluxes $H_{NS}=H$ and
$H_{RR} = H'$ which, when combined\foot{Our conventions are
slightly different from \beckerD, as we take ${\cal T}$ as our
torsion and not $-{\cal T}$. Therefore, our definition of $G_3$
will be the same as in \kst.} to form $G_3$, \eqn\gthree{G_3 = H'
- \varphi H = 2~i{\rm Im}~\varphi~ ({\bar A}~dz^1\wedge d{\bar
z}^2 \wedge dz^3 + {\bar B}~ d{\bar z}^1 \wedge dz^2 \wedge
dz^3),} clearly indicates, that only the ($2,1$) piece survives.
Here $\varphi$ is the usual axion-dilaton combination in Type IIB.
This is consistent with the constraints imposed by supersymmetry,
as these predict the vanishing of the $(3,0), (0,3)$ and $(1,2)$
parts of $G_3$. Also notice, that the background previously
studied in \beckerD\ can actually be mapped to this one by an
$SL(2,\IZ)$ matrix \eqn\sltwoz{\pmatrix{0 & 1\cr -1&0},} which
has a fixed point at $\varphi = i$. Furthermore, the definition
of $G_3$ in \gthree\ can be inferred directly from ${\cal
M}$-theory. If we define a generic (1,2) form as $\omega$, then
the $G$-flux in ${\cal M}$-theory is \eqn\gfluxinm{G = \omega
\wedge dz^4 - \omega^\ast \wedge d{\bar z}^4 = {{\bar G}_3 \over
\varphi - {\bar \varphi}}\wedge dz^4 - {G_3 \over \varphi - {\bar
\varphi} } \wedge d{\bar z}^4,} where $dz^4 = dx^{10} + \varphi
dx^{11}$. Now wedging this with a holomorphic (4,0) form and
using the relation: $\int dz^4 \wedge d{\bar z}^4 = 2i~{\rm Im}
\varphi~dx^{10} \wedge dx^{11}$, we can easily reproduce the IIB
superpotential $\int G_3 \wedge \Omega$ and hence infer $G_3$
from there.

As discussed in \beckerD, the localized flux in \gflux\ which
involve the $F^i$ fields contribute to the world volume
gauge fluxes and the non-localized fluxes are responsible for the
${\cal H}$-torsion and the geometry in the heterotic description.
However, the analysis of \beckerD\ took mostly the non-localized
fluxes into account and ignored more or less the localized
fluxes. This led to \eqn\hvalue{d{\cal H} = 0,} as the condition
on the ${\cal H}$ field. However, due to supersymmetry and anomaly
cancellation, it is well known that ${\cal H}$ cannot be a closed
form. It was noted in \beckerD, that the localized fluxes, when
taken into account, should give the right anomaly condition on
the heterotic side. In this section, among other things, we will
verify this fact. The constraint on the ${\cal H}$-flux will be
\eqn\condH{d{\cal H} = {\alpha'\over 2}[p_1(R) - p_1(F)],} where $p_i$
are the Pontryagin classes for the spin and gauge bundles
respectively. Since ${\cal H}$ and $J_{a \bar b}$ (the fundamental
form) are related by the torsional constraints \HULL, \rstrom,
\SmitD, the fundamental form is very constrained because of this
and non-K\"ahlerity. These issues have been already discussed in
\beckerD, where it was shown, how the fundamental form satisfies
the torsional constraints {\it ignoring} the gauge fluxes. In
this paper we shall take these into account and study their
effects.

\subsec{The Background Geometry}

Let us begin by writing the background geometry explicitly. The
notations used here are the same as in \beckerD. We shall also
take the Type IIB coupling $g_B = 1$. The heterotic background
is: \eqn\backgeom{\eqalign{ & ds^2 = 4 \Delta^2 ~{\tilde g}_{1
\bar 1} {\tilde g}_{3 \bar 3}~dz^1 d{\bar z}^1  + 4 \Delta^2 ~
{\tilde g}_{2 \bar 2} {\tilde g}_{3 \bar 3}~dz^2 d{\bar z}^2 +
|dz^3 + 2 B {\bar z}^2 dz^1 - 2 A {\bar z}^1 dz^2 |^2, \cr &
{\cal H} = A ~dz^2 \wedge d{\bar z}^1 \wedge d {\bar z}^3 - B
dz^1 \wedge d {\bar z}^2 \wedge d {\bar z}^3 + {\bar A} ~d{\bar
z}^2 \wedge dz^1 \wedge dz^3 - {\bar B}~d{\bar z}^1 \wedge d{z}^2
\wedge d {z}^3 ,\cr & g_{het} = 2 \Delta ~ {\tilde g}_{3 \bar
3},}} where $\Delta$ is the warp factor and $g_{het}$ is the
heterotic coupling constant and the constant values of $A$ and $B$
are given below. The three-form background ${\cal H}$ is a {\it real} 
quantity which satisfy the torsional constraints \beckerD.
The unwarped metric $\tilde g_{ij}$ is basically
flat everywhere except at some points, where there are
singularities. In fact in ${\cal M}$-theory the contributions to
the $X_8$ polynomial come from those singular points. Ignoring
them will lead us to trivial background fluxes. Taking this into
account, the metric can be simplified to \eqn\metinhetone{ g_{a
\bar b} = \pmatrix{ \Delta^2 + |B z^2|^2 & - AB z^1 {\bar z}^2 &
{B {\bar z}^2\over 2}\cr \noalign{\vskip -0.20 cm}  \cr -A {\bar
B} {\bar z}^1 z^2 & \Delta^2 + |A z^1|^2 &  -{A {\bar z}^1\over
2}\cr \noalign{\vskip -0.20 cm}  \cr {\bar B z^2 \over 2} &
-{\bar A z^1 \over 2} & {1 \over 4}},} where we have ignored an
overall factor of 4. In the presence of localized fluxes the
metric will be more involved, as we shall derive towards the end
of this section. We are also assuming, that the warp factor is
independent of the $z_3$ direction. In the following we shall
show that \eqn\warpy{ \Delta \equiv \Delta (|z_1|, |z_2|),} is a
consistent assumption. {}From \backgeom\ we see that the base,
with coordinates $z^1, z^2$ is stretched by $\Delta^2 {\tilde
g}_{3 \bar 3}$ and the torus parametrized by $z^3$ is non
trivially fibred over the base. The above metric can be generated
from ${\cal M}$-theory on a four-fold by making a series of
$U$-duality transformations. If we denote the volume of any
typical four-cycle of the four-fold by $v$, then the constants
$A, B$ appearing in \backgeom\ have definite values of
\eqn\valueAB{ A = {2+i \over v}, ~~~~ {\rm and} ~~~~ B = {i \over
v}.} In deriving this we have to be careful about the total
volume of our space\foot{We do not necessarily mean that {\it
all} four-cycles have equal volumes. The above analysis goes
through easily with arbitrary sizes of the four-cycles.}.
As it turns out (and is also discussed in
\beckerD) the volume of our space is reduced by $1/4$ and hence
all the quantization rules are changed slightly. Furthermore,
notice that the previous values for $A$ and $B$ differ by a
factor $1\over v$ from the results of \beckerD, where $v$ was set
to one for convenience. The above background is not complete,
unless we specify the gauge bundle. The gauge bundle for our case
is $D_4^4 = SO(8)^4$ which, as discussed in \beckerD, comes from
stabilizing the seven brane moduli. This fixing also guarantees,
that there are no non-perturbative corrections to the moduli
space.

\subsec{Solution to the Warp Factor Equation}

The solution to the warp factor equation can be found in two
stages. We will attack this question from the Type IIB point of
view, where we have the advantage of writing the warp factor
equation linearly. Therefore, let us first assume our manifold is
{\it non-compact}. In this way we can concentrate on regions far
from the orbifold singularities and write the warp factor
equation as \eqn\wfactor{ {\del \Delta^2 \over \del z^2} +
{|B|^2} {\bar z}^2  = {\del \Delta^2 \over \del z^1} + {|A|^2}
{\bar z}^1  = 0.} As discussed in \beckerD, the above equations
have an origin on the Type IIB side from the self-duality
relation for the five-form $F_5$. The solution to \wfactor\ is
\eqn\solwap{ \Delta^2 = {\rm c}_o - {5~|z^1|^2 + |z^2|^2 \over
v^2},}
where $v$ is the volume of any generic four-cycle of the ${\cal
M}$-theory four-fold. We will fix $c_{\rm o}$ below.
Also observe, that the warp factor is independent of the $z^3$
direction and depends only on $|z^i|$, for $i=1,2$.

Let us now bring back the singularities by making the Type IIB
manifold compact. The term, that will now contribute to the warp
factor equation is the $X_8$ polynomial of ${\cal M}$-theory as
\eqn\xeight{ X_8 = {3\over 32} \sum_{z^i, w^j}~\delta^4 (z - z^i)
\delta^4 (w - w^j),} where $z^i= (z^1_i, z^2_i)$ and $w^j =
(z^3_j, z^4_j)$ are the fixed points of $T^4/{\cal I}_4 \times
T^4/{\cal I}_4$. This polynomial, which is a bulk term in ${\cal
M}$-theory, will appear on the Type IIB side as gravitational
couplings on the $D7$ branes and $O7$ planes and will give no
contribution to the bulk. This will modify the Type IIB warp
factor equation to \eqn\warpnow{\quabla~\Delta^2 = {-2( |A|^2 +
|B|^2)} + \sum_{i,j}~\delta^2 (w - w^i)\delta^4 (z - z^j),} where
the operator $\quabla$ has been introduced in \beckerD\ and we
have ignored the contributions to the warp factor from gauge
bundles. Observe that \warpnow\ doesn't take the delocalization
along $w \equiv z^3$ into account, because this is derived
directly from \xeight. If we take this effect into account, the
solution of the Type IIB warp factor equation becomes
\eqn\solwarpp{ \Delta^2 = {\rm c}_o - {5~|z^1|^2 + |z^2|^2 \over
v^2} + \sum_{a,b}~{{\rm c}_1 \over |z^1 - a|^2 ~ |z^2 - b|^2},}
where $a,b$ are the fixed points on the base and ${\rm c}_1$ is
defined below. The constant ${\rm c}_0$ is the boundary value of
the warp factor, i.e it's value, when the size of the compact
manifold is very large. In other words, \eqn\czero{ {\rm c}_o =
{}^{\ol}_{a,b,v \to \infty}~ \Delta^2,} and the manifold therefore
becomes unwarped. Observe that the warp factor \solwarpp\ is well
behaved everywhere except at the orbifold points. This is because
we have taken ${\rm tr} (R \wedge R) = \sum_j ~\delta^4 (z -
z^j)$. For a resolved orbifold (i.e for a $K3 \times K3$
compactification) the warp factor will be well defined
everywhere. We can incorporate this by assuming \eqn\cone{{\rm
c}_1 \equiv {\rm f}(|z^1 - a|, |z^2 - b|) = 1,~~ {\rm and} ~~
{\rm f}(0,0) = 0,} where the first equality holds away from the
fixed points. Similar arguments can be given, when we include the
gauge fluxes in the warp factor equation, the form of which is
given in \beckerD. In fact the warp factor is well defined everywhere
because it is well defined on every {\it patches} of our manifold even
though
the two-form field $B$ is not globally defined \beckerD. This also resolves
another issue regarding the periodicity of the $z^i$ coordinates. The form
of the warp factor presented above in \solwarpp\ doesn't seem to
have the required periodicity. This is because we have defined the source of
the warp factor (the two form $B$) on patches.
Another point to note here is, that for a
large sized manifold, even though the warp factor vanishes, there
are still non-vanishing fluxes. The constant flux density goes to
zero (because $A, B \to 0$ for $v \to \infty$) but the localized
fluxes remain non-zero at the fixed points (which are shifted to
infinity). This is where we face a contradiction, because the
manifold seems to behave as a Calabi-Yau but there are still
fluxes left, which cannot be supported on a Calabi-Yau. The
resolution, as discussed in \rstrom, \beckerD, is to assume, that
these manifolds do not have a large radius limit. We'll discuss
more about this in section 4.4. Further details on the function
$f$ will be presented elsewhere.

\subsec{The Torsional Connection and Chern-Class}

Another interesting question is the choice of connection for the
compactifications considered herein. This is important because our
manifolds will be complex with a covariantly constant complex
structure $J^i_j$, i.e \eqn\compstr{\nabla_i J^j_k \equiv \del_i
J^j_k + \Omega_{il}^j~ J^l_k - \Omega^l_{ik}~ J^j_l = 0,} where
$\Omega^i_{jk}$ is the connection used. We cannot use the
Christoffel connection because, as we show below, this will lead
to a contradiction. The connection, that will be consistent for
us is the torsional connection defined as
\eqn\torconn{\Omega^i_{jk} = \Gamma^i_{jk} \pm {\cal T}^i_{jk},}
where $\Gamma^i_{jk}$ is the Christoffel connection and ${\cal
T}_{ijk}$ is the torsion. The ambiguity of the sign will be
clarified in the next sub-section. The torsion is related to the
background ${\cal H}$ by the following relation:
\eqn\torrel{{\cal T}_{ijk} = {\cal T}^l_{ij}~g_{lk} = {1\over
2}{\cal H}_{ijk}.} As discussed in detail in \HULL,\rstrom, and
\beckerD, the covariantly constant complex structure \compstr\
implies the following important relation between the ${\cal H}$
field and the metric (in terms of complex coordinates)
\eqn\torequn{ {\cal H}_{ab \bar c} = 2 g_{\bar c [a,b]},} which
is called the torsional constraint. In \beckerD\ we showed
explicitly, how the background \backgeom\ satisfies the equations
\torequn\ (and its complex conjugates). In fact, the derivation
of \torequn\ from \compstr\ requires the fact that the Nijenhaus
tensor \eqn\nijen{N_{mnp} = {\cal H}_{mnp} - 3 J^q_{[m} ~J^r_n
~{\cal H}_{p]qr},} vanishes. This can be easily shown, by taking
into account the dilatino supersymmetry equation for the
heterotic background. Inserting \torequn\ in \torrel, we can
determine the connection we should use. For example, with the
choice of minus sign in \torconn\ it is given by a simple relation
\eqn\bactorsion{\Omega^a_{bc} = g^{a \bar d}~g_{{\bar d}c, b} =
{\bar \Omega}^{\bar a}_{\bar b \bar c},} with all other
components zero. For our case the preferred connection will
include a plus sign in \torconn, as we shall see later on. One can
show, that for K\"ahler manifolds the above connection reduces to
the Christoffel connection\foot{In the next section (section 3),
where we discuss the mathematical properties of non-K\"ahler
manifolds, we will refer to the torsional connection simply as
$\omega$. This will symbolize the {\it preferred} connection for
our case. For the later sections we will distinguish between the
affine connection $\omega_i^{ab}$ (or the Christoffel connection
$\Gamma_{ij}^k$) and the torsion ${\cal H}_{ijk}$, unless
mentioned otherwise.}.

The above choice of the connection solves one of the apparent
puzzles related to the form of the ${\rm tr} R \wedge R $ term. As
explained in \rstrom, as a consequence of the anomaly cancellation
condition the fundamental form of a manifold with torsion has to
satisfy
\eqn\fulleqn{
i \partial \bar \partial J =  {\rm tr }\left( F \wedge F - R \wedge R
\right).
}
In general, $ {\rm tr} R \wedge R$ will contain a $(3,1)$ and
$(1,3)$ piece, if we use the Christoffel connection, giving in
most cases Calabi-Yau manifolds as backgrounds. By using the
connection given in \bactorsion, it is possible to show, that
${\rm tr} R \wedge R$ will always be a $(2,2)$-form and in this
case we expect to find a manifold with torsion as a solution.

Such a manifold with torsion will, in general, be non-K\"ahler.
As
has been explained in \HULL, if the vector
\eqn\vector{
v^m = J^{mn} J^{kl} {\cal H}_{nkl},
}
vanishes the manifold is semi-K\"ahler, while if ${\cal H}_{ijk}$
vanishes the manifold is K\"ahler. For the background geometry
\backgeom\ it is easy to see, that
\eqn\nonka{
v_m = \partial_m (\log \det g),
}
where $\det~g=\det g_{a\bar b}= {1\over 2} \Delta^4$. This
implies a non-vanishing value for $v^m$ and therefore the
manifold defined in \backgeom\ is not even semi-K\"ahler.

It is also easy to show, that the manifold \backgeom\ has a
vanishing first Chern class. Indeed the trace of the curvature
2-form is given by
\eqn\curvtwo{
{\cal R} = R_{mnkl}J^{kl} dx^m \wedge dx^n = i \partial \bar \partial
(\log \det g) = d( \partial -\bar \partial) \parallel \Omega\parallel^2,
}
since the norm of the holomorphic $(3,0)$-form is given in terms
of the warp factor. In order to get the above result, we have used
the connection defined in \bactorsion, instead of the Christoffel
connection. The first Chern class, given by $c_1 =\int {\cal R}$
vanishes, since $\Omega$ is globally defined. An alternative way
to see this would be to use \compstr. Contracting \compstr\ by
$J_m^n$  we get (in complex coordinates): \eqn\weget{J_{a}^b ~
{\tilde \nabla}_c J^c_b = {\cal H}_{~~a d}^{d} - {\cal H}_{~~a
\bar d}^{\bar d} = \del_a \phi,} where $\phi$ is the dilaton,
${\tilde \nabla}$ is the torsion free covariant derivative
 and the last equality comes from the supersymmetry
transformation of the dilatino, as explained in \beckerD. The
above expression is a total derivative.

\subsec{Sigma Model Description}

\noindent~(a)~ $\underline{\rm Basic~Concepts}$

\vskip.2in

Let us now discuss the sigma model description of our model. Most
of the details, which are well known, can be extracted from
\hetcit, \HULL, \rstrom\ and therefore we shall be brief. To the
lowest order in $\alpha'$, the light-cone-gauge action for the
heterotic string in the presence of a non-zero two-form potential
${B}$ and a gauge field $F = dA$ is given by
\eqn\hetsigma{\eqalign{S = & {1 \over 8 \pi \alpha'} \int~d\sigma
d\tau ~ [\del_+ X^i \del_- X^j(g_{ij} + { B}_{ij}) + iS^n (D_+
S)^n + i \Psi^A (D_- \Psi)^A\cr
 & + {1\over 2} F_{ijAB} \sigma^{ij}_{mn} S^m S^n \Psi^A \Psi^B +
{\cal O}(\alpha')],}} where $X^{\mu} = (X^+, X^-, X^i), ~(i =
1,..., 8)$ are the bosonic fields and $S^m$ with $m=1,\dots, 8$,
describe spinors in the vector representation of $SO(8)$, which
are the superpartners of $X^i$. The remaining degrees of freedom
are the anti-commuting world-sheet spinors $\Psi^A, ~A = (1,...,
32)$, which transform as scalars under $SO(8)$. On-shell we also
impose \eqn\onshell{ \del_+ S^m = 0 = \del_- \Psi^A,} so that
$S^m$ are right moving, while $\Psi^A$ are left moving. If we
denote the Yang-Mills field representations by $T^q_{AB}$, where
$q$ labels the adjoint of the gauge group, then $F_{ijAB} =
F^q_{ij} T^q_{AB}$ (with a similar notation for the corresponding
one-form potential $A_i$), while the anti-symmetrized product of
gamma matrices is written as $\sigma^{ij}_{mn}$. Finally, the
covariant derivatives appearing in \hetsigma\ are given by
\eqn\covderiv{\eqalign{& (D_- \Psi)^A = \del_- \Psi^A +
A_i^{AB}~(\del_- X^i) \Psi^B, \cr &(D_+ S)^m = \del_+ S^m +
{1\over 2} (\omega_i^{ab} - {\cal T}_i^{ab})
\sigma^{mn}_{ab}(\del_+ X^i)S^n.}} Observe the way the spin
connection $\omega_i^{ab}$ and the torsion ${\cal T}_{ijk}$
appear in the covariant derivative of $S^m$. If $e^a_i(X)$
represent some orthonormal frame, such that $e^a_ie^b_j
\delta_{ab} = g_{ij}$, then the spin connection satisfies the
torsion {\it free} equation \eqn\torfree{ de^a + \omega^{ab}
\wedge e^b = 0.} If we denote the local gauge rotations by
$\alpha^{AB}$ (under which $\Psi^A$ rotates) and the local
$SO(8)$ Lorentz-rotations by $\beta^{mn}$ (under which $S^m$
transforms), then in general these symmetries are anomalous. As is
well known, these anomalies cancel, if we impose the constraint
\condH\ on the ${\cal H}$ field. In fact the $B$ field has to
transform in the following way \hullwitten\
 \eqn\bfieldtr{\delta B_{ij} =
\alpha'[\gamma_{[i}\cdot \del_{j]}\beta -  A_{[i}\cdot
\del_{j]}\alpha],} to cancel the anomalies. Here $\gamma$ is used
to define the curvature $R$ as $R = d\gamma + \gamma \wedge
\gamma$. In the next sub-section we will show how this is related
to the usual spin connection. We will also discuss later how to
actually realize this from our ${\cal M}$-theory set-up.

The background metric and the $B$ field of the particular example
constructed in \beckerD\ are already given in \backgeom. In this
case the unwarped metric is flat ${\tilde g}_{a \bar b} = \eta_{a
\bar b}$. Using this we can calculate the inverse of the complete
metric, $g^{a \bar b}$, and obtain
\eqn\invmet{ g^{a \bar b} = {1\over  \Delta^2} \pmatrix{ {1\over
2}&0& -B \bar z^2\cr \noalign{\vskip -0.20 cm}  \cr 0&{1\over
2}&  A\bar  z^1 \cr \noalign{\vskip -0.20 cm}  \cr - \bar B  z^2
&  {\bar A}   z^1 & 2(\Delta^2 + |A z^1|^2+ |B z^2|^2)},}
where $\Delta$ is as usual the warp factor. Taking the inverse of
the metric and the expression for ${\cal H}$ given in \backgeom,
one can easily confirm \eqn\confirmwhat{{\cal H}_{\bar a b \bar
c}g^{b \bar c} = \del_{\bar a} \phi.} This in turn is consistent
with the fact that we have a vanishing first Chern-class, as
discussed in the previous sub-section.

We should point out, that because of \bfieldtr, which leads to
$d{\cal H} \ne 0$ we cannot write the metric in terms of any {\it
local} potential $K$ because \eqn\locpot{ g_{a[\bar b, \bar c]d}
- g_{d[\bar b, \bar c] a} = \alpha' f(A, \gamma) ~~~ \Rightarrow
~~~ J = i(\del \bar K - \bar \del K) + {\cal O}(\alpha').} Here
$f(A, \gamma)$ is a function, that can be determined from
\bfieldtr. Therefore, the internal manifold is not K\"ahler.
However, if $d{\cal H}=0$, we can have a local potential in terms
of which we can write the metric, as the function $f(A,\g)$ will
be vanishing in this case. But, as briefly mentioned
 in \beckerD, this will make the warp factor constant and therefore
there will be no warped solution. A way to see this (at leading
order in the $\alpha'$ expansion) is as follows. If we take the
torsional connection into account, the supersymmetry variation of
the gravitino can be written concisely as
\eqn\susyvar{ {\cal D}_m \epsilon  =
(\del\!\!\!/ \phi -{1\over 6}{\cal H}) \epsilon=0.}
After a few simple but cumbersome manipulations \susyvar\
gives rise to the following condition \SmitD
\eqn\become{
{4\over 3}\Gamma^{mnpq}
{\tilde \nabla}_m {\cal H}_{npq}\epsilon  =
\left[R + {16\over 3}{\cal H}^2_{mnp} +
20 (\del_m \phi)^2 +6 \partial^m \partial_m \phi
\right] \epsilon, }
where ${\tilde \nabla}$ is the torsion free covariant derivative.
Embedding the spin connection into the gauge connection makes the
left hand side of \become\ to vanish. The right hand side of
\become\ then takes the form
\eqn\becomea{
R + {16\over 3}{\cal H}^2_{mnp} +
20 (\del_m \phi)^2 +6 \partial^m \partial_m \phi
 =0,
} which is consistent with the fact, that supersymmetry imposes
the classical equation of motion (assuming, of course, that the
transformation laws are on-shell). The contracted Einstein
equation and the equation for the three-form field ${\cal H}$
come out directly from the supersymmetry transformations. In the
absence of sources and singularities this equation can be
integrated over the internal manifold giving the condition
\eqn\becomeb{
\int d^6 x \sqrt{g}\left[ R + {16\over 3}{\cal H}^2_{mnp} +14
(\del_m \phi)^2 \right]=0.}
If supersymmetry is unbroken, all the terms in the above sum are
positive and therefore embedding the spin connection into the
gauge connection implies (at this order in the $\alpha'$ expansion)
\eqn\zero{
{\cal H}_{mnp}=0 \qquad {\rm and} \qquad \phi={\rm const}.
}
It then follows from \zero\ that
\eqn\nonwarp{ {\tilde \nabla}_m \epsilon = 0,}
which implies a Ricci-flat K\"ahler metric as our solution. An
alternative proof based on the non-existence of a holomorphic
$(3,0)$ form, when we make the assumption $d{\cal H} = 0$ is
given in \PapaDI.

\vskip.2in

\noindent~(b)~ $\underline{\rm Anomaly~Cancellation}$

\vskip.2in

Something discussed above in \zero\ and \nonwarp\ may seem a
little puzzling. Let us go back again to \covderiv\ and make the
following identifications \eqn\folidentif{\eqalign{& A_i^{ab}~
\leftrightarrow ~ (\omega_i^{ab} - {\cal T}_i^{ab}), \cr & \Psi^A
~\leftrightarrow ~S^{\dot p},~~A = 1,..., 8,}} where the rest of
the $\Psi^A$ for $A = 9,...., 32$ do not couple to the background
fields and $\dot p$ labels the components of a spinor in the
$8_c$ representation of $SO(8)$. In fact, the above
transformations can be used to rewrite our heterotic string sigma
model lagrangian \hetsigma\ with {\it all} the left-movers
replaced in the following form \eqn\leftrepl{\int~d\sigma
d\tau~{\Big [}iS^{\dot p}(D_-S)^{\dot p} + {1\over 4} R_{ijkl}
\sigma^{ij}_{\dot p\dot q} \sigma^{kl}_{mn} S^{\dot p}S^{\dot q}
S^m S^n{\Big ]}.} In this form the above lagrangian resembles the
Green-Schwarz superstring moving in a curved background. This
theory is anomaly free, which would imply that it doesn't receive
any Chern-Simons corrections. In the above action the curvature
tensor $R_{ijkl}$ is defined with respect to the connection
\eqn\conncovde{{\tilde \omega}_i^{ab}= \omega_i^{ab} - {\cal
T}_i^{ab} \equiv \omega_i^{ab} - {1\over 2} {\cal H}_i^{ab},}
and
contractions are done with $e^a_i$. As discussed by
 \HULL, there would, in general, be higher order $\alpha'$ corrections to
\hetsigma,
which in fact vanish, if we embed the gauge field in the
torsional-spin connection ${\tilde \omega}_i^{ab}$, as done above.
In other words, we identify $\gamma$ in \bfieldtr\ as
\eqn\lambu{\gamma_i^{ab} = \omega_i^{ab} - {1\over 2} {\cal
H}_i^{ab}.} The three-form field strength, that is invariant
under the transformation \hullwitten\ is \eqn\threeform{{\cal H}
= dB - \alpha'{\Big [}\Omega_3(A) -\Omega_3(\omega - {1\over 2}{\cal
H}){\Big ]},} where we have defined \eqn\omedef{\Omega_3(A) = A \wedge F
- {1\over 3} A \wedge A \wedge A,} and $\Omega_3(\omega - {1\over
2}{\cal H})$ is given by a similar equation with $A$ replaced by
$\omega - {1\over 2}{\cal H}$. The above ${\cal H}$ field
satisfies the Bianchi identity \condH. At this point it seems from
the analysis done in the previous sub-section, that we would not
have a warped solution because $d{\cal H} = 0$, once we embed the
gauge connection into the torsional-spin connection. However, we
can assume that our identification $\gamma = A = \tilde \omega$ is
only to the lowest order in $\alpha'$, as we do not want to embed
the gauge connection into the torsional-spin connection. This
would mean, that there would be corrections to the three-form
equation \threeform, implying that the sigma-model action would
also receive corrections\foot{Note that on the right hand side of \threeform\
the $\omega$ and ${\cal H}$ are actually one-forms as in \lambu,
whereas the left hand side ${\cal H}$ is a three-form. Since $\Omega_3$
eventually makes a three-form, we hope that this will not confuse
the readers.}. Observe, that all these corrections are of order
${\Big (}{t^2 \over 4 \pi \alpha'}{\Big )}^{-1}$, where $t$ is
the radius of our manifold. Therefore, they are small, if the
radius is big. Thus we will assume \eqn\weasfor{A = \tilde\omega
+ {\cal O}(\alpha'),} which leads to $d{\cal H}\ne 0$. In the
following we shall use \threeform\ to study the constraints on
the size of our six-manifold. Details on this will be addressed
in section 4.

However, there is still an ambiguity in defining the 
connection\foot{The Chern-Simons term in \threeform\ arises from 
anomaly cancellation condition and there can, in principle, be {\it any}
connection to cancel the anomalies. The difference, however, is that, by 
choosing another connection we have to introduce set of counterterms in the
theory. Therefore we will stick to the case when the connection in 
\threeform\ is \lambu\ and elsewhere it is given by (2.50). This aspect have
been discussed, for example, in \hullambig. We thank Chris Hull for 
correspondences on this issue.}.
This appears explicitly, when we demand, that our sigma-model
action \hetsigma\ to be invariant under world-sheet supersymmetry
transformations. If the supersymmetry transformation parameter is
$\epsilon^p$, then we require ${\hat D}_+ \epsilon^p = 0$, where
${\hat D}_+$ is the same operator as in the second equation of
\covderiv, but now defined with respect to the connection
$\hat\omega$ instead of $\tilde\omega$, where
\eqn\hatome{{\hat\omega}_i^{ab} \equiv {\omega}_i^{ab} + {1\over
2}{\cal H}_i^{ab}.} In fact this connection is more relevant for
deriving the torsional constraints of our model\foot{In \rstrom\
the connection used was in fact $\omega - {\cal H}$, which differs
from \hatome\ by a relative sign and also a relative factor of one
half. Thus, the torsional equation of \rstrom\ has an extra minus
sign and a missing factor of 2, as was also pointed out in
\beckerD. However, since the torsion used there was $-{\cal T}$
this discrepancy doesn't alter any results.}. Therefore, this
seems to leave us with the ambiguity in calculating the curvature
tensor appearing in \condH. However, observing \threeform\ we see
that we can, in fact, shift this ambiguity into a redefinition of
the $B$ field and keep the curvature tensor appearing in \condH\
unambiguous. Transforming $B$ to \eqn\tranbto{ B ~ \to ~ B -
{\alpha'\over 2}~{\omega} \wedge {{\cal H}},} where the one-forms
${\omega}$ and ${{\cal H}}$ are created from the corresponding
three-forms by contracting with the vielbeins $e^{ai}e_a^{j}$,
the curvature tensor is now defined with respect to $\omega$. Therefore, the
obstruction \eqn\obstr{ \int~{\Big [}{\rm tr}~(R \wedge R)-{1\over
30}{\rm Tr}~(F \wedge F){\Big ]} = 0,} is defined with respect to
$\omega$ and is independent of the choice of connection. In
passing, note that in \threeform\ even though we have shifted the
ambiguity into a redefinition of $B$, the three-form ${\cal H}$
still appears on both sides of \threeform.

Finally, observe that even for a non-zero $d{\cal H}$, i.e $A \ne
{\tilde\omega}$, as long as we have the Killing spinor equation
${\hat D}_+ \epsilon^p = 0$, the two loop beta function (at least
for the metric and to the lowest order in $\alpha'$) is trivial.
As argued in \witteno, this may still be true to higher orders in
$\alpha'$. However, this argument relies on having a non-K\"ahler
manifold, which is a {\it deformation} of the usual Calabi-Yau
space and as such supports torsion, which is ${\cal O}(\alpha')$.
Also as argued in \xenwit, the conformal invariance of these
backgrounds may be spoiled by non-perturbative effects like
world-sheet instantons. For our case, the torsion is of order 1
and the backgrounds receive ${\cal O}(\alpha')$ corrections. This
case is different from \witteno\ and even though we might have
one-loop finiteness the fact that we demand a vanishing beta
function for our case is a little subtle because of the size
constraints of our six-manifold \DineSB. In particular, it can be
shown that the corrections to the two loop beta function are
suppressed, as long as the size of the manifold is sufficiently
large. More details on this and whether we can have a large sized
internal manifold will be addressed in section 4.

\subsec{The Anomaly Relation in the Heterotic Theory}

Until now we haven't taken the localized fluxes in \gflux\ too
seriously. In the ${\cal M}$-theory setup the localized part of
the $G$-flux is of the form \eqn\loc{ {G \over 2 \pi} = \sum_{i =
1}^4 ~ F^i(z^1, z^2, \bar z^1, \bar z^2) \wedge \Omega^i(z^3,
z^4, \bar z^3, \bar z^4),} where the supersymmetry constraints on
$G$ imply the primitivity condition $g^{a \bar b}F^i_{a \bar b} =
0$ on the gauge fields. In \loc\ we are summing over four fixed
points, where at every fixed point there are four singularities.
The two forms $\Omega^i_{a \bar b}$ are defined on the first
$T^4/{\cal I}_4$ in ${\cal M}$-theory, while $F^i_{a \bar b}$ are
located on the second $T^4/{\cal I}_4$, around which the $D7$
branes and $O7$ planes wrap. As discussed in \beckerD, these
localized fluxes are responsible for the gauge fields on the
seven-branes on the Type IIB side. We are ignoring the constant
fluxes for the time being.

The localized forms $\Omega^i$ are given by the $16$ non-trivial
($1,1$) forms on $T^4/{\cal I}_4$. Let us concentrate to a region
near one of the singularities. This space will look like a
Taub-NUT (TN) space and the (1,1) form is the unique (1,1) form
of the TN space. This space has one radial coordinate $r$ and
three angular coordinates $(\theta, \varphi, \psi)$ with
identification \eqn\iden{(r, \theta, \varphi, \psi) ~~ \equiv ~~
(r, \pi - \theta, \pi - \varphi, - \psi).} The metric on this
space is very well known. There is one singular point at $r = 0$,
where the circle parametrized by $\psi$ shrinks to zero. This
circle is non-trivially fibred over the base $(r, \theta,
\varphi)$. The anti-self-dual harmonic form on this space can be
written in terms of real coordinates as \harmoni
\eqn\harmonic{\Omega = {\rm f}_o~e^{-{\rm f}_1(r)}{(} d\sigma -
{\rm f}_2(r)~ dr \wedge \sigma {)},} where, as discussed in
\harmoni, at large distances, i.e $r \to \infty$, ${\rm f}_1(r) =
{1\over 2} r,~~ {\rm f}_2(r) = {1\over 2}$, while ${\rm f}_o$ is a
constant. Hence $\Omega$ is normalizable. The one-from $\sigma$
is defined as \eqn\defsigma{ \sigma = - sin~\psi~d\theta + cos ~
\psi~sin ~\theta~d\varphi.} Inserting this in \loc\ we
immediately face a problem. {}From the form of \harmonic\ we see,
that $\Omega$ depends on $z^i, {\bar z}^j, ~~i,j = 3,4$. But to
go to the Type-I theory (or heterotic) we require it to be
independent of these coordinates! Therefore, we have to study
this space locally, near $r = 0$ and ignore all other fixed
points. This in particular means, that we can only trace the
behavior of {\it one} gauge field in the heterotic theory.

Near $r \to 0$, $\Omega$ is well defined and the behavior can be
written explicitly as \harmoni \eqn\behavewell{\Omega = {\rm
f}_o~ {\Big (}d \sigma_o - {2\over \pi}~dr \wedge \sigma_o{\Big
)}.} The above form is near the so-called bolt singularity. The
one form $\sigma_o$ is now different and is given by
\eqn\diffoneform{ \sigma_o =  d\psi_o + cos~\theta_o~d\varphi_o,}
where ($\psi_o, \theta_o, \varphi_o$) are the appropriate
coordinates near the bolt \harmoni. The above is basically a
constant form, if we ignore the angular dependences. Parametrizing
appropriately, we can in fact write \behavewell\ as a constant
($1,1$) form \eqn\consform{\Omega = a~dz^3 \wedge d{\bar z}^4 +
b~d{\bar z}^3 \wedge dz^4,} where $a,b$ are constants, that can be
determined from \behavewell. Plugging this in \loc, and using the
fact that $dz^4 = dx^{10} + {\rm i} dx^{11}$, we can determine the
three form fields $H \equiv H_{NS}$ and $H' = H_{RR}$ in the Type
IIB theory as \eqn\handhprime{\eqalign{ & H = a~F_{1 \bar 2} dz^1
\wedge d{\bar z}^2 \wedge dz^3 + b~ F_{\bar 1 2} d{\bar z}^1
\wedge dz^2 \wedge d{\bar z}^3, \cr & H' = -{i}a~F_{1 \bar 2} dz^1
\wedge d{\bar z}^2 \wedge dz^3 + {i}b~ F_{\bar 1 2} d{\bar z}^1
\wedge dz^2 \wedge d{\bar z}^3,}} where we have ignored the
contribution from the constant fluxes. One can view the above as
if we had taken a large sized manifold, where the flux densities
\valueAB\ are essentially zero. The resulting Type IIB three-form
potential is\foot{We are choosing the axion-dilaton $\varphi =
i$, as this is still the solution, even in the presence of
localized fluxes. This will be clear soon, when we show, that the
torsional constraints are satisfied for this case too.}
\eqn\fluxverfy{G_3 = -2ia~F_{1\bar 2}~ dz^1 \wedge d{\bar z}^2
\wedge dz^3.} Now as discussed in \beckerD, to go to the Type I
theory we have to make two T-dualities along the toroidal
directions. Even with the choice of extra localized fields
\handhprime, the only non-vanishing components of ${\cal H}$ have
one leg along the $z_3$ or $\bar z_3$ direction. Indeed, it is
easy to verify that
\eqn\bmncomp{{\cal B}_{mn} = 6 B_{[8m}B'_{n9]} = 0,\qquad {\rm
for} \qquad m,n=4,\dots,7.}
The equation of motion of the six-form potential $C^{(6)}$ of the
Type I theory will lead to the Bianchi identity of the heterotic
string as follows. First, observe that
\eqn\hhprimeg{H \wedge H' = 2ab~F\wedge F~\wedge dx^8 \wedge dx^9,}
where we have defined $F$ as a ($1,1$) form, such that $g^{a\bar
b}F_{a \bar b}= 0$ and $F_{ab} = F_{\bar a \bar b} = 0$. These are
basically the Donaldson-Uhlenbeck-Yau (DUY) equations for the
gauge bundle. The explicit form of $F$ will be discussed in
\toappear. For the time being we take $F$ to be of the form
\eqn\fdefin{ F = F_{1\bar 2}~dz^1 \wedge d{\bar z}^2 +
F_{\bar 1 2}~d{\bar z}^1 \wedge dz^2.}
Inserting \hhprimeg\ into the Type IIB Chern-Simons coupling $D^+
\wedge H \wedge H'$, where $D^{+}$ is the four-form potential,
results in an interaction on the world-volume of the D7-brane,
since the part of $H \wedge H'$ containing the Yang-Mills fields
is localized. After T-duality, the Chern-Simons interaction maps
to the $C^{(6)}\wedge F \wedge F$ interaction of the Type I
theory. On the other hand the non-localized piece of $H \wedge
H'$ gives rise to the kinetic term for $C^{(6)}$ in the bulk. A
quick way to see this would be as follows. Let us denote the
coordinates of the fiber $T^2$ as 8,9; $K3$ as $X, Y, Z, W$ and
Minkowski as $\mu, \nu, \rho, \sigma$. Thus in our notation
\eqn\spcadim{ {\cal M}^{10}_{[A,B,...]} = K3_{[X,Y,Z,W]} \times
{T^2_{[8,9]}\over \IZ_2} \times \IR^4_{[\mu,\nu,\rho,\sigma]}.}
All the coordinates are real and $A, B, C, ...$ run over full ten
dimensional indices. Also the non zero components of $B$  and
$B'$ are of the form $(i, X)$ where $i = 8,9$.

Consider the interaction $D^+ \wedge H\wedge H'$. After
integrating by parts we can write one of the contributions as
\eqn\oneterm{\int_{{\cal M}^{10}} ~dD^+_{W\mu\nu\rho\sigma}~
B_{8X}~ H'_{9YZ}.} This interaction results from the kinetic term
of $C^{(6)}$ in the Type I theory \eqn\Iterm{\eqalign{\int ~ {\cal
H}^I\wedge *{\cal H}^I & = \int ~ {\cal H}^I_{ABC}{\cal
H}^I_{A'B'C'} ~~ \epsilon^{A'B'C'}_{~~~~~~~~DEFGHIJ}\cr & =\int~
g^I_{DD'} g^I_{EE'}~ {\cal H}^I_{ABC} ~ {\cal H}^I_{A'B'C'} ~~
\epsilon^{A'B'C'D'E'}_{~~~~~~~~~~~~~FGHIJ},}} by using two
T-dualities. We will use the fact, that the ${\cal H}^I$ field has
one component along the $T^2$ direction (here $I$ denotes Type I
components). For definiteness, consider the component
\eqn\compI{\int~g^I_{8X} g^I_{99}~ {\cal H}^I_{8YZ} ~ {\cal
H}^I_{X'Y'Z'}~~ \epsilon^{89X'Y'Z'}_{~~~~~~~~~~~Wabcd}.} We can
use the T-duality rules given in \mesO\ taking into account, that
the $\epsilon$ symbol of Type I is not the same as the $\epsilon$
symbol of Type IIB. In particular, the volume of $T^2$ has been
inverted, and there are additional terms involving $B$ fields.
But we can write
\eqn\epsilonIIB{\epsilon^{(I)89X'Y'Z'}_{~~~~~~~~~~~~~Wabcd} =
g_{88}~ g_{99}\epsilon^{89X'Y'Z'}_{~~~~~~~~~~~Wabcd} + .....}
where $...$ are the terms involving $B, B'$. Inserting
\epsilonIIB\ into \compI\ and using the T-duality rules and the
fact that the five-form is not closed, we get \eqn\termIIB{\int ~
dD^+_{Wabcd}~ B_{8X}~ H'_{9YZ} +....} modulo signs and numerical
factors. This is what we wanted to show.

Taking the previous interactions into account, the equation of
motion for $C^{(6)}$ after U-dualities will be
\eqn\anomone{d{\cal H} = - \alpha' F \wedge F,}
where ${\cal H} = \ast dC^{(6)}$ and we have normalized $a,b$
appropriately. The $\alpha'$ factor appearing above comes from the
seven-brane. For the full non-abelian case \anomone\ will take
the form ${\rm tr}(F \wedge F)$. In deriving \anomone\ we have
used the orthogonality property of the ($1,1$) forms\foot{This is
easy to work out. Here we simply sketch the details: Let us
denote the twenty ($1,1$) forms on $T^4/{\cal I}_4$ as
$\Omega^i_{(1,1)}, i = 1,...,h^{1,1}$ (replacing the orbifold by
a smooth $K3$, we have to interpret the singularities as the
points where the fiber torus degenerates on the base. As it is
known, there is no simple way to write the metric or the harmonic
forms for $K3$, and therefore the discussion for smooth $K3$ will
be more involved. In this paper we will stick to the orbifold
limit only). The G-flux in ${\cal M}$-theory can be written as :
$${G\over 2\pi} = \sum_{i=1}^4~ A_{ij}~\Omega_{(1,1)}^i(z^k, \bar
z^k)\wedge \Omega_{(1,1)}^j(z^l, \bar z^l) +
\sum_{p=1}^{h^{1,1}-4}~F^p(z^k, \bar z^k) \wedge
\Omega_{(1,1)}^p(z^l, \bar z^l),$$ where $A_{ij}$ are constant
numbers, $k = 1,2$ and $l = 3,4$. In the above equation the first
sum is over the constant (1,1) forms and the other sum is over
$16$ localized forms. We have ignored the other two (2,0) and
(0,2) forms (which are also constant forms). Using now the
orthogonality property:
$$\int_{T^4/{\cal I}_4}~\Omega_{(1,1)}^p(z^l, \bar
z^l) \wedge \Omega_{(1,1)}^q(z^l, \bar z^l) = 16 \pi^2
\delta^{pq},$$ we can confirm the terms on the branes and the
bulk. In deriving this, we have ignored the mass parameter of the
TN space. These details do not alter any of the generic
discussion, that we give here. Restoring it will simply add one
more (determinable) parameter in the theory. This issue has been
discussed earlier in \beckerD.}.

Observe that in our calculation we couldn't reproduce the complete
anomaly relation for the heterotic string. This is because we
considered the supergravity approximation, where we can see the
contributions from the gauge bundle but not the contributions from
the spin bundle. To get the complete anomaly, we will lift our
solution to ${\cal M}$-theory, where the gravitational part of
the anomaly is well known. To do so, let us consider the
following chain of dualities \rDJM \eqn\chaindual{{\cal M}~{\rm
on}~{T^4\over {\cal I}_4}~~~\to~~~ {\rm IIB~on}~ {T^2\over
\Omega\cdot (-1)^{F_L}\cdot {\cal I}_2}~~~ \to~~~ {\rm
IIA~on}~{T^3 \over \Omega\cdot (-1)^{F_L}\cdot {\cal I}_3}.}
{}From the above chain we see, that if in the Type IIB theory we
have sixteen $D7$-branes plus four $O7$-planes, then we get in the
Type IIA theory sixteen $D6$-branes plus eight $O6$-planes. From
the Type IIA point of view every $D6$ branes lifts up in ${\cal
M}$-theory to a Taub-NUT space and every $O6$ plane lifts as an
Atiyah-Hitchin space satisfying \rDJM \eqn\tnahit{\int_{{\rm
Taub-NUT}}~p_1 = \int_{{\rm Atiyah-Hitchin}}~p_1 = {1\over
24}\int_{T^4/{\cal I}_4}~ p_1,} where $p_1$ is the first
Pontryagin class of the spin bundle. The above relation directly
translates to the Type I case by doing a series of U-duality
transformations. The gravitational terms are in fact related to
$\int_{T^4/{\cal I}_4}~X_8$ in such a way, that the heterotic
anomaly equation is \eqn\hetander{d{\cal H} = {\alpha'\over
2}[p_1(R) - p_1(F)],} where $R = d\gamma + \gamma \wedge \gamma$,
$F = dA + A \wedge A$ as defined earlier. As previously
emphasized in \beckerD, the two terms in \hetander\ have
different origins in ${\cal M}$-theory.

Finally, we should address the important question of torsional
constraints in the presence of \handhprime. Due to the
anti-symmetrization in \handhprime, we can write the NS-NS $B$
field as \eqn\bnsns{B = a~A_{\bar 2}~d\bar z^2 \wedge dz^3 +
b~A_2~dz^2 \wedge d\bar z^3.} This in particular will affect the
heterotic metric because under U-duality NS-NS B-fields contribute
to the metric. If we {\it ignore} the constant fluxes and take
only the localized fluxes into account, then the heterotic metric
will become \eqn\metinhet{ g_{a \bar b} = \pmatrix{ \Delta^2 &0&
0\cr \noalign{\vskip -0.20 cm}  \cr 0& \Delta^2 + A_2 A_{\bar 2}&
-{A_2\over 2} \cr \noalign{\vskip -0.20 cm}  \cr 0 & -{A_{\bar 2}
\over 2} & {1 \over 4}},} where we have ignored an overall factor
of 2 and also assumed that the Type IIB coupling is $g_B = 1$, as
this is a consistent solution \beckerD. For our case it is easy
to see, that all the torsional constraints are satisfied. The
only non-trivial relation is \eqn\trorelnont{{\cal H}_{2 1 \bar
2} = 2~g_{\bar 2 [1,2]}.} As shown earlier in \bmncomp, ${\cal
H}_{2 1 \bar 2} = 0$ and therefore, the metric derivative implies
the following linear equation for the warp factor \eqn\warpline{
{\del \Delta^2 \over \del z^1} + A_2~\del_1 A_{\bar 2} = 0.} This
is precisely the expected dependence because, from the Type IIB
point of view, the five-form equation is \eqn\fivefor{ F_5 =
{1\over 2} (B \wedge H' - B' \wedge H) + {\rm sources},} where
$B, H'$ are the contributions from the {\it constant} fluxes
(which we are ignoring for the time being) and the {\it sources}
contribute precisely $A \wedge F$. The above derivation shows,
that the torsional constraints are satisfied to the lowest order
in $\alpha'$.

\subsec{Enhanced Gauge Symmetry}

In \beckerD\ it was argued, that the full non-abelian symmetry in
our model can actually appear from wrapped membranes on
two-cycles of the ${\cal M}$-theory four-fold. Here we shall
provide some details generalizing earlier work of \senM. This
construction relies basically on the duality chain \chaindual. We
are also viewing the orbifold $T^4/{\cal I}_4$ as a $T^2/{\cal
I}_2$ fibration over every fixed point of $T^2/{\cal I}_2$ and a
$T^2$ fibration elsewhere. The metric can be written as
\eqn\mmet{ ds^2 = \Delta^{2\over 3}~(g_{3\bar 3}|dz^3|^2 +
g_{4\bar 4}|dz^4|^2),} where $\Delta$ is the warp factor. In this
space it is easy to construct a two-sphere as a cylinder shrunk at
two points. A membrane wrapped on this sphere will have a mass
given by \eqn\memmass{m_{ij} = {\alpha}~\int_{~r_i}^{~r_j}~
\Delta^{4\over 3} d\psi ~dr,} where $\psi, r$ are defined in
\iden. We denote the two points at which the cylinder shrinks as
$r_i, r_j$, and ${\alpha}$ takes into account the dimensionful
parameters. The above relation is not the full story. As argued
in \senM, there are additional two cycles, which contribute states
with masses: \eqn\massma{n_{ij} =
{\beta}~\int_{~r_i}^{-r_j}~\Delta^{4\over 3} d\psi ~dr,} which, in
the Type IIB theory will be interpreted as states appearing due
to the orientifold reflection. To see what kind of algebra we
generate from above, we have to study the {\it intersection
matrix}. This is constructed by considering the possible number of
points where two spheres can intersect. For our case, when we
concentrate near the region, where we have $T^2/{\cal I}_2$ as our
fiber, the intersection matrix ${\cal I}$ is easily shown to be
\eqn\interse{{\cal I} = \pmatrix{ 2&-1&0& 0\cr \noalign{\vskip
-0.20 cm}  \cr -1& 2 & -1&0 \cr \noalign{\vskip -0.20 cm}  \cr
0&-1& 2 &  -1\cr \noalign{\vskip -0.20 cm}  \cr 0 & -1& 0 & 2}.}
This is the Cartan matrix for the $D_4$ algebra and therefore
globally we have a $D^4_4$ algebra and the enhanced gauge symmetry
is realized when \eqn\mn{m_{ij} = n_{ij} = 0, ~~~\forall ~i,j}
and from the symmetry of the system we also have ${\alpha} =
{\beta}$. In the language of Type IIB theory this is the case,
when the $D7$ branes lie on top of an $O7$ planes and therefore
the masses of the strings are zero.

\newsec{Topological Properties of Non-K\"ahler Manifolds}

In this section we will study topological properties of Non-K\"ahler
manifolds.
We shall discuss various algebraic geometric properties of these manifolds
and
give a generic construction of a large class of them which are compact
and also complex along with their complete Betti numbers etc. The basic
algebraic geometric techniques used here can be extracted from \grifhar.
Some
of the details presented in section 3.2 are also obtained by \GP.

 We begin with a brief description of some of the lesser
known properties of these manifolds.

\subsec{Basic Features of Non-K\"ahler Manifolds}

\vskip.2in

\noindent (a)~$\underline{\rm
Failure~of~identities~on~Hodge~numbers }$

\vskip.2in

K\"ahler manifolds have a great deal of structure that general complex
manifolds have not, although the Hodge numbers $h^{pq}$ are still defined
for general compact complex manifolds. Here are some properties of Hodge
 numbers for
K\"ahler manifolds that do not, in general, hold for compact complex
manifolds.

The Hodge numbers do not completely determine the Betti numbers,the
topological numbers,
which characterize a general (real or complex)
manifold $M$.
The $p$-th Betti number $b_p$ is
the dimension of the $p$-th De Rham cohomology group $H^P(M)$, which
depends only on the topology of $M$. By the Hodge theorem $b_p$ counts
the number of harmonic $p$-forms ${\omega}_p$ on $M$
\eqn\harf{{\Delta}_d{\omega}_p=0,}
with ${\Delta}=*d*d+d*d*$, where $d$ is the exterior derivative as usual.

For a complex manifold one introduces the Dolbeault cohomology (or
$\bar \del $-cohomology) $H^{p,q}_{\bar \del}(M)$, whose
dimension is the Hodge number $h^{pq}$. The number of harmonic
$(p,q)$ \eqn\harft{{\Delta}_{\bar \del}{\omega}_{p,q}=0,} which
is (in general) not a topological invariant and is determined in
terms of the Hodge numbers. Here one defines the operator
\eqn\defoper{{\Delta}_{\bar \del}={\bar \del}{\bar
\del}^{\dagger}+ {\bar \del}^{\dagger}{\bar \del}.} Only in the
case of K\"ahler manifolds do the Laplacians agree
${\Delta}_d=2{\Delta}_{\bar \del}$ and the Betti numbers become
sums of Hodge numbers. In the non-K\"ahler case we still have the
inequality
$${\sum_{p+q=n}}h^{pq}\geq b_n,$$
but equality does not hold in general.
Also the Hodge numbers are not in general symmetric; one
may not assume that $h^{pq}=h^{qp}$.

Here is another important property of K\"ahler manifolds that can
fail for general compact complex manifolds. If $N$ is a compact
complex submanifold of $M$, then a K\"ahler metric on $M$ restricts
to one on $N$. If $\omega$ is the K\"ahler class, and $N$ has complex
dimension $n$, it follows that $\int_N\omega^n>0$. If $M$ is not
K\"ahler, it is possible that there is no closed $2n$ form on
$M$ whose integral over $N$ does not vanish.

\vskip.2in

\noindent (b)~$\underline{\rm
\bar\del-cohomology~and~the~Frolicher~spectral ~sequence}$

\vskip.2in

For any complex manifold $M$ and any holomorphic bundle $E$ over $M$,
${\bar \del}$ is well defined as an operator on sections of $E$ and,
more generally, $\{0,q\}$ forms with values in $E$. This gives a complex
whose cohomology is called the ${\bar \del}$-cohomology with values in $E$.
If we specialize to the case where $E$ is the bundle of holomorphic
$\{p,0\}$
forms, then the $h^{pq}$ is defined to be the rank of the ${\bar \del}$-
cohomology. This is the meaning of the Hodge numbers for a general complex
manifold. What is true in the K\"ahler case but fails in the general case is
that every ${\bar \del}$-cohomology  class has a representative cycle that
is $\partial$-closed as well. Instead, $\partial$ operates on the
${\bar \del}$-cohomology, generating a new complex, and only the cohomology
of this complex contributes to the cohomology of $M$. This is the first
stage
of what is called the Fr\"olicher spectral sequence.  In principle, there
could be even further cancellations, but nobody seems to know of a case
where this actually takes place.

\vskip.2in

\noindent (c)~$\underline{\rm The~Serre~spectral~sequence~of~a~fibration}$

\vskip.2in

For the sake of simplicity, we will work throughout in the context
of real cohomology.
Let $B$ and $F$ be topological spaces. Then the K\"unneth formula
asserts that
\eqn\kunneth{H^m(B\times F)=\sum_{p+q=m}~H^p(B)\otimes H^q(F),}
for each non-negative integer $m$ or, more compactly,
$H^*(B\times F)=H^*(B)\otimes H^*(F)$.

We recall that a fiber bundle over $B$ with fibre $F$ is a
space $E$ with a projection
$\pi:E\rightarrow B$  that is locally, but in general not globally, the
product of
$B$ with $F$. The definition of a fibration with fiber $F$ is more
 technical, but
somewhat less restrictive. In either case, there is a spectral
 sequence $\{E^{pq}_r,d_r\}$ with
\eqn\specseq{E^{pq}_2=H^p(B)\otimes H^q(F), ~~~
d_r:E^{pq}_r\rightarrow E^{p+r\, q-r+1}}
 is a homomorphism with $d_r^2=0$, and, suppressing the
 superscripts,  $E_{r+1}$ is the
  cohomology of the complex $\{E_r,d_r\}$. This is called
  the Serre (or Leray-Serre) spectral
  sequence. Because $d_r$ has negative degree in the second
   superscript, $E^{pq}_r$
  eventually stablizes for each $pq$, even if $B$ and $F$ are
  not finite dimensional.
  This limit is called $E^{pq}_\infty$ and
$H^m(E)$ has the same rank as $\sum_{p+q=m}E^{pq}_\infty$.

In particular, $E^{p,0}_\infty$ can be identified with a quotient
of $H^p(B)$, namely
\eqn\quotofh{
\pi*(H^p(B))\subseteq H^p(E).}
The kernel of $\pi^*$ is generated by the images of all $d_r$,
the ``eventual coboundaries'' in
the spectral sequence. Similarly
\eqn\perncocy{E^{0q}_\infty=\bigcap_r\ker(d_r),}
 otherwise known as the ``permanent cocycles"  can be identified
 with  a subgroup of $H^q(F)$, namely
 $j^*(H^q(E))$, where $j$ represents the inclusion of $F$ in $E$ as a fiber.

One sees that the  Betti numbers of the product $B\times F$ provide
upper bounds for those of $E$.
Moreover, $d_r$ always has total degree +1, so that the Euler number
 of $E$ is the product of those
of $B$ and $F$ whether or not the fibration is actually a product.

We  recall that the K\"unneth formula can actually be interpreted
as the tensor product of
the two cohomology rings. This gives a ring structure to $E_2$. It
turns out that $E_r$ has
a ring structure for each value of $r$, for which $d_r$ is a derivation
so that $E_{r+1}$ inherits
its ring structure from $E_r$.  In practice, this greatly facilitates
computations.

However, the ring structure of $H^*(E)$ need not be identical
with that of $E_\infty$. What is really the case is that there is
a filtration of $H^*(E)$ with $E_\infty$ as the associated graded
ring.
  While a filtered real vector space is necessarily isomorphic
  (although not canonically so) to its associated graded space, this is not
  the case for real algebras. The point is that rank is the only
  invariant of finite dimensional
   real vector spaces, but the structure of real algebras is more
complicated.

There is also a Serre homology spectral sequence, $\{E_{pq}^r,d^r\}$
for which $E_{pq}^r$ and
$d^r$ are the real duals respectively of $E^{pq}_r$ and $d_r$, and
 $E^\infty$ is the associated
graded group to a filtration of $H_*(E)$.
What is particularly important for our purposes is that an element
 of $H_*(F)$ maps to zero
in $H_*(E)$ (is ``homologous to zero'') if and only if it is an eventual
boundary
in the Serre homology sequence.

It should be noted that, throughout this section, we have
suppressed a complication involving the action of the fundamental
group of the base on the homology or cohomology of the fibre.
However this action is trivial in the case of a principal fiber
bundle with connected group, and this is the only case we will be
considering.

\vskip.2in

\noindent~(d)~$\underline{\rm The~case~of~a~torus~bundle}$

\vskip.2in

A special case of the Serre sequence, and the only one that will
concern us in the sequel, is that in which $F$ is a torus
$T=S\times S'$, where $S$ and $S'$ are copies of $S^1$, each with
a fixed orientation.  Then in the Serre sequence $E^{pq}_2$ is
non-vanishing only for $q=0$, $q=1$ and $q=2$. It is immediate
{\it a priori} that the only non-trivial $d_r$ are $d_{2}$, and
$d_3$ , and $E_{4}=E_\infty$. Moreover, it follows from the ring
structure, that if $d_2$ vanishes identically, then so does $d_3$.
If we write $s$ and $s'$ for integral generators of $H^1(S)$ and
$H^1(S')$ respectively, we may identify \eqn\idenwit{\eqalign{&
E^{p0}_2 ~~~\mapsto ~~~H^p(B) \cr & E^{p1}_2 ~~~ \mapsto ~~~
s\otimes H^p(B)\oplus s'\otimes H^p(B) \cr & E^{p2}_2 ~~~ \mapsto
~~~ s\otimes s'\otimes H^p(B)}} We write  $c$ and $c'$
respectively for $d_{2}(s)$ and $d_2(s')$ in $H^{2}(B)$, and will
refer to $(c,c')$ as the double Chern class of $T$ with respect
to the decomposition $S\times S'$. Then we have
\eqn\decomp{d_2(s\otimes s')=s'\otimes c- s\otimes c',} where
there cannot be any cancellation, since the two summands are in
different ``sectors'' of $E_2^{p1s}$, and it is  a
straightforward consequence  that if either $c$ or $c'$ is
distinct from zero in $H^2(B,R)$,  any generator of $H_2(T)$ is a
boundary in the Serre homology sequence and so is homologous to
zero in the total space. {}From this, it follows that the integral
of any closed two-form over any fiber vanishes. In particular, if
the total space is a complex manifold and the fibers are
submanifolds, it follows that the total space cannot be K\"ahler.

If $B$ is a simply-connected four dimensional manifold, we can
compute the Betti numbers of  $E$ very easily. Since the
 Euler characteristic $\chi(T)=0$, it follows that $\chi(E)=0$. Since
 $E$ will also be an orientable manifold and
therefore satisfy Poincar\'e duality,  $b_1(E)$ and $b_2(E)$ will determine
all the other Betti numbers of $E$. But we have
\eqn\betti{\eqalign{&  H^1(E)=\ker(d^{0,1}_2)\subseteq H^1(T), ~~~~{\rm and}
 \cr
& H^2(E)= H^2(B)/d_2^{0,1}(H^1(T))}}
since $E_3^{01}=E_\infty^{01}$ and $E_3^{20}=E_\infty^{20}$ are the only
non-vanishing components of $E_\infty$ of total  degree $1$ and $2$
respectively.

This allows us to compute the Betti numbers of $E$, ~$b_i(E)
\equiv b_i$,~ as \eqn\mumbetti{(b_0, b_1, b_2, b_3, b_4, b_5,
b_6) ~ = ~ 1,~ 0,~ b_2(B)-2,~ 2b_2(B)-2,~ b_2(B)-2,~  0,~ 1} if
$c$ and $c'$ are linearly independent, and \eqn\bettinow{(b_0,
b_1, b_2, b_3, b_4, b_5, b_6) ~ = ~ 1,~ 1,~ b_2(B)-1,~ 2b_2(B)-2,~
b_2(B)-1,~  1, ~ 1} if $c$ and $c'$ are linearly dependent but do
not both vanish as real cohomology classes. If $c$ and $c'$ are
both trivial as real cohomology classes, then $E_2=E_\infty$ and
$E$ has the same Betti numbers as $B\times T$.

Thus for the example studied in \backgeom, since the tori are all
square, the real and imaginary part of the curvatures will be
linearly independent and therefore the Betti numbers will be given
by \mumbetti,~ i.e. \eqn\betforus{ \{~ b_i~ \} = 1,~ 0,~ 20,~
42,~ 20,~ 0,~ 1.} {}It follows from the remarks following
equation \decomp, that the manifold is non-K\"ahler and has Euler
characteristics, $\chi = 0$.

\subsec{Holomorphic Torus Bundles}

The examples of non-K\"ahler complex manifolds
we shall consider are generalizations
 of the Hopf surface, which is the quotient of $C^2-\{(0,0)\}$
by the multiplicative group of scalars generated by 2. The Hopf
surface is a torus bundle over $CP^1$ since  $C^2-\{(0,0)\}$ is
fibred over $CP^1$ by $C^\times$, the multiplicative group of
non-zero complex numbers. The action of the discrete doubling and
halving group is fiberwise and the quotient fibers are tori. We
shall generalize this in two stages.

\vskip.2in

\noindent~(a)~$\underline{\rm Torus~bundles~from~holomorphic~line~bundles}$

\vskip.2in

Let $M$ be any complex manifold and $L$ a holomorphic line
bundle over  $M$. Let $L^\times$
be the complement of the zero-section in the total space of $L$.
Then $C^\times$, the
 multiplicative group of non-zero complex  numbers acts
 holomorphically on $L^\times$. We
 divide out by the cyclic subgroup generated by any
 non-unimodular complex number, $\kappa$, to
 obtain a complex manifold $N$ that is fibred over $M$ with
 toral fibers. This is a direct
 generalization of the Hopf surface, for which $M=CP^2$, $L$ is the
 Hopf bundle and $\kappa=2$.

\vskip.2in

\noindent~(b)~$\underline{\rm More~general~torus~bundles}$

\vskip.2in

Since a torus is a complex manifold, one can consider the most general
holomorphic principal (meaning that the group of the bundle is the torus,
acting on itself by translation) torus bundles. The usual construction of a
bundle from
transition functions applies in this case. Let $T$ be a torus, and let
$M$ be a complex manifold.
 A holomorphic torus bundle can be obtained from a covering of $M$ by
 open sets, $U_\alpha$ together
 with holomorphic maps
\eqn\holmap{\phi_{\alpha\beta}:U_\alpha\cap U_\beta:\rightarrow T,}
satisfying the usual requirements, namely $\phi_{\alpha\alpha}$ is the
identity
and $\phi_{\alpha\beta}\circ\phi_{\beta\gamma}=\phi_{\alpha\gamma}$
wherever both sides are defined.

\vskip.2in

\noindent~(c)~$\underline{\rm An~example}$

\vskip.2in

Let $M$ be the product of two copies of $CP^1$. Let $z_1$ and $z_2$ be
homogeneous
coordinates on the first factor, and let $w_1$ and $w_2$ be
homogeneous coordinates on the second factor. Let $T$ be defined as
$C/\Gamma$,
where $\Gamma$ is a lattice of rank 2. Let $U_{ij}$ be the open
subset of $M$ defined by $z_i\neq 0;\quad w_j\neq 0$.

Let $\alpha$ and $\beta$ be a basis of $\Gamma$. Define
$\exp_{\alpha}$ and $\exp_{\beta}$ as maps from $C$ to $C^\times$
defined respectively by $\exp_{\alpha}(z)=\exp(2\pi i\frac
z{\alpha})$ and $\exp_{\beta}(z)=\exp(2\pi i{z\over {\beta}})$.
Then the quotient map $C\rightarrow C/\Gamma=T$ factor through
each of the maps $\exp_{\alpha}$ and $\exp_{\beta}$. We will write
$\pi_{\alpha}$ and $\pi_{\beta}$ for the other factor in each
case, so that the quotient map can be expressed both as
$\pi_{\alpha}\circ\exp_{\alpha}$ and
$\pi_{\beta}\circ\exp_{\beta}$. With this notation, we define
\eqn\wedefmap{\phi_{ij\,\,i'j'}((z_1,z_2),(w_1,w_2))=
\pi_{\alpha}{\Big (}{z_i \over z_{i'}}{\Big )}\pi_{\beta} {\Big
(}{w_j\over w_{j'}}{\Big )}.} It is straightforward to see, that
the transition functions so defined satisfy the required
relations. The total space of the torus bundle in this example
turns out to be $S^3\times S^3$, so that we obtain a family of
complex structures on $S^3\times S^3$, parametrized by the moduli
space of tori. However, this example generalizes easily in
several significant directions.

\noindent {\bf 1.} With no change of notation whatsoever, $M$ can
be replaced by the
product of two complex projective spaces of arbitrary dimension,
yielding complex
structures on the product of any two odd dimensional spheres.

\noindent {\bf 2.} The transition functions $\phi_{ij\,i'j'}$ can
be generalized by
\eqn\tranfu{\phi_{ij\,\,i'j'}((z_1,z_2),(w_1,w_2))=
\pi_{\alpha}{\Big (}{z_i\over z_{i'}}{\Big )}^m\pi_{\beta} {\Big
(}{w_j\over w_{j'}}{\Big )}^n} for any integers $m$ and $n$;
somewhat more generally, we can instead relax the requirement
that $\alpha$ and $\beta$ form a basis of the lattice, taking them
to be any lattice elements at all.

\noindent {\bf 3.} We can increase the number of projective factors,
associating a
lattice element to each factor and proceeding
 as above.

\noindent {\bf 4.} We can increase the dimension of the torus. In
this case, we have a lattice of rank $2n$ in $C^n$. If $\alpha$ is
the lattice element associated to a projective factor, then the
factor of the transition functions associated to that projective
factor takes its values in the portion of the torus covered by
the  complex span of $\alpha$. This will, in general, be a
noncompact, possibly even dense, subset of the torus, but the
construction remains valid.

\noindent {\bf 5.} Holomorphic torus bundles, like bundles in any category,
transform
contravariantly with respect to holomorphic
maps so that any holomorphic map from a complex manifold $M$ to any product
of
projective spaces  induces many torus bundles over $M$.

\vskip.2in

\noindent~(d)~$\underline{\rm Characteristic~classes~of~holomorphic~torus~
bundles~of~one~complex~dimension}$
\smallskip\noindent~$\underline{\rm over~\hbox{K\"ahler}~manifolds}$

\vskip.2in

Let $T=C/\Gamma$ as in the previous section, and let $M$ be a K\"ahler
manifold.
We write, as usual, $\O$ for the sheaf of germs of holomorphic complex
valued functions of $M$.
We write $\O_T$ for the sheaf of germs of holomorphic $T$-valued functions
on
$M$. We
can identify $\Gamma$ with the sheaf of locally constant functions from
$M$ to $\Gamma$. This
gives the exact sequence of sheaves
\eqn\sheave{0\rightarrow\Gamma\rightarrow \O\rightarrow \O_T\rightarrow 0}
 which in turn yields a long exact cohomology sequence in which the terms
 of interest are
\eqn\shcoho{\rightarrow H^1(M,\Gamma)\rightarrow H^1(M,\O)\rightarrow
H^1(M,\O_T)\rightarrow
H^2(M,\Gamma)\rightarrow H^2(M,\O)\rightarrow}
where holomorphic principal $T$-bundles over $M$ are parametrized by
$H^1(M,\O_T)$. The map
\eqn\anmaop{H^2(M,\Gamma)\rightarrow H^2(M,\O)}
factors through $H^2(M,C)$, and the composition
\eqn\cgammap{H^1(M,\O_T)\rightarrow H^2(M,\Gamma)\rightarrow H^2(M,C)}
permits us to define what we will call the complex chern class
$c_\Gamma\in H^2(M,C)$ of the holomorphic principal $T$ bundle. $c_\Gamma$
is a
linear combination of integral cohomology classes with coefficients in
$\Gamma$.
Because $\Gamma$ has rank $2$ over the integers, $c_\Gamma$
is a linear
combination of at most two integral classes. The
map $H^2(M,C)\rightarrow H^2(M,\O)$
(locally constant cochains to locally exact cochains) corresponds
to projection of a
complex cohomology class on its $(0,2)$ component, so that exactness
of \shcoho\ at
the penultimate term implies that the $(0,2)$ component of $c_\Gamma$
must vanish.
This is a necessary and sufficient condition for a complex linear
combination of two
integral cohomology classes to be the complex Chern class of a
holomorphic principal torus bundle.

\vskip.2in

\noindent~(e)~$\underline{\rm Connections~and~curvature}$

\vskip.2in

Let $E$ be the total space of a principal $T$ bundle over the K\"ahler
manifold $M$,
where $T=C/\Gamma$. We observe that $T$ acts on $E$ as a transformation
group with $M$ as orbit space.
If $z$ is an affine coordinate on the universal covering space of $T$
then $dz$ is well defined as
a one-form along the fibers, but not as a one-form on $E$. A holomorphic
connection is defined to
be a $T$ invariant $(1,0)$-form $\omega$ on $E$ whose restriction to each
fiber coincides with $dz$.
The adjective ``holomorphic'' in this context refers to the fact $\omega$
has no $(0,1)$
component; $\omega$ cannot be chosen to be a holomorphic form unless the
 bundle is
flat ({\it i.e.} the transition functions are locally constant). There
is no single
 natural choice for $\omega$, but existence can be established using a
 smooth partition of unity on $M$
 together with holomorphic one-forms induced by local trivializations.

Two different holomorphic connections differ by  $\pi^*(\theta)$
where $\pi:E\rightarrow M$ is the projection map and $\theta$ is
a $(1,0)$ form on $M$. This observation makes the space of
holomorphic connections an affine space over the space of
$(1,0)$-forms on $M$.

The next observation is that if $\omega$ is a holomorphic
connection on a torus bundle, then
$d\omega$ is a closed two-form that descends to $M$ and whose
de Rham cohomology class is
independent of $\omega$. We will call it $R_\omega$ the curvature
form of the connection.
Although $R_\omega$ is not, in general, homogeneous with respect
to the Hodge decomposition, it
 clearly has no $(0,2)$ component.

We will call $\omega$ a preferred connection, if $R_\omega$ is
$d$-harmonic. Note that
 since $M$ is K\"ahler, this implies that $R_\omega$ is harmonic
 with respect to all three
 of $d$, $\partial$ and $\bar{\partial}$, and
 that the $(1,1)$- and $(2,0)$- components of
 $R_\omega$ are separately harmonic, and therefore closed.
 To see that a preferred connection exists,
 let $\omega$ be any holomorphic connection and let $R_E$ be
 the harmonic representative of $R_\omega$.
 Let $\theta$ be a one-form with $d\theta=R_E-R_\omega$. Then the
 $(0,1)$-component
 of $\theta$ is $\bar{\partial}$-closed, therefore anti-holomorphic,
 and therefore closed
 since $M$ is K\"ahler. Hence it makes no contribution to $d\theta$,
 and may be assumed to be zero.
 Then $\omega +\pi^*(\theta)$ is a preferred connection. Moreover,
 a preferred connection is unique
 up to a holomorphic one-form on $M$.

We note next that for any local section $\sigma$ of the torus bundle, any
point of the $E$ lying over the support $U$ of the section is uniquely
expressible as $t\sigma(m)$.
This identifies $\pi^{-1}(U)$ holomorphically with $U\times T$ and,
with respect to this decomposition,
any holomorphic connection $\omega$, preferred or not, has the form
$dz + \theta$, where $\theta$
is a $(1,0)$-form on $U$. Changing the holomorphic section has the
effect of changing $\theta$
by a holomorphic summand. This both provides a local recognition
principle for holomorphic connections
and identifies the space of holomorphic connections on a holomorphic
torus bundle with $H^0(M,A^{1,0}/\Omega^1)$,
where $A^{1,0}$ denotes the sheaf of smooth $(1,0)$-forms and
$\Omega^1$ denotes the sheaf of holomorphic one-forms.

To obtain a global recognition principal for holomorphic
connections, we need to identify the cohomology class on $M$
represented by the curvature form $R_E$. Let us begin by choosing
an integral basis $\{\alpha,\beta\}$ for $\Gamma$ and making the
observation that $T$ as a topological group is the product
$S^1_\alpha\times S^1_\beta$, where $S^1_\alpha$ and $S^1_\beta$
are respectively the subgroups of $T$ covered by real multiples
of $\alpha$ and $\beta$. Then there are integral cohomology
classes $c_\alpha$ and $c_\beta$ for which $c_\Gamma=\alpha
c_\alpha+\beta c_\beta$. Moreover $(c_\alpha, c_\beta)$ is the
double Chern class of the underlying topological torus bundle
with respect to the decomposition $T=S^1_\alpha\times S^1_\beta$.
We may, incidentally, conclude that $E$ is not a K\"ahler
manifold unless $c_\Gamma=0$.

We write $E_\beta=E/S^1_\alpha$ and $E_\alpha=E/S^1_\beta$. Then
$E_\alpha$ and $E_\beta$ are circle bundles over $M$ and $E$ can
be recovered as the fiber product of $E_\alpha$ and $E_\beta$.
When we look at the torus bundle from this point of view, the
components $c_\alpha$ and $c_\beta$ of the double Chern class
become respectively the Chern classes of the circle bundles
$E_\alpha$ and $E_\beta$ respectively.

Now let  $\omega$ be a holomorphic connection and consider the forms
\eqn\conformn{\eqalign{&\omega_\beta={1 \over \bar{\alpha}\beta-
\alpha\bar{\beta}}(\bar{\alpha}\omega-\alpha\bar{\omega}), ~~~{\rm and} \cr
&\omega_\alpha=-{1 \over \bar{\alpha}\beta-\alpha\bar{\beta}}
(\bar{\beta}\omega-\beta\bar{\omega})}}
$\omega_\alpha$ and $\omega_\beta$ are real, since, in each case,
 both numerator and denominator are imaginary.
Moreover, $\omega=\alpha\omega_\alpha+\beta\omega_\beta$. $\omega_\alpha$
and $\omega_\beta$ vanish along
the orbits of $S^1_\beta$ and $S^1_\alpha$ respectively and
so descend respectively to one-forms on $E_\alpha$
and $E_\beta$. We can now see that $d\omega_\alpha=c_\alpha$
and $d\omega_\beta=c_\beta$. The usual
factors of $2\pi$ and $i$ are missing because $\omega_\alpha$
and $\omega_\beta$ are real and have
period $1$ instead of $2\pi$ or $2\pi i$ on the fibers of
 $E_\alpha$ and $E_\beta$  respectively.
The conclusion is that $d\omega$ represents
$\alpha c_\alpha + \beta c_\beta=c_\Gamma$.

This observation gives a necessary and sufficient condition for a
harmonic two-form $R$ on $M$ to be a curvature form for some
torus bundle: $R$ must represent a complex linear combination of
two integral cohomology classes, and the $(0,2)$-projection of
$R$ must vanish.

\vskip.2in

\noindent~(f)~$\underline{\rm Metric,~torsion~ and~ holomorphic~ forms}$

\vskip.2in

For any holomorphic connection $\omega$, $\frac 1 i\omega\wedge\bar{\omega}$
is
a real $(1,1)$-form corresponding to a positive semi-definite Hermitian
inner
product on the tangent bundle of the total space that is definite
along the fibers.
Let $\pi$ be the projection from $E$ to $M$, and let  $J_M$ be
a K\"ahler form on $M$. It follows that
\eqn\projline{J_E=\frac f i\omega\wedge\bar{\omega} +g\pi*(J_M)}
is a real $(1,1)$-form corresponding to a positive definite
Hermitian metric on the total
space for any identically positive functions $f$ and $g$.

The torsion is simplest to compute if we take $f$ and $g$ to
be constant; clearly
 variable $f$ and $g$ will create additional terms in the torsion.
  We choose both to
 be constant and set $f=1$. Then since $J_M$ is closed, the
 contributions to $\partial J_E$
 and $\bar{\partial}J_E$ will come only from the first term.
 We further assume for simplicity
 that $\omega$ is a preferred connection so that
 $\partial\bar{\partial}\omega=0$ and $\partial\omega$
 and $\bar{\partial}\omega$ are separately harmonic. Then

\eqn\linebund{\eqalign{&\partial J_E=\frac 1
i(\partial\omega\wedge\bar{\omega}-\omega\wedge\partial\bar{\omega})\cr
& \bar{\partial} J_E=\frac 1 i(\bar{\partial}
\omega\wedge\bar{\omega}-\omega\wedge\bar{\partial}\bar{\omega})\cr
&\partial\bar{\partial}J_E=-\frac 1 i\bar{\partial}\omega\wedge
\partial\bar{\omega}.}}
Finally, we observe that if $M$ has complex dimension $n$
and $\Omega$ is a holomorphic
$(n,0)$-form on $M$, then $\omega\wedge\pi^*(\Omega)$ is
 a holomorphic $(n+1,0)$ form
on $E$, and is independent of the choice of $\omega$. This
guarantees that if $M$ has a
nowhere vanishing top dimensional holomorphic form, then so does $E$.

\vskip.2in

\noindent~(g)~$\underline{\rm The~ case~ in~ which~
M~ is~ a~ product~ of~ two~
tori:~ the~ metric~ of~\backgeom}$

\vskip.2in

If $M$ is a product of two tori, $H^{2,0}(M)$ and $H^{0,2}(M)$
are each one dimensional. It follows that any two linearly
independent integral cohomology classes in $H^2(M,Z)$ admit a
complex linear combination without a $(0,2)$-component, so that
there is  no lack of holomorphic torus bundles over any product
of two tori, or even over a more general torus of two complex
dimensions.

However the case, as in the metric of \backgeom, where the curvature is
a pure $(1,1)$-form, is considerably more restrictive. In that case, we
need to find a class in $H^{1, 1}(M)$ that is a complex linear combination
of two integral cohomology classes. Moreover, we have
\eqn\hommap{H^{1,1}(T_1\times T_2)=H^{1,1}(T_1)\oplus H^{1,1}(T_2)\oplus
H^{1,0}
(T_1)\otimes H^{0,1}(T_2)\oplus H^{0,1}(T_1)\otimes H^{1,0}(T_2)}
and the form of the metric in \backgeom\
 tells us that the class we need is orthogonal to the first two summands.
 For a generic choice of $T_1$ and $T_2$, such a class will not exist.

We note that the metric of \backgeom\ implies the connection
\eqn\imcone{\omega=dz^3+2i~\bar{z}^2 dz^1-(4+2i)~\bar{z}^1 dz^2,}
where $dz^3$ is the standard holomorphic one-form along the fiber. This
expression assumes the origin of the coordinates $(z^1,z^2)$ to be at one
of the fixed points of an involution of $T_1\times T_2$. Because of its
form,
$\omega$ descends to the quotient of $T_1\times T_2$ by the involution and
coincides with  $dz^3$ along the blowup of the fixed point. Changing fixed
points entails adding a semi-period to each coordinate. This changes the
form of
$\omega$ by a holomorphic one form  that is a linear combination of
$\bar{s}_1 dz^2$
and $\bar{s}_2 dz^1$ where $s_i$ is a semi-period
of $T_i$. Moreover there is a sign
ambiguity in the semi-period that corresponds to the sign
ambiguity in the coordinates, so
that the new form of $\omega$ still  descends to the quotient.
It follows from the foregoing
that $\omega$ satisfies the local condition to be a  holomorphic
connection on a holomorphic
torus  bundle over the desingularization of $(T_1\times T_2)/Z_2$.

To check the global condition, we note that the curvature form
for the torus bundle implied by the metric of \backgeom\ is
\eqn\ricmet{R=2i~dz^1\wedge d\bar{z}^2-(4+2i)~dz^2\wedge
d\bar{z}^1} apart from an irrelevant real factor of ${1 \over
v}$. Let us investigate what choices for the three tori ($T_1$,
$T_2$, and the fiber $T$) involved are consistent with this
choice.

We will assume that $T_i$ has lattice generators $1, \mu_i$. It
will always be possible to rescale without invalidating the
following analysis, provided the tori are not rescaled
independently. We choose integral cohomology generators $a_i$ and
$b_i$ on $T_i$ on the understanding that $a_ib_i$ is a positive
multiple of the orientation class. Then we have $dz_i=a_i + \mu_i
b_i$ and the curvature form for the torus is given, in terms of
the integral  generators  as \eqn\intgen{\eqalign{& R =
A_1~a_1b_1+ A_2~a_1b_2+ A_3~b_1a_2 + A_4~a_2b_2, ~~~{\rm where}
\cr & A_1 = 4+4i, \qquad A_4 =
2i\mu_1\bar{\mu_2}+(4+2i)\bar{\mu_1}\mu_2 \cr &A_2 =
2i\bar{\mu_2}+(4+2i)\mu_1, \qquad A_3 =
2i\mu_1+(4+2i)\bar{\mu_2}}} It is immediate by inspection that
$R$ will certainly be a complex linear combination of two
integral classes if both $\mu_1$ and $\mu_2$ are Gaussian
integers (complex numbers with integral real and imaginary
parts). In that case, $T$ the torus of the fiber can be taken to
have the Gaussian integers as its lattice. This is the standard
``square'' torus.

\newsec{Background Superpotential}
In the following we will be deriving the form of the
superpotential, that is induced by the non-vanishing ${\cal
H}$-flux in the heterotic theory compactified on a manifold with
torsion. This superpotential gives rise to a potential for the
moduli fields of the internal manifold.
In the supergravity approximation we will basically follow a similar
approach as in \BeckerNN, for compactifications of ${\cal M}$-theory
on Calabi-Yau four-folds.
Thus, instead of extracting
the \sp\ from the potential for the scalars, we will be
considering the dimensional reduction of a term quadratic in the
gravitino appearing in the ten-dimensional action of ${\cal N}=1$
supergravity coupled to Super-Yang Mills theory. In four
dimensions the term quadratic in the gravitino has a coefficient
proportional to the \sp\ (see for example equation (25.24) of
\weba), which makes the calculation easy. This is in contrast to
the scalar potential, which is quadratic in the \sp\ and also
contains the derivatives $D_{\a} W$. Equivalently, we could have
considered the dimensional reduction of the gravitino
supersymmetry transformation to obtain the form of the
superpotential, as has been done in \becons\ for the case of
compactifications of the heterotic string on a Calabi-Yau
three-fold.

But before we go into deriving the form of the superpotential ,
let us first discuss the existence of the holomorphic three-form
for our background manifold \backgeom.

\subsec{Holomorphic Three-Form}

The superpotential discussed in the Type IIB theory \kst\ is
constructed from the holomorphic three-form $\Omega$ and $G_3$,
made from the anti-symmetric tensors and the dilaton-axion. The
three-form $\Omega$ is constructed from covariantly constant Weyl
spinors $\eta_-$ on the manifold and can be written as
\eqn\thromegs{ \Omega = \eta_-^\top \Gamma_{123}\eta_- ~dz^1
\wedge dz^2 \wedge dz^3,} where $\Gamma_{123}$ is the
anti-symmetrized product of three gamma matrices on the internal
manifold as usual. The superpotential $\int G_3 \wedge \Omega$
survives on $T^4/{\cal I}_4 \times T^2/\IZ_2$ and, as discussed in
detail in \beckerD, minimizing this superpotential with respect to the
complex structure and the dilaton-axion we get the background
constraints.

Now to go to the Type I theory we have to make two T-dualities
along the $T^2$ directions. Under a single T-duality along, say,
direction $x^a$, the chiral fermions transform as \eqn\chiral{
\eta_- ~ \to ~ {\tilde\eta}_- = \sqrt{g^{-1}_{aa}} ~ \Gamma_{11}
~ \Gamma_a ~ \eta_-,} where $\Gamma_{11}$ is the ten-dimensional
chirality operator. The holomorphic three-form in the Type
I/heterotic theory is then given by  \eqn\holoinhet{ \Omega =
{1\over \Delta^2} ~ \eta_-^\top \Gamma_{123} \eta_- ~dz^1 \wedge
dz^2 \wedge dz^3,} where $\Delta$ is the warp factor. Notice, that
the coefficient is $e^{-2\phi}$ as shown in \beckerD, as the
dilaton $\phi$ in the heterotic theory is proportional to the warp
factor. We have also taken into account, that the gamma matrices
in the heterotic theory differ from the corresponding gamma
matrices in the Type IIB theory by some powers of the warp factor.
As discussed in \rstrom\ and \beckerD, even in the presence of the
dilaton in $\Omega$ we get \eqn\dilindep{ \del_{\phi} \Omega = 0 =
\del_{g_{mn}} \Omega,} and the holomorphic three-form is a
non-trivial function only of the complex structure $\tau_{ij}$.
For a specific choice of complex structure, $\tau_{ij} = i
\delta_{ij}$, we can use \confirmwhat\ to show, that $\Omega$
satisfies ${\bar\del} \Omega = 0$, i.e it is a ${\bar\del}$ closed
($3,0$) form with everywhere non-vanishing $\int \Omega \wedge
{\bar\Omega} > 0$. Here ${\bar\Omega}$ is the complex conjugate
of $\Omega$. And since $\Omega$ has no zeroes, we cannot multiply
it with a meromorphic form to get linearly independent forms.
Thus $\Omega$ is unique. Existence of a unique $\Omega$ is also
consistent with the Bianchi identity of the three-form ${\cal H}$.
The norm of $\Omega$ is given in terms of warp factor as
\eqn\normofome{\eqalign{|| \Omega || & = {\Big (}\Omega_{123}
{\bar \Omega}_{\bar 1 \bar 2 \bar 3} g^{1\bar 1} g^{2 \bar 2}
g^{3 \bar 3}{\Big )}^{1\over 2} \cr & = {1\over \Delta^4}{\Big (}
1 + {5 |z^1|^2 + |z^2|^2 \over \Delta^2} {\Big )}^{1\over 2},}}
where we have used the inverse of the metric \invmet\ and the
normalization $\eta^\dagger_- \eta_- = 1$ for the covariantly
constant spinors. In fact, we can interpret the norm of the
holomorphic three-form in terms of {\it torsion classes}.
Recently it was argued in the context of string theory
compactifications, that the torsion of an $SU(3)$ structure falls
into five different classes called ${\cal W}_i, ~ i = 1, .., 5$
\carluest\ and \louisL. In the classification done by \carluest\
and \louisL, the ${\cal W}_i$ for our torsional background
\backgeom\ is given by \eqn\loutor{\eqalign{& [{\cal W}_3] =
{\cal T}_{ijk}, ~~~ [{\cal W}_1] \oplus [{\cal W}_2] = 0 \cr &
[{\cal W}_5] = -2[{\cal W}_4] = a~ {d\Delta \over \Delta} + b~ {d
{\tilde\Delta} \over{\tilde\Delta}}},} where ${\cal T}_{ijk}$ is
the torsion, $a,b$ are constants and ${\tilde\Delta}$ is the
modified warp factor. If we define $f(|z|)=5 |z^1|^2 + |z^2|^2$,
then for our case \eqn\backours{a = -4, ~~~ b = {1\over 2}, ~~~
{\tilde\Delta} = 1 + \Delta^{-2} f(|z|).} More detailed analysis
of torsion classes can be extracted from \louisL. In fact the
vanishing of Nijenhuis tensor is related to the vanishing of 1,2
torsion classes.

\subsec{Superpotential}

In ten dimensions the low-energy effective action describing
${\cal N}=1$ supergravity coupled to Super-Yang-Mills theory
contains (to leading order in $\alpha'$) a quadratic term
in the gravitino $\Psi$
\eqn\zi{ \Delta S_{10} = \int d^{10}x \sqrt{-g} (\bar \Psi_M
\Gamma^{MNPQR} \Psi_R ){\cal H}_{NPQ}. }
Here $\Gamma^{MNPQR}$ is the anti-symmetrized product of
ten-dimensional gamma matrices and $\cal H$ is the heterotic
three-form, which in the supergravity approximation is given by
${\cal H}=dB-{\alpha}'{\Omega}_3(A)$, where $A$ is the one-form potential
of the non-abelian two-form and ${\Omega}_3$ describes the Chern
Simons form as usual.

We would like to compactify this interaction on the six-dimensional
non-K\"ahler manifolds, which were described in the previous sections.
In order to derive the form of the four-dimensional
superpotential, the ten-dimensional Majorana-Weyl gravitino
$\Psi_{\mu}$ is decomposed according to
\eqn\zii{ \Psi_{\mu } = \psi_\mu \otimes \eta_- + {\rm
c.c.}+\dots }
where $\psi_\mu$ is the four-dimensional gravitino and $\eta_-$
is the covariantly constant Weyl spinor of the internal manifold.
The dots contain terms, that are needed in order to diagonalize
the kinetic terms of the gravitino on the external and internal
spaces.

Inserting \zii\ into \zi\ and decomposing the ten-dimensional
gamma matrices, it is easy to see, that there appears a term in
the four-dimensional effective action of the form
\eqn\ziii{ \Delta S_4 = \int d^4x \sqrt{-g} (\bar \psi_\mu
\gamma^{\mu \nu} \psi_\nu ) \cdot \int {\cal H} \wedge \Omega, }
where $\Omega$ is the holomorphic $(3,0)$ form of the internal
manifold and $\cal H$ describes the internal components of the flux.
Therefore, in the supergravity approximation the superpotential takes
the form
\eqn\ziiixx{ W=\int {\cal H} \wedge \Omega, }
where we are integrating over the six-dimensional internal manifold.
To this order the superpotential takes a similar form as the one
derived in \becons\ for compactifications of the heterotic string
on a Calabi-Yau three-fold. However, this superpotential is not the complete
result for manifolds with torsion, as there are higher order contributions
to
the superpotential, which cannot be seen in the previous supergravity
approximation.

One way to proceed to obtain the complete result is by using T-duality
arguments, as follows.
First, $dB$ is basically related to the T-dual of the corresponding R-R
form $H'$ in the Type IIB theory (we are assuming a vanishing axion field).
This is
clearly given in \kstt, where they also argue that the NS-NS three-form
of the Type IIB theory
contributes to the spin-connection $\omega$. Such contributions cannot be
seen in the supergravity approximation previously done.
Therefore, besides the obvious contribution from $dB$,
T-duality rules can be used to show, that the superpotential receives
further contributions from the curvature part living on the Type
IIB branes. In fact, a single Type IIB $O7$-plane and $D7$-brane,
will give, after T-duality a contribution to the Type I
superpotential of the form
\eqn\kscontri{W_1 =  - e^{-\phi^I}~\omega \wedge
d\omega,} where $\phi^I$ is the Type I dilaton. However, as we
discussed earlier, this is not the complete result. Due to the
large number of branes and planes (and non-perturbative effects)
the contribution from the curvature part is a little more
involved, than the expression, that we get from a single
system.
The complete result can be written as\foot{A way to see this would be to
look at the gravitational couplings on branes and planes in the Type IIB
theory \rDJM. These couplings are distributed in some specific way for the
$D7$- and $O7$-planes. As shown in the second reference of \rDJM, this
distribution is consistent with the full non-perturbative corrections to the
brane-plane background.}
\eqn\ourcontrib{W_2 = -
e^{-\phi^I}~(\omega \wedge d\omega + {2\over 3} \omega \wedge
\omega \wedge \omega),} which in fact using our earlier notation
is $\Omega_3(\omega)$, the Chern- Simons term. At this point we
should take into account that our background has torsion. This
will modify $\omega$ to \eqn\tarrg{\omega_{\mu}^{ab} ~~\to ~~
{\tilde \omega}_{\mu}^{ab} = \omega_{\mu}^{ab} - {\cal
T}_{\mu}^{ab},} where ${\cal T}$ is the torsion. Therefore, if we
now take the gauge fields also into account and make an S-duality
to go to the heterotic theory, we can identify the form of \sp\ as
\eqn\ziv{ W = \int {\cal H} \wedge \Omega, }
where ${\cal H}$ is not only given by $dB$ but also contains the
Yang-Mills and gravitational Chern-Simons terms, as discussed
earlier in \threeform. We have also identified the torsion with
the ${\cal H}$ field as in \conncovde, and therefore is real and positive 
definite. However we will soon argue that there is another choice of 
superpotential which is {\it complex} and is useful to study backgrounds 
with non-K\"ahler geometry. 
The dilaton factor
appearing in the holomorphic three-form above can be easily seen
to appear in the  string frame.

An alternative way to derive the superpotential for the heterotic
string on a manifold with torsion is to start with the
superpotential of the heterotic string compactified on a
Calabi-Yau three-fold \becons\ or equivalently with the
superpotential derived in the supergravity approximation \ziiixx.
In this approximation the flux involves the complete Chern-Simons
term for the gauge field ${\cal H}=dB-{\alpha}'{\Omega}_3(A)$. We
can now proceed with a similar logic as before. The only gauge
invariant anomaly free expression for the flux appearing in the
superpotential, should involve the Cherns-Simons term involving
the spin connection ${\Omega}_3({\omega})$, which comes from a
higher order term in the heterotic theory and cannot be seen in
the supergravity approximation. If we now take into account, that
our background has torsion, then the spin connection gets
replaced by the torsional-spin connection $\tilde \omega$ and we
obtain the same result for the superpotential, as the one we
derived taking the D7-branes and O7-branes from the Type IIB
theory into account \ziv.

 Comparing \ziv\ and \threeform\ we see, that ${\cal H}$ appears
on {\it both} sides of \threeform. Therefore, we have to solve
\threeform\ for ${\cal H}$ at every order in $\alpha'$ and plug
the result into \ziv. First, we observe, that $d{\cal H}$
contributes an ${\cal O}(\alpha'^2)$. Secondly, the Chern-Simons
term $\Omega_3({\tilde \omega})$ can contribute (to a particular
order in $\alpha'$) terms linear, quadratic and cubic in ${\cal
H}$. Therefore, to {\it first} order in $\alpha'$, lowest order
in the vielbein $e^a$ and to a {\it linear} order in ${\cal H}$ we
recover from \threeform\ the known relation \eqn\bghfield{ {\cal
H} = dB
 + \alpha'[\Omega_3(\omega) -\Omega_3(A)] + .... \equiv f + ....}
where the dotted terms involve, to a given order in $\alpha'$, the
pullback ${\cal H}$ as a one-form to the right hand side of \threeform.

The next order is $\alpha'$ is more involved, because we have to
solve a differential equation to determine the value of ${\cal
H}$. We will discuss about this soon. But before we go into the
details of moduli stabilization we should mention a puzzle.

\vskip.2in

\noindent (a) $\underline{\rm A~puzzle~related~to~T{\rm
-}duality}$

\vskip.2in

As we had mentioned in the introduction, an independent
calculation of the heterotic superpotential on manifolds with
torsion has been performed by Tripathy and Trivedi \pktspt. The
superpotential derived here and the one presented in the second
reference of \pktspt\ do not look identical at first sight. In
our case we see, that since ${\cal H}$ in \bghfield\ is real we
obtain a real superpotential after inserting ${\cal H}$ into our
superpotential \ziv, whereas the potential obtained by Tripathy
and Trivedi is complex. The origin of the complex superpotential
in their approach can be traced back from the Type IIB (or ${\cal
M}$-theory) set-up. From \gfluxinm\ we see, that the $G_3$ flux of
the Type IIB theory has a real and an imaginary part. The form of
the superpotential of the heterotic theory can be calculated by
performing two T-dualities on the Type IIB superpotential. The
three-form $G_3$ has a NS-NS part and a R-R part. The $H_{NS}$
part is responsible for the gravitational (and partly) the
Yang-Mills Chern-Simons terms on the heterotic side. The $H_{RR}$
part gives us the three-form ${dB}$ contribution to the \sp\ and
also contributes to the Yang-Mills Chern-Simons terms. Therefore,
T-duality rules will give us a complex three-form for the Type
I/heterotic theory.

The resolution to this discrepancy comes from the fact, that the
three-form ${\cal H}$ defined in \threeform\ appears on both
sides of this equation. Therefore, to obtain the exact form of
${\cal H}$ we have to solve for ${\cal H}$ in \threeform. The
analysis done in \bghfield\ is only to lowest order in $\alpha'$
and linear in ${\cal H}$. In fact, the
background can be  complex because, as we will
argue in the next section, taking all contributions in $\alpha'$
into account gives a cubic equation from \threeform, which in
general has real {\it and} complex roots. 
{\it This however doesn't mean 
that the heterotic three-form, that satisfies torsional equation and appears 
in the susy relations, is a complex quantity. By heterotic three-form we 
will always mean the real root of the cubic equation. The other two roots, that
are complex conjugates of each other, are constructed from the real three-form
plus a complex twist}.
For the simplest case
(where we only take the dependence on the radial modulus into
account) the superpotential will look like (for simplicity we set
$\alpha' = 1$ here, while in the rest of the paper we will keep
track of $\alpha'$) \eqn\suplooklike{ W = \pm \int (f + i~b
\omega + \sum_{m,n~\in ~{\IZ /2}} i~c_{mn}~f^m t^n) \wedge \Omega,}
where $b$ is a function of $t$ and $t$ is the size parameter of
the non-K\"ahler manifold. The constants $c_{mn}$ depend on $\alpha'$.
We have already defined $f$ in
\bghfield, which contains $dB +$ Chern-Simons terms. In the
next few sections we will determine the above summation in
detail\foot{A few points to consider here (more details will appear in the next
sections): In the above formula for the superpotential \suplooklike\ the
terms in the summation are arranged so that  $f^m t^n$ is
{\it dimensionally} the same as $f$. For example, (as we will see soon)
$\sqrt{t^3/\alpha'}$ is dimensionally the same as $f$. We will explicitly
determine the first few terms in \suplooklike\ when $m = 0, n = {3\over 2}$ and
$m = 2, n = -{3\over 2}$. The rest of the terms can be easily determined from
our general analysis. The $\pm$ sign in front of \suplooklike\ represents the
two possible choices of ${\cal H}$ that we have.}.
In fact, we will give a precise method, by which the
generic heterotic superpotential can be determined to all orders.
The simplest form, that we presented above in \suplooklike\ is
enough, to see how the radial modulus, would get stabilized.
The $\omega$ dependent term is responsible for {\it twisting} the
torus fiber, and the explicit appearance of $t$ in \suplooklike\
will fix the radial modulus. We can also see, how various terms in
\suplooklike\ are related to the corresponding T-dual Type IIB
picture. The $H_{NS}$ three-form flux of the Type IIB theory is
responsible for the spin-connection $\omega$ in \suplooklike\ and
the $dB$ part of $f$ comes from the T-dual of $H_{RR}$ flux
\kstt,\pktspt. These are exactly the terms appearing in the
superpotential of the second reference in \pktspt. Of course,
T-duality rules are approximate and therefore, it will be
difficult to obtain the complete result for the superpotential
derived in this paper by using T-duality rules applied to the
Type IIB superpotential. Most of the earlier works have ignored
the gauge fluxes and do not see the Chern-Simons part of the
superpotential.

\subsec{Potentials for the Moduli}

In this section we will argue, that many of the moduli for the
heterotic compactifications are lifted by switching on ${\cal H}$
fluxes\foot{An alternative way to fix moduli are discussed in \eva. Here
use have been made of asymmetric orientifolds or duality twists to 
fix the K\"ahler moduli. It will be interesting to find the 
connection between our approach and theirs.}. 
We will show this for the particular example constructed
in \beckerD. The basic strategy of the argument is as follows.

{\bf 1.} We compactify the heterotic theory on a six manifold,
which is given by $T^4/{\cal I}_4 \times T^2$ with a vanishing
expectation value for the fluxes. The important hodge numbers for
our manifold are: $h^{11} = 21 = h^{21}$. This will determine the
K\"ahler moduli and the complex structure moduli respectively. The
metric for our manifold is well known and is given in terms of
flat coordinates.

{\bf 2.} Now we turn on the three-form fluxes. This will back-react on
the geometry by ``twisting'' the fiber torus, so that it is now
non-trivially fibred over the $T^4/{\cal I}_4$ base\foot{A
somewhat indirect reasoning to see this is the following. We start
with the Type IIB theory on $K3 \times T^2 / \IZ_2$ without any
fluxes. T-dualizing twice we get to the Type I on $K3 \times T^2$.
Now on the Type IIB side, after switching on $H_{NS}$ and $H_{RR}$
fluxes, the geometry gets warped in the way shown in
\sav,\beckerD\ with no other changes. T-dualizing this
configuration twice we get the Type I theory on a non-K\"ahler
manifold. {\it Alternatively} one could think, that we switch on
$B$ fluxes in the Type I set-up. To {\it preserve} some
supersymmetry the background has to be necessarily twisted. This
is the ``back-reaction'' of the three-form fluxes on the geometry. In
fact, it will soon be clear, that the superpotential does
incorporate this twisting via the three-form fluxes defined in terms of the
$B$ fields plus the twist as explained briefly in \suplooklike. }.
The line
element $dz^3$ of the fiber will change as \eqn\lineelem{ dz^3 ~
\to ~ dz^3 + 2i~ {\bar z}^2 dz^1 - (4 + 2i)~ {\bar z}^1 dz^2,}
where we are ignoring the irrelevant factor of ${1\over v}$
appearing
in \valueAB. The base $T^4/{\cal I}_4$ will also
be changed by the warp factor $\Delta$. The supersymmetry will
reduce from ${\cal N} = 2$ to ${\cal N} = 1$.
As discussed in \kst\ and \kstt, the
fact that $G_3$ is of type (2,1) guarantees, that we have {\it at
least} an ${\cal N} = 1$ supersymmetry. However in writing
the three-form flux in the Type IIB theory as in \gthree,
we have ignored two other contributions. First, the $(2,0) \oplus
(0,2)$ choice of the G-flux in ${\cal M}$-theory and second the
localized fluxes. In the presence of any (or both) of these contributions we
can
preserve exactly an ${\cal N} = 1$ supersymmetry, as we would have expected.

{\bf 3.} The three-from flux will actually change from $dB$ to the
value derived in \bghfield\ including higher order corrections.
The kinetic term for the
heterotic three-form flux, $\int
{\cal H} \wedge \ast {\cal H}$, will give a potential for some of the
moduli because it depends, first of all, on the complex structure.
This comes from the
definition of the coordinates $dz^i = dx^i + \tau^{ij} dy^j$, where
$x^i, y^j$ are the real coordinates and $\tau^{ij}$ are the
complex structure parameters and from the choice of the harmonic forms.
Second, due to the presence of the hodge star
the potential depends on the metric and third, as briefly mentioned in
\suplooklike, on the size $t$ of the manifold.
Let us illustrate the procedure by an example.

\vskip.2in

\noindent (a) $\underline{\rm A~toroidal~example}$

\vskip.2in

The toroidal compactification of $SO(32)$ heterotic string broken
down to a suitable subgroup has a Narain moduli space ${\cal
M}_1$ given by \narainJ \eqn\narain{{\cal M}_1 = {SO(n,n+16) \over
SO(n) \times SO(16+n)}~~~{\rm mod} ~~\Gamma,} where $\Gamma$ is
the T-duality group $SO(n,n+16,\IZ)$. For a generic
compactification of the heterotic string on a $T^n$, we can
illustrate the procedure mentioned above by which the moduli pick
up masses, when suitable fluxes are switched on. This has been
discussed to some extent in \nemanja.

The toroidal compactification of the heterotic string to $10-n$
dimensions has a gravity multiplet and some vector multiplets,
that take the form
\eqn\multivec{(g_{\mu\nu}, B_{\mu\nu}, \phi, nA_{\mu})~\oplus~
(n+16) (A_{\mu}, n \phi),} with a moduli space ${\cal M}$ of dimension
 $[{\cal M}]=1 +n(n+16)$. Apart from the dilaton, the K\"ahler
moduli and complex
structure moduli contribute ${1\over 2}n(n+1)$, the anti-symmetric
tensor contributes ${1\over 2}n(n-1)$ and the sixteen abelian vectors
contribute $16n$. The action for the moduli fields is given by
\eqn\actformod{\int d^{10-n}{\rm x} {\sqrt{-g}} e^{-\phi}{\Big
[}(\del_{\mu} \phi)^2 + {1\over 8} \del_{\mu}M \cdot \del^{\mu}M
{\Big ]},} where the matrix $M$ is given in \narainJ. If we denote
the scalars coming from the K\"ahler structure and complex
structure moduli as $\sigma_i$, the ones coming
from the $B$ field moduli as
$b_i$ and the ones from the vectors as $a_i$, then one can easily identify
\eqn\idenMwithmod{{1\over 2}\del_{\mu}M \cdot \del^{\mu}M =
\sum_{i = 1}^{n(n+1)/2} (\del_{\mu} \sigma_i)^2 - 2 \sum_{j =
1}^{16n} (\del_{\mu} a_j)^2 - \sum_{k = 1}^{n(n-1)/2} (\del_{\mu}
b_k)^2 + ...} Observe, that in
\idenMwithmod\ all the scalars are massless, as they should and that
the supersymmetry is ${\cal N}=16$.

Let us now switch on fluxes. These fluxes are generically
internal and appear on both, the Yang-Mills sector as well as the
tensor sector. These fluxes take the following concrete form \eqn\fluxefor{
F^a_{mn} = \alpha^a_{mn}, ~~~~ {\cal H}_{mnp} = \beta_{mnp} + ...}
The
$\alpha$ fluxes are not arbitrary, but determined in terms
of the fluxes $\beta$, because of the anomaly relation
$d{\cal H} + tr~F^a \wedge F^a = 0$. The reader can extract more
details on this from \nemanja.
Also, since we are considering a flat torus, all the curvature polynomials
vanish.
In particular ${\rm tr} R
\wedge R$ would be zero.

In the presence of fluxes the lagrangian takes the form
\nemanja \eqn\lagchagto{\int
d^{10-n}{\rm x} {\sqrt{-g}} e^{-\phi}{\Bigg [}{1\over 4}
 \sum_{i = 1}^{n(n+1)/2}(\del_{\mu} \sigma_i)^2 - {1\over 2}\sum_{j =
1}^{16n}
(\nabla_{\mu} a_j)^2 - {1\over 4}\sum_{k = 1}^{n(n-1)/2}
(\nabla_{\mu} b_k)^2 - V(\sigma){\Bigg ]}.} Here we make the following
observations:

{\bf 1.} The scalars coming from the two form anti-symmetric
tensor field and the gauge fields have become charged. Therefore, their
kinetic terms are given by covariant derivatives, defined as
\eqn\covmod{\eqalign{& \nabla_{\mu} a_i = \del_{\mu} a_i -
\alpha_i \cdot {\cal A}_{\mu}, \cr & \nabla_{\mu} b_i = \del_{\mu}
b_i - \beta_i \cdot{\cal A}_{\mu} + ...},} where ${\cal A}_{\mu}$
are the Kaluza-Klein gauge fields.

{\bf 2.} The scalars representing the K\"ahler structure and complex
structure moduli are not charged, but a {\it potential}
$V(\sigma)$ is developed from the kinetic term of the three-form
fluxes ${\cal H} \wedge * {\cal H}$, as well as from the gauge fluxes
$F^a \wedge * F^a$. The explicit form of the potential is given
by \eqn\expforpot{ V = g^{mm'}g^{qq'}g^{pp'} \beta_{mqp}
\beta_{m'q'p'} + \sum_{a = 1}^{16n} g^{mm'}g^{qq'}
\alpha^a_{mq}\alpha^a_{m'q'},} where $g^{mp}$ is related to the scalars
$\sigma_i$ describing the K\"ahler and complex structure moduli (as
these are determined from the metric). In fact \expforpot, actually
fixes {\it all} the complex structure moduli and some of the
K\"ahler structure moduli.

{\bf 3.} As we shall soon see, the above consequences are
quite generic. Switching on tensor fluxes, would give charges to
the scalars coming from the Kaluza-Klein reduction of the tensor
fields and would
give a potential to all the complex structure moduli and some of
the K\"ahler moduli. In the toroidal case studied above,
the fluxes will also reduce the supersymmetry to some smaller
value and would convert the $T^n$ to some {\it twisted} $T^n$.
This twist can, of course, be explicitly determined, if there exists a
Type IIB or ${\cal M}$-theory dual of the model following the
procedure of \beckerD. In the absence of a Type IIB (or ${\cal
M}$-theory) dual the procedure to determine the twist is
complicated. For $n = 6$, the case is subtle because sometimes,
even if there would exist a Type IIB dual, the existence of fluxes may
become forbidden, if the corresponding four-fold in ${\cal
M}$-theory has zero Euler-characteristics \rBB.
For other values
of $n$, there are no known obstructions. However, if we demand no
supersymmetry for our background, then again there would be no
obstruction for any values of $n$.

{\bf 4.} Notice also, that the dilaton moduli remains unfixed by
the fluxes. This is in {\it contrast} to the Type IIB case studied in
\kst, \beckerD\ and \kstt. In the Type IIB examples the potential is
determined from the kinetic term $G_3 \wedge * G_3$, where $G_3 =
H' - \varphi H$. Therefore, the dilaton-axion appears explicitly
in the potential. For the heterotic theory this is not the case.
However, as we shall soon see, for the heterotic case the {\it
radial} modulus can, in fact, be fixed at tree level, which was not
possible for the Type IIB case. In the later case higher order
$\alpha'$ corrections have to be taken into account, in order to generate
a potential for the radial modulus \BeckerNN.

{\bf 5.} To summarize, the complete action for the heterotic theory in
$10-n$
dimensions with gauge group broken to $U(1)^{2n + 16}$ at any
generic point will be given by \eqn\hetactnowdet{ \int
d^{10-n}{\rm x}{\sqrt{-g}} e^{-\phi} {\Big [} R + (\del \phi)^2 -
{1\over 2}{\cal H}^2 {\Big ]} + {\rm S_{gauge}} + {\rm
S_{moduli}},} where ${\rm S_{gauge}}$ is the action for the $16 +
2n$ gauge fields and ${\rm S_{moduli}}$ is the action for the
moduli fields.

\vskip.2in

\noindent (b) $\underline{\rm Moduli~stabilization~on~the~K3
\times T^2~manifold}$

\vskip.2in

When heterotic string is compactified on a K3 manifold, the low
energy effective theory is ${\cal N} = 1$ supergravity (8
supercharges) coupled to tensor multiplets, hypermultiplets and
vector multiplets as
\eqn\vechyper{(g_{\mu\nu},B^+_{\mu\nu},\psi^+_\mu)~ \oplus~
(B^-_{\mu\nu}, \psi^-,\phi)~ \oplus ~ 20 (4\phi, \chi^-)~ \oplus
~ 16 (A_\mu,\lambda^+),} where $\pm$ denotes the chirality of Weyl
spinors in six dimensions (in the subsequent discussion we shall
ignore the fermions), while for the anti-symmetric tensor field
this symbol indicates, if the field is self-dual or
anti-self-dual. The sixteen vector multiplets are the
contributions of the Cartan subalgebra of the gauge
group\foot{For the $E_8 \times E_8$ heterotic string on $K3$, if
we set the gauge connection to the $SU(2)$ spin connection, then
the 10d Yang-Mills multiplet will contribute
$$[(133,1) + (1,248)](A_\mu,\lambda^+)~\oplus~ [10(56,1) + 45 (1,1)]
(4\phi, \chi^-),$$ to the massless spectrum in six dimensions,
where the terms in the bracket are the representations of the
unbroken gauge group $E_7\times E_8$. The ten-dimensional gravity
multiplet will contribute in the same way as before.}.
Compactifying further on a torus and keeping a generic choice of
the ${\rm n_V}$ vector multiplets and ${\rm n_H}$ hypermultiplets
in four dimensions, we get a massless spectrum with ${\cal N} = 2$
supersymmetry (the spinors are Majorana)
\eqn\hypvefour{(g_{\mu\nu}, A_\mu, 2\psi_\mu)~\oplus~ ({\rm n_V}
+ 1) (A_\mu, 2\phi, 2\lambda)~\oplus~ {\rm n_H} (4\phi, \chi).}
The extra vector multiplet can be traced from the six dimensional
perspective. This is called the vector-tensor multiplet having
$(\phi, B_{\mu\nu}, A_\mu, 2\lambda)$, where $\phi$ is the
dilaton \kapulov. The anti-symmetric tensor is the axion in four
dimensions and therefore this multiplet can be identified to be an
{\it abelian} vector multiplet. As discussed in \kapulov, the
off-shell structure of this multiplet differs from that of the
vector multiplet by the presence of a central charge, which
vanishes on-shell. Furthermore, the axion-dilaton has a continuous
Peccei-Quinn symmetry, which helps to determine the tree-level
prepotential for all heterotic vacua. The loop corrections are
severely restricted accordingly. We shall denote the
axion-dilaton by ${\tau = {a} - i e^{-\phi}}$. As before, the
action for the moduli is given by \eqn\acfomodnow{ \int d^4{\rm
x} \sqrt{-g}~{\Bigg [} {\del_\mu \tau \del^\mu {\bar \tau}\over
(\tau - {\bar \tau})^2} + {1\over 4} \sum_{i =1}^{\rm n_V}
\del_\mu z^i \del^\mu {\bar z}^i - {\rm h}_{ab} \del_\mu \sigma^a
\del^\mu \sigma^b {\Bigg ]},} where $z^i$ are the complex scalars
in the vector multiplets and the scalars $\sigma^a$ are the
scalars in the hypermultiplets, with ${\rm h}_{ab}$ as the metric
on the moduli space. In fact, for our case when $D_4^4$ is broken
to $U(1)^{16}$, we can easily check, that ${\rm n_V} = 16, a = 1,
...., 21$. Of course , not all the scalars in the hypermultiplet
are related to the K\"ahler moduli and complex structure moduli of
the six manifold $K3 \times T^2$, as 22 of these scalars come
from the anti-symmetric tensor. The moduli fields of the vector
multiplets are contained in the $M$, appearing in \actformod.
However, since the instanton number is constrained to be 24, the
numbers of vectors and hypers depends on this constraint. The
moduli space of hypermultiplets ${\cal M}_{\rm H}$ is a
submanifold spanned by the moduli of the K3 surface ${\cal
M}_{K3}$ and the moduli space of vector multiplets ${\cal M}_{\rm
V}$ which, along with $\tau$, span a special K\"ahler manifold.
These moduli spaces are given respectively by \nati, \kapulov
\eqn\kthree{{\cal M}_{K3} = {SO(4,20) \over SO(4) \times SO(20)},
~~~ {\cal M}_{\rm V} = {SU(1,1)\over U(1)} \otimes {SO(2, {\rm
n_V}) \over U(1) \times SO({\rm n_V})},} where the $\tau$ spans
the moduli space ${SU(1,1)\over U(1)}$. More details on the
moduli space structure are given in \nemanja.

Let us now switch on a three-form flux, which breaks the
space-time supersymmetry to ${\cal N} = 1$, and would convert the
six-dimensional manifold $K3 \times T^2$ to the non-K\"ahler
complex manifold discussed in \beckerD. The three-form flux takes
the form \eqn\thforflux{{\cal H} = a~ \Omega + \sum_{i =
1}^{h_{21}} b^i~ \chi_i + {\rm c.c.},} where $\Omega$ is the
holomorphic (3,0) form and $\chi_i$ are the
 (2,1) forms of the internal manifold. Observe,
that the background three-form flux, that we switched on ignores
the Chern-Simons contribution, and therefore it is related to the
$dB$ part of ${\cal H}$. For the subsequent discussion this
doesn't affect much. We have also kept the background gauge
fluxes to be zero for simplicity. Switching on the above
three-form flux, the story should be the same as discussed
earlier. The action for the moduli field will become
\eqn\modfienowq{ \int d^4{\rm x} \sqrt{-g}~{\Bigg [} \sum_{i
=1}^{\rm n_V + 1} G_{ij} \del_\mu Z^i \del^\mu {\bar Z}^j -
\sum_{a,b = 1}^{22} {\rm h}_{ab} \nabla_\mu \sigma^a \nabla^\mu
\sigma^b +\sum_{c,d = 1}^{61} {\rm h}_{cd} \del_\mu \sigma^c
\del^\mu \sigma^d  - V (\sigma) {\Bigg ]},} where we have combined
the scalars $\tau$ and $z^i$ into $Z^i$ and defined a metric
$G_{ij}$ accordingly. The components of the metric can be easily
ascertained from \acfomodnow. Let us summarize the situation.

\noindent {\bf 1.} Some of the scalars of the hypermultiplet have
become charged in the presence of the background fluxes. In fact,
these scalars are precisely obtained by the dimensional reduction
of the anti-symmetric two-form. The covariant derivatives for
these scalars are defined with respect to $a_i,b_i$ as before.

\noindent {\bf 2.} The scalars from the vector multiplet and the
vector-tensor multiplet will remain massless, as we are not
switching on the gauge fluxes. In the presence of gauge fluxes
these scalars would also be charged.

\noindent {\bf 3.} Some of the scalars in the hypermultiplet,
which come from the K\"ahler and complex structure of the
six-manifold, will develop a potential $V$ from the kinetic term
of the three-form flux, written in terms of complex coordinates as
$\int {\cal H} \wedge * {\bar {\cal H}}$. From the above choice
of the three-form \thforflux, the potential can be written
explicitly as \taylor,\GukovYA,\kst \eqn\potwritexpl{V = {i\over
2 {\rm Im}~\tau~ (\int \Omega \wedge {\bar \Omega})^2} {\Bigg [}
\int \Omega \wedge {\bar \Omega} \int {\cal H} \wedge {\bar
\Omega}\int {\bar {\cal H}} \wedge { \Omega} +\int \chi^i \wedge
{\bar \chi}^j \int {\cal H} \wedge {\chi_i}\int {\bar {\cal H}}
\wedge {\bar {\chi_j}} {\Bigg ]}.} The reason why this would fix
some of the scalars in the hypermultiplet is, because the complex
structure appears implicitly in writing the harmonic forms
$\Omega$ and $\chi$, when we decompose the three-forms in terms of
$h_{21}$ harmonic forms. This implies, that all the complex
structure moduli will get fixed in the process. Fixing the radius
moduli is however subtle, as we shall discuss in the next section.

\noindent {\bf 4.} Observe, that even though we turn on the gauge
fluxes, making the scalars in the gauge multiplet charged, we {\it
cannot} give a potential to the axion-dilaton $\tau$. Hence, in
this setup the axion-dilaton remains unfixed. Also, as we
mentioned briefly at the beginning of the section, the presence of
the $B$ field twists the fiber. To preserve supersymmetry this is
a consistent operation. However, this effect of the twisting
should also be apparent from the form of superpotential, that we
have in the heterotic theory. In fact, there is an additive term
to the superpotential, that is proportional to the twisting, which
does the job. This additive term is directly correlated with the
existence of a complex contribution to the superpotential. We
will discuss about this issue, when we determine the complete
superpotential for the heterotic theory in the next section.
This additive term in the superpotential is responsible for fixing
some of the K\"ahler structure moduli.
Therefore, to summarize the total number of moduli, that we could
fix are the 22 scalars in the hypermultiplets, that become
charged and the 21 complex structure moduli. Some K\"ahler moduli
and the overall radial modulus also get stabilized.

\subsec{Radial Modulus}

In Type IIB compactifications with fluxes the condition for
unbroken supersymmetry is, that the flux $G_3$ should be a (2,1)
form and that the (3,0), (0,3) and (1,2) parts should vanish. This
requirement allows us to fix all the complex structure moduli and in some special cases
some of the K\"ahler moduli. The condition that $G_3$ is a (2,1)
form can be recast in the form 
\eqn\pring{ J \wedge G_3 = 0,} 
which means that $G_3$ is a primitive form.
As discussed in \kst, a rescaling the metric or the fundamental
form as
$$
J \to t J,
$$
leaves the primitivity condition invariant and therefore the
radial modulus will remain a free parameter, at least at the tree
level. Therefore, the superpotential determines all the complex
structure moduli and some of the K\"ahler structure moduli but
not the radial modulus.

Making two T-dualities and an S-duality we go to the heterotic
picture. On the heterotic side many of the Type IIB moduli have a
different interpretation. Therefore, it is not surprising, that
some moduli that remain unfixed in the Type IIB theory might be
fixed in the heterotic setup. In this section we will argue, that
the radial modulus can actually be determined in heterotic
compactifications with torsion. Below we will give two reasons,
that support this claim.

First, in the infinite radius limit the internal manifold cannot
support the non-vanishing fluxes anymore and this leads to a
contradiction. Indeed, imagine that the radius of the internal
manifold could become arbitrarily large. In this limit the
constant fluxes tend to vanish. In fact, from the choice of flux
densities in \valueAB, we see that as 
\eqn\vtozer{v \to \infty,\qquad A \to 0\qquad {\rm and } \qquad  B
\to 0,} 
even though the total flux integral is still non-zero. Therefore,
the contribution to the total flux has to come from its localized
part. In the large radius limit the fixed points go to infinity,
but the fluxes in \handhprime\ still remain non-zero. From the
warp factor equation \solwarpp\ we see, that in the limit \vtozer\
the warp factor tends to be a constant 
\eqn\seefromwa{\Delta \to \sqrt{c_o},} 
implying that the manifold is just a product $T^4/{\cal I}_4
\times T^2$. In this limit the torus is no longer non-trivially
fibred over the base.
Using now the torsional constraints \warpline, this would imply $
A \wedge F = 0$, which is a contradiction because we have just shown, that
localized sources survive in the large radius limit.

A second reason of why the radial modulus should be stabilized uses
the torsional equation, which relates the background three-form
field ${\cal H}$ to the two-form $J$ as 
\eqn\toreqrel{ {\cal H} = i(\del - {\bar\del})J.} 
We would like to see, how the left hand side of this equation
transforms under $J \to t J$, i.e. we would like to study the
behavior of ${\cal H}$ in \threeform\ under this transformation.
Defining
$$
{\tilde{\cal H}} = {\cal H} e^{-2} \equiv {\cal H}_{ijk}e^{aj}
e^{bk},
$$
one can easily show, that the Chern-Simons term related to the
torsional-spin connection $\tilde\omega$ is given by 
\eqn\torspcone{\Omega_3(\tilde\omega) = \Omega_3(\omega) + {1\over
4} \Omega_3(\tilde{\cal H}) -{1\over 2} (\omega \wedge {\cal
R}_{\tilde{\cal H}} + {\tilde{\cal H}}\wedge {\cal R}_\omega),} 
where we define $\Omega_3(\tilde{\cal H})$ in somewhat
similar way as $\Omega_3(A)$
or $\Omega_3(\omega)$:
$$
\Omega_3(\tilde{\cal H})=\tilde{\cal H} \wedge d{ \tilde{\cal H}}
- {1\over 3} \tilde{\cal H}\wedge \tilde{\cal H} \wedge
\tilde{\cal H}.
$$
The quantity ${\cal R}_{\tilde{\cal H}}$ is the curvature
polynomial due to the torsion and is defined as
$$
{\cal R}_{\tilde{\cal H}}= d\tilde{\cal H} - {1\over 3}
\tilde{\cal H} \wedge \tilde{\cal H},
$$
whereas ${\cal R}_\omega$ differs from the usual curvature
polynomial by $-{1\over 3}\omega \wedge \omega$. In fact, we can
write \torspcone\ in a more compact form as 
\eqn\comform{\Omega_3(\tilde\omega) = {\Big (}\omega - {1\over 2}
\tilde{\cal H}{\Big )} {\Big (}{\cal R}_\omega - {1\over 2} {\cal
R}_{\tilde{\cal H}}{\Big )},} 
with the curvature polynomials defined
above\foot{In this form it is instructive to compare with the other
choice of torsional-spin connection ${\hat\omega}$
$$\Omega_3(\hat\omega) ={\Big (}\omega +{1\over 2}
\tilde{\cal H}{\Big )} {\Big (}{\cal R}_\omega + {1\over 2}{\cal
R}_{\tilde{\cal H}} + {1\over 3} \tilde{\cal H} \wedge \tilde{\cal
H}{\Big ),}$$
which differs from \comform\ in relative signs and an additional
term. As discussed before, for some purposes we need
${\tilde\omega}$ and for others ${\hat\omega}$. In fact, for
deriving the torsional equations we need the connection
${\hat\omega}$, as this is more appropriate (see section $2.4$).
However, for the time being we take the connection ${\tilde\omega}$,
as this appears in the anomaly relation. We shall point out the
consequence of the $\pm$ ambiguity in the connection later on, but one
immediate thing to notice appears from
 comparing \torspcone\ with \tranbto.
In fact, we see, that when we shift the ambiguity in the connection into a
redefinition of
the two-form $B$, the
three-form could be written completely in terms of the gauge fields
$A$ and the affine-connection $\omega$ without recourse to any
torsional-connection.
Calling the shifted field strength of the $B$ field in \tranbto\ as
$h_{ijk}$, we can easily confirm
$${\cal H} = h +
\alpha'[\Omega_3 (\omega) - \Omega_3 (A)] + {\rm covariant
~terms}.$$
These covariant terms can be determined directly from \torspcone\
 and are globally defined, whereas the
spin connection $\omega$ and gauge fields $A$ have to be defined
on patches. This clearly tells us, that for any given four-cycle ${\cal
C}$, the constraint
$$\int_{\cal C}~d{{\cal H}} ~=~
\alpha' \int_{\cal C} ~ [d\Omega_3 (\omega) - d\Omega_3 (A)] ~=~
0,$$
which leads to \obstr, is defined with respect to
either of the connections
$\omega,~ {\tilde\omega}~ {\rm or}~ {\hat\omega}$ and is therefore
unambiguous.}.

{}From the above analysis it is easy to infer, what the background
torsion is. If we concentrate only to the lowest order in
$\alpha'$ and linear order in ${\cal H}$, the three-form
background is given by 
\eqn\thfobgis{ {\cal H} = dB {\Big (} 1 - {\alpha'\over 2} {\cal
R}_\omega e^{-2}  {\Big )} + \alpha' [ \Omega_3(\omega) -
\Omega_3(A)] + {\cal O}(\alpha'^2).} 
Comparing \toreqrel\ and \thfobgis\ it is easy to see, that
$J \to t J$ is no longer a symmetry. Indeed, the two sides of
\thfobgis\ transform as
$$
{\cal H} \to t {\cal H} \qquad {\rm and} \qquad {\cal R}_\omega
e^{-2} \to t^{-1} {\cal R}_{\omega}e^{-2}.
$$
Observe that in  \thfobgis\ the term involving ${\cal R}_\omega
e^{-2}$ is rank zero. To the {\it same} order in ${\cal H}$ we
have ignored $-{\alpha'\over 2} \omega \wedge d{\tilde{\cal H}}$,
which could also contribute. But since $d{\tilde{\cal H}}\sim
{\cal O}(\alpha')$ this term is irrelevant.

Since the terms in \thfobgis\ scale differently under rescaling of
the metric we conclude, that there must exist an upper limit for
the size of the torsional manifold. To all orders in ${\cal H}$
and $\alpha'$ the equation, that we need to solve is 
\eqn\mexi{{\cal H} +{\alpha'\over 2}{\Bigg [} \omega \wedge {\cal
R}_{\tilde{\cal H}} + {\tilde{\cal H}}\wedge {\cal R}_\omega
-{1\over 2} {\tilde{\cal H}} \wedge {\cal R}_{\tilde{\cal
H}}{\Bigg ]}
 = dB + \alpha'[\Omega_3(\omega)-\Omega_3(A) ] .}

{}From the above two reasons we find, that the radial modulus can
in fact be controlled in this setup. As a result, there should be a
potential for this modulus. This potential would follow from the
form of the Lagrangian, after taking the non-vanishing fluxes into
account. This Lagrangian involves the conventional Einstein term
coupled to the kinetic term for the flux. After expanding the
Einstein term the action takes the form  
\eqn\lagor{ \int \sqrt{{\rm det}~g}~g^{-1}~[ \del(g^{-1} \del g) +
g^{-2} (\del g)^2] + \int \sqrt{{\rm det}~g}~g^{-3}~{\cal H}^2, }
where we integrate over the compact non-K\"ahler six-manifold. In
this form, the Lagrangian will be {\it invariant} under $g \to t
g$, if we also set ${\cal H} = dB$, implying that the radial
modulus does not receive a potential. This is what usually
happens. However, as we have seen in the previous sections
${\cal H}$ is no longer $dB$, but is
given by \thfobgis. Does this give a potential to the radial
modulus?

It turns out, that to the order in \thfobgis, this {\it fails} to
give a potential for the radial modulus. To see this, let us denote
the radius by $t$ and write locally the metric components as
$g_{\mu\nu} = t~\eta_{\mu\nu}$, where $\mu,\nu$ span the six
compact directions. The first terms in \lagor\ give us the kinetic
term for $t$, i.e $\del_\mu t ~\del^\mu t$. The spin connection one-form
$\omega$ scales as $t^{-1}\del t$. Therefore, to lowest order in
$\alpha'$ and linear order in $\cal H$, the ${\cal H}$ term does
not give a potential for the radius. In order to get a
non-vanishing potential, we will have to consider the exact equation
for ${\cal H}$, given by \mexi.

The reason of why the above argument failed is,
that to the approximation taken in \thfobgis\
all terms besides ${\cal H}$ contain either $\omega$ or ${\cal R}_\omega$ and therefore
involve derivatives of $t$. But we also have terms cubic in
${\cal H}$, which originate from the third term in the bracket
on the left hand side of \mexi, so that this equation becomes
\eqn\simchoi{ {\cal H} +
{\alpha'\over 12} {\tilde{\cal H}} \wedge {\tilde{\cal H}} \wedge
{\tilde{\cal H}} + \dots = dB +\alpha'[\Omega_3(\omega) - \Omega_3(A)],}
where $\dots$ represent the $\omega$ dependent terms. We can write the
above equation as a cubic equation for ${\cal H}$
\eqn\cubic{
{\tilde{\cal H}}^3 + a ~{\cal H} + b = 0.}
The variables ${a, b}$
can be determined from \simchoi. In general, the equation \cubic\
is not a typical cubic equation, because ${\cal H}$ defined above
are forms on a manifold and not real numbers\foot{The precise
value of the cubic wedge products of ${\tilde{\cal H}}$ can be
shown to be equal to
$$
{\tilde{\cal H}} \wedge {\tilde{\cal H}} \wedge {\tilde{\cal H}} =
{\tilde{\cal H}}_i^{ab}~ {\tilde{\cal H}}_j^{cd}~ {\tilde{\cal
H}}_k^{ef}~ {\rm Tr}(M^{ab} M^{cd} M^{ef})~dx^i \wedge dx^j \wedge
dx^k,
$$
where the one form
${\tilde{\cal H}}_i^{ab} = -
{\tilde{\cal H}}_i^{ba} = {{\cal H}}_{ijk}e^{ai} e^{bk}
$
and
$M^{ab}$ are the tensors of the holonomy group of the manifold. As
can be seen  this doesn't vanish generically and therefore gives
rise to a cubic equation as \cubic. The one form
${\tilde{\cal H}}_i^{ab}$ shares similar properties with
the spin connection $\omega_i^{ab} = \omega_{ijk}e^{aj}e^{bk}$.}.
Therefore, let us first take a simple
situation, which can convert \cubic\ into a cubic equation with
real variables. For the present purpose, this will suffice to
clarify the basic ideas involved here. We will soon generalize
this to our six dimensional non-K\"ahler manifold.

\vskip.1in

\noindent (a) $\underline{\rm A~simple~toy~example}$

\vskip.1in

Let us choose a background, that can convert equation \cubic\
into a simple cubic equation in terms of functions and not
forms. The simplest background will be
\eqn\sibg{{\cal H}_{ijk} = h{\cal C}_{ijk}, \qquad e_{ai} =
t^{1/2}e^o_{ai} \qquad {\rm and} \qquad  e^o\cdot e^o = \eta,}
where ${\cal C}_{ijk}$ is an antisymetric in six dimensions,
$e_{ai}$ are the vielbeins and $\eta \equiv \eta_{ij}$ is the
flat metric. The size of the six manifold is given by $t$, as
defined earlier. We will also assume, that the only non-zero
component of ${\cal H}$ is ${\cal H}_{1 2 \bar 3}=h$. It is easy
to see, that \cubic\ becomes a simple cubic equation
\eqn\cueqbc{h^3 +
ph + q = 0 \qquad {\rm with} \qquad p, q \in \IR,}
where $p,q$ are given below. The solution to this equation
is well known
using Vieta's substitution\foot{Define ${\cal H} = w - {p\over
3w}$ and substitute it in \cubic. The equation will become a
quadratic equation in $w^3 \equiv \lambda$, given by $\lambda^2 +
q \lambda - {p^3 \over 27}= 0$, whose solutions are $w^3 = {1\over
2}{\Big (} - q \pm \sqrt{q^2 + {4p^3 \over 27}}{\Big )}$. Calling
these roots as $\delta^3_{\pm}$, $w$ can be written as $w =
\delta_{\pm}(1, \omega, \omega^2)$, where $\omega$ is the
cube-root of unity. The solutions are therefore $\delta_+ +
\delta_-, \delta_+ \omega + \delta_- \omega^2$ and
 $\delta_+ \omega^2 + \delta_- \omega$. This is Cardan's solution for a
cubic
equation. For a generic cubic equation of the form $h^3 + a_2 h^2
+ a_1 h + a_0 = 0$, we can define $h = y - {a_2 \over 3}$. In this
form the equation looks in the same way as \cueqbc\ with $p = a_1 - {1\over
3} a_2^2$ and $q = {1\over 3} (3 a_0 - a_1 a_2) + {2\over 27}
a_2^3$. Calling the roots as $\alpha_i$, we get: $\sum~ \alpha_i
= - a_2, ~ \sum~\alpha_i \alpha_j = a_1, ~ \sum~\alpha_i \alpha_j
\alpha_k = - a_0$. Observe also, that since ${p^3 \over 27} + {q^2
\over 4} > 0$, we have one real root and pair of complex
conjugate roots.}. Observe, that for simplicity we have ignored
all $\omega$ dependent terms and ${\cal O}(\alpha'^2)$ terms
appearing in \mexi.
To leading
order in $\alpha'$ the presence of $\omega$ would induce
quadratic and  linear terms in ${\cal H}$ in the previous equation.
This does not change the analysis, that we perform here, because
{\it any} generic cubic equation can always be brought to the
form \cubic\ with additive contributions to $p,q$. For the
simplest case we can specify $p,q$ as
\eqn\pandq{p \sim  \pm
{t^3 \over \alpha'},\qquad q \sim  \mp {t^3 f \over \alpha'},
\qquad {\rm and} \qquad {\big <}dB +
\alpha'\Omega_3(\omega) - \alpha' \Omega_3(A) {\big >} _{abc} \equiv
f\epsilon_{abc},}
where the sign ambiguity reflects the sign ambiguity in the
torsional-spin connection, $\omega \mp {1\over 2} {\cal H}$, i.e
the choice of either ${\tilde\omega}$ or ${\hat\omega}$ in terms
of earlier notations. The reader might be concerned by the fact,
that since  \simchoi\ involves {\it wedge} products, it will be
difficult to get a simple cubic equation. But it is easy to see,
that all the other factors due to wedging can be absorbed into the
definition of $p$ and $q$ and therefore, we can express \simchoi\
as a cubic equation in $h$ with $p,q$ {\it proportional} to
\pandq, as we are not very concerned about precise factors.
Nevertheless, the exact factors appearing here can be worked out
easily. In this section we take a simple example, where the
background is just ${\cal H}_{12 \bar 3}$. This implies
\eqn\imnev{
{\tilde{\cal H}}_1^{ab} = {\cal H}_{1 \alpha \beta} e^{a \alpha}
e^{b \beta}
= 2h ~{\cal C}_{12 \bar 3} e^{2[a} e^{b]\bar 3} \equiv {2}
t^{-1}~ h {\cal C}_{12 \bar 3} e_{\rm o}^{2[a} e_{\rm o}^{b]\bar
3} = t^{-1} h \alpha_1^{ab},}
where $\alpha_1^{ab} = -
\alpha_1^{ba} =
 2 {\cal C}_{1 2 \bar 3}~e_{\rm o}^{2[a} e_{\rm o}^{b]\bar 3}$ and
$e^{a i} = t^{-{1\over 2}}e^{a i}_{\rm o}$. We have also defined the
anti-symmetrization between $a, b$ as $[ ...]$, as usual. The
above formula can easily be generalized to the case, in which
other components of ${\cal H}$
are turned on and not all ${\cal C}$ are equal. 
Some aspects of this  will be dealt with
in the next sub-section. We are also
using complex coordinates, but it is easy to infer the corresponding
case with real coordinates. Similarly to the above equation we can obtain
the other components of $\tilde {\cal H}_i^{ab}$.

\eqn\defaga{{\tilde{\cal H}}_2^{cd} =
 2h~ {\cal C}_{2 \bar 3 1} e^{\bar 3 [c} e^{d]1} =
t^{-1}h \alpha_2^{cd}\qquad {\rm and} \qquad {\tilde{\cal H}}_{\bar
3}^{ef} = t^{-1} h \alpha_{\bar 3}^{ef}.}
Now we can determine
the cubic wedge product between the ${\cal H}$'s, using the result
of the earlier footnote. We again use $M^{ab}$ as the tensors of
the holonomy group of the manifold. The result will be
\eqn\precresult{[{\tilde{\cal H}}\wedge {\tilde{\cal H}}\wedge
{\tilde{\cal H}}]_{12 \bar 3} = t^{-3} h^3
\alpha_1^{ab}\alpha_2^{cd} \alpha_{\bar 3}^{ef}{\rm Tr} (M^{ab}
M^{cd} M^{ef}) = t^{-3} h^3{\cal Q},}
where ${\cal Q}$ is
an integer determined in terms of $\alpha_i$ and $M$ in an
obvious way. In general, ${\cal Q}$ is a non-zero integer and for
the analysis done in this sub-section and the next one, we shall
normalize ${\cal Q}$ to 1 by defining variables appropriately. This
will not alter any of our results because we only require the
form of the moduli and not the precise factors.
(Later we will consider the case when ${\cal Q}$
vanishes.)
 From here
the set of equations \pandq\ can be derived.
To determine the solution for
a cubic equation, we define, as usual, two variables $A$ and $B$ as
\eqn\aandbas{ A = {\Bigg (} - {q\over 2} + \sqrt{{q^2\over 4} +
{p^3 \over 27}} {\Bigg )}^{1/ 3} \qquad {\rm and } \qquad  B =
{\Bigg (} - {q\over 2} - \sqrt{{q^2\over 4} + {p^3 \over
27}}{\Bigg )}^{1/ 3}, }
where $p,q$ have been introduced above. Observe, that if $p \to
\infty, ~q \to \infty, ~{q/ p} = {\rm constant}$, we recover the
situation, where no torsion is switched on. This implies, that
expanding in ${1/ p}$ is a legitimate thing to do (we will give
further justification later). In fact, for the limit mentioned
above we get $h \sim -{q/ p}$, which is a reasonable estimate in
the absence of torsion, because this implies ${\cal H} = dB +
{\cal O} (\alpha')$. Therefore, the solution of the cubic equation
for ${\cal H}$ \cubic\ will be typically
\eqn\cubsol{
A+B, \qquad  -{A+B \over 2} + i {\sqrt{3}(A-B)\over 2}, \qquad
{\rm and} \qquad -{A+B
\over 2} - i {\sqrt{3}(A-B)\over 2 },}
giving two complex and one real solution.

This by itself is interesting, because it implies, that we can
actually have a complex three-form in the heterotic string
theory. This means, that the superpotential written in terms of
$\int {\cal H} \wedge \Omega$ can have a complex part and this is
precisely, what we have been looking for, when we mentioned the
apparent $i$-puzzle in an earlier section. To see, whether the
present solution is related to the solution obtained by performing
T-dualities from a Type IIB background, we have to study the set
of solutions \cubsol\ carefully. Let us define a quantity
$$
s \equiv \sqrt{1 + z^2} -1 \qquad {\rm with} \qquad z =
\sqrt{27q^2 \over  4p^{3}},
$$
where $z$ is a small quantity and therefore $s$ can be expanded in powers of
$z$.
We can express $A$ and $B$ in terms of these variables as
\eqn\absim{ A =
a (1 + s - z)^{1/ 3}\qquad {\rm and } \qquad B = -a(1 + s + z)^{1/ 3},
}
where $a = \sqrt{p/ 3}$. Expanding the above relations in powers
of ${1/ p}$, we can determine the solutions of \cubsol\ order by
order. The solutions we obtain are
\eqn\weneed{ \eqalign{ & A + B = - {2 a z\over 3} + {4 a s z \over
9} - {10 a z^3 \over 81} - {10 a s^2 z \over 27} + {80 a s^3 z
\over 243} + \dots,\cr & A - B = 2a + {2 a s \over 3} - {2 a (s^2
+ z^2) \over 9} + {10 a s (s^2 + 3 z^2) \over 81} + \dots }}
{} What we need to know is, how the background three-form field
${\cal H}_{1 2 \bar 3} = h$ depends on the size $t$ of the
six-manifold. For the real solution we get
\eqn\realhis{
h = f - {\alpha'f^3\over t^{3}}  + ....}
Comparing with \pandq\ we see, that to the lowest order in
$\alpha'$ this is exactly, what we expect. Observe also that, for the 
usual case where is the radius is not fixed $-$ due to Dine-Seiberg runaway
problem \DineSB $-$ the radius will tend to go to infinity and therefore 
$h \to f$. Therefore this three-form $h$ is the {\it real} three-form of the 
heterotic theory that satisfies the torsional equations and  appear in the 
anomaly and susy relations.  
The other two complex
solutions are
\eqn\comsoar{ h = - {f \over 2}~ \pm~ i ~
\sqrt{t^{3}\over \alpha'} + \dots .}
These solutions do not satisfy the torsional equations because they are 
complex, but are anomaly free and gauge invariant. The real part of them are 
proportional to the usual heterotic three-form and the complex part provide
the necessary {\it twist} to change the topology of our space from $b_1 = 2$ to
$b_1 = 0$ (this will be more apparent when we will also include the 
spin-connection $\omega$ later in the section). This is a non-trivial 
constraint because by switching on a supergravity
 three-form we cannot change the topology of any space. 
Therefore we need some twist. This is precisely provided by 
our choice of the complex three-form!  
Thus, for the
connection, that is relevant for the physics of non-K\"ahler
manifolds, the background three-form should be real and complex.
As seen from the analysis of \kstt, T-duality rules actually
chooses the complex three-form. 

It now remains to see, what the potential for the radial modulus
in our toy model is. We shall choose the complex $h$ in \comsoar.
It is easy to check, that the resulting potential takes the form
\eqn\potofh{V(t)=\sqrt{g}g^{1\bar1}g^{2\bar2}g^{3\bar3}~{\cal
H}_{12\bar 3} ~{\cal H}_{\bar 1 \bar 2 3} = {t^3 \over \alpha'} +
{\cal O}(t^{-n}),}
where the metric components appearing above have been scaled by
$t$, i.e. $g \to t g$. It can also be checked, that the real parts
of \comsoar\ always contribute a potential of order ${\cal
O}(t^{-n})$, where $n$ is an even integer. In fact, the only
positive power of $t$ comes from the second term of \comsoar.
Therefore, we can now express the action for the radial modulus as
\eqn\radmod{
{\cal L} = \del_\mu t~\del^\mu t ~+~ {t^3\over \alpha'} +\dots,}
implying, that the size of the six-manifold is determined by the
scale $\alpha'$, appearing in the anomaly relation. We will give
an estimate for the size of the six-manifold soon. But first we
need to see, whether we can extend the above calculations to the
realistic case of non-K\"ahler manifolds.

\vskip.2in

\noindent (b) $\underline{\rm
Extension~to~non-\hbox{K\"ahler}~manifolds}$

\vskip.2in

{}From the above analysis we concluded, that the size of the
six-manifold can indeed be stabilized by a potential generated by
the complex three-form flux. However, the model discussed above
is not realistic, since we have chosen very simple fluxes. In this
section we will consider the extension of the above analysis to
non-K\"ahler six-dimensional manifolds.

The first difficulty we find, when dealing with the background
\backgeom\ is, that ${\cal H}$ has many components. Thus, when we
consider the equation \simchoi\ we have to be careful with the
wedge products involved. As a consequence ${\tilde{\cal H}}$ will
also be more complicated. In the toy example ${\tilde{\cal H}}$
was simply proportional to $t^{-1} h$, because only one component
${\cal H} = {\cal H}_{12\bar 3}$ was non-vanishing.

{}From \backgeom\ we can see that the possible
components of ${\cal H}$ are
\eqn\posback{{\cal H}_{\bar 1 2
\bar 3} \equiv h_1 ~{\cal C}_{\bar 1 2 \bar 3}\qquad
{\rm and } \qquad {\cal H}_{1
\bar 2 \bar 3} \equiv h_2 ~{\cal C}_{1 \bar 2 \bar 3},}
and their complex conjugates, which we denote as $h_3$ and $h_4$
respectively. In the analysis below we shall ignore the spin
connection dependent terms, as they do not change the form of the
cubic equation. The one-form ${\cal H}_i^{ab}$, for example, can
be shown to be proportional to
\eqn\oneforiseqto{\eqalign{&{\cal
H}_{\bar 1} \equiv {\cal H}_{\bar 1}^{ab} = {\cal H}_{\bar 1 j
k}~ e^{ja}~e^{kb} = t^{-1} (\alpha_{\bar 1}^{ab}~ h_1 +
\beta_{\bar 1}^{ab}~ h_4),\cr & {\cal H}_2 = t^{-1}(\alpha_2^{ab}~
h_1 + \beta_2^{ab}~ h_4), ~~~ {\cal H}_{\bar 3} = t^{-1}
(\alpha_{\bar 3}^{ab}~ h_1 + \beta_{\bar 3}^{ab}~ h_2),}}
where $\alpha_i^{ab}, \beta_i^{ab}$ are constants, whose exact
form can be determined from the vielbeins. Due to \oneforiseqto\
the cubic equation will be far more complicated, than the one
discussed before. For simplicity we shall assume all components
of $dB + {\cal O}(\alpha')$ to be proportional to a function $f$,
as before. We will show later, that this simplification does not
substantially affect the results. We will also concentrate mainly
on the lowest order in $\alpha'$. The generic cubic equation for
the ${\cal H}_{\bar 1 2 \bar 3}$ component of the three-form is
\eqn\gencube{h_1^3 + m h_1^2 +
n h_1 + s  = 0,}
 where $m, n, s$ depend on $h_4, h_2, t$ and
$\alpha'$. These are identified as
$$
m = (ah_2 + bh_4),\qquad n =
{\Big (}{t^3\over \alpha'} + ch_4 h_2 +dh_4^2 {\Big )}
\qquad {\rm and } \qquad s
= {\Big (}-{ ft^3 \over \alpha'}+ eh_4^2 h_2 {\Big ),}
$$
where $a, b, ...$ are integers. These expressions can be easily
determined, but for the analysis below we do not need the explicit
form for these constants.

The equation above can now be written as \cueqbc\ by
using the shift technique for cubic equations. If we identify
$h_1$ with $h - {m / 3}$, then the $p, q$ variables appearing in
\cueqbc\ can be given in terms of $m,n$ and $s$ as
\eqn\mns{p = n - {m^2 \over 3} = {t^3\over \alpha'} + B
\qquad {\rm and} \qquad
 q = s - {mn\over 3} + {2 m^3 \over 27} = -{ ft^3 \over
\alpha'} + D,
}
where $B, D$ are {\it independent} of $\alpha'$.

This is very important because it immediately implies, that to
lowest order in $\alpha'$ the solution given in \weneed,
\realhis\ and \comsoar\ is still satisfied! Thus every component
of ${\cal H}$ is proportional to the already determined solutions
in \realhis\ and \comsoar. Therefore, we expect a similar
potential for the radial modulus in the most generic case as well.
This completes the argument\foot{The higher order terms are more involved
and the simple analysis that we presented here gets modified. However we
still expect to see similar behavior for {\it all} orders because
the scaling arguments that we presented earlier continue to hold at
every order and therefore the radial modulus gets fixed at any arbitrary
order. We have not demonstrated this because the mathematics, though
straightforward, becomes very involved at higher orders in $\alpha'$. We
hope to tackle this is future.}.

\vskip.2in

\noindent~(c)~$\underline{\rm The~{\cal Q}=0~case}$

\vskip .2in

So far we have discussed the case, when the cubic contribution
${\tilde {\cal H}}^3$ is non-zero. This is the usual case for
some choices of holonomies. However, we could also have situations
for which
\eqn\calqq{{\cal Q} = \alpha_1^{ab}~\alpha_2^{cd}~ \alpha_{\bar 3}^{ef}~
{\rm Tr}(M^{ab} M^{cd} M^{ef}) ~=~ 0.}
In this case, the cubic equation simplifies very much and to
lowest order in $\alpha'$ we have $h = -q/ p$. In fact, the relations
for the Chern-Simons form
will change to a much simpler relation between the
spin-connection $\omega$ and the background three-form ${\cal H}$ as
\eqn\chaofcom{
\eqalign{& \Omega_3(\tilde\omega) = \tilde\omega \wedge
d\tilde\omega,
\cr
& \Omega_3(\hat\omega) = \hat\omega \wedge d\hat\omega,}}
where now comparing to the earlier expressions of $\Omega_3$ we see that
the forms are similar (in the usual case the form of
$\Omega_3(\hat\omega)$ differs from $\Omega_3(\tilde\omega)$).

Let us consider now a simpler situation in which we ignore the spin connection
$\omega$. In this case,
we can show that the three-form equation
will eventually be
\eqn\thrforlook{{\cal H} - {\alpha'\over 4}
\tilde{\cal H} \wedge d\tilde{\cal H} = f,}
where $f$ is defined as before, but in this case it has no
$\omega$ dependence. A brief reflection shows us, that this case
is more complicated, than the ${\cal Q} \ne 0$ case. Of course, to
the zeroth order in $\alpha'$ we simply obtain ${\cal H} = f$,
which is the real root obtained earlier. Having ${\cal Q}=0$
means, that there are no terms of order $\alpha'$ and therefore
the next non-trivial equation is of order $\alpha'^2$. This is
simply because \threeform\ takes now the form
\eqn\simpbeca{d{\cal H} = \alpha'{\Big [} d\omega \wedge d\omega -
d\omega \wedge
d\tilde{\cal H} + {1\over 4} d\tilde{\cal H} \wedge d\tilde{\cal H} -
{\rm tr}~F \wedge F{\Big ].}}
This makes the situation a little involved and therefore
\thrforlook\ can only be solved iteratively. However, in the next
sections we will not consider this case. To the zeroth order it
is clear, that the only root is the real one and the second Chern
class for the gauge bundle satisfies $c_2(F) = \int d\omega \wedge
d\omega$. It is also clear, that even for this case the radius
would be stabilized, because from \simpbeca\ we see, that the left
hand side scales as $t$, whereas the right hand side scales as
$a~t^{-1} + b~t^{-2}$, where $a,b$ can be easily ascertained from
\simpbeca\foot{Furthermore in \simpbeca\ we see that even though $\omega$
term fails to give a potential (discussed earlier in 4.4 (a)) the term with
$d{\tilde{\cal H}}$ could in principle be responsible for the potential.
We will address this issue in more detail elsewhere.}.

\vskip.2in

\noindent~(d)~$\underline{\rm
Superpotential~for~the~heterotic~theory~and~radius~
of~the~six~manifold}$

\vskip.2in

In the above sections we argued, how a {\it complex} three-form
could arise in the heterotic theory. In this section we would like to
compare the
superpotential, that we get from this setup to the superpotential
obtained by doing U-dualities from the Type IIB theory.
In order to do this, we must first
incorporate back the spin connection $\omega$ into the
calculations.

The main equation, which takes into account all the variables has
already been spelled out in \mexi. Let us concentrate on the
component $h_1$ for the time being. Using the earlier definitions,
we can show, that the cubic equation, that we get at this stage is
the same as \gencube\ with $m, n, s$ defined as
\eqn\custage{m =
a_1 + a_2 \omega, \qquad n = {t^3 \over \alpha'} - a_3 \omega + a_4
(d\omega + \omega^2) + a_5 \qquad {\rm and} \qquad
s = -{f t^3\over \alpha'} +
{\cal O} (\omega^4),
}
where $\omega$ is the one-form spin connection
$\omega_\mu^{ab}$ and $a_i$ are constants independent of
$\alpha'$. We can now define the shift $\beta$ in $h_1 = h -
\beta$ as $\beta = {1\over 3}(a_1 + a_2 \omega)$. This results,
as before, in a cubic equation of the form \cueqbc\ with $p$ and $q$
given by
\eqn\pqagain{\eqalign{& p = {t^3 \over \alpha'} + {\cal A}~\omega
+ {\cal O} (\omega^2), \cr & q = - {f~t^3 \over \alpha'} + {\cal B}~\omega
+ {\cal
O}(\omega^2),}}
 where the $\alpha'$ dependence is shown explicitly and ${\cal A}$ 
and ${\cal B}$ can be
determined from \custage.
All the arguments dealt in the earlier sub-sections will go
through without any subtlety. In particular $A \pm B$ are defined
as usual and from there we can extract the three-form $h_1$
explicitly. The superpotential will now have the generic form, to
lowest order in $\alpha'$ 
\eqn\supgenfor{ W
=\pm\int (f + i b~\omega + \sum_{m,n~\in~\IZ/2} i~c_{mn}~f^m t^n )
\wedge \Omega,}
where $b$ depends on $t$ in some specific way (the detailed
dependence is not very important for our present purpose) and $c_{mn}$ are
constants that depend on $\alpha'$.\foot{There is yet another real 
contribution to the superpotential which appears from the heterotic 
gauge field $F$ as $F\wedge J \wedge J$ where $J$ is the fundamental two 
form for the manifold \becsuper. 
This is distinct from $\Omega_3 (A) \wedge \Omega$ 
where $\Omega_3 (A)$ is the gauge Chern-Simons term. This term will be 
responsible for generating the Donaldson-Uhlenbeck-Yau kind of equations for 
gauge bundles. Furthermore the above choice of superpotential does indeed
reproduce the torsional equations as is shown in \becsuper. More details on
the superpotential and how to see the masses of the KK monoples etc will 
be addressed in part II of this 
paper.}
Comparing the $U$-duality results of \kstt\ we can see, that this
form of the superpotential is what we expect, at least to this
order. The shift of the superpotential proportional to the spin
connection $\omega$ is, what is responsible for the twisting, as
we mentioned earlier. This aspect has also been pointed out in
\kstt\ and also in the second reference of \pktspt, where the
detailed derivation of the superpotential using T-duality rules
for the heterotic theory is given to the lowest order in
$\alpha'$. The higher powers of $t$ terms in the superpotential
are responsible for fixing the radial modulus of our manifold, as
we show below. Therefore, to all orders in $\alpha'$ the
superpotential will have the generic form $W = \pm \int {\cal H}
\wedge \Omega$, with ${\cal H}$ given by \eqn\suptoallorders{{\cal
H} = \sum_{m,n,p,q~\in~\IZ/2} b_{mnpq}~\alpha'^m f^n (-t)^p \omega^q,}
where $f,t,\omega$ are arranged (for every term) so that they are
dimensionally same as $f$ as we saw before and $b_{mnpq}$ are constants.
To
obtain these terms from the Type IIB theory using T-duality, we
need to look for higher order corrections to the T-duality rules.
It will be interesting to do this, but this is at present beyond
the scope of this work.

Finally, we should use all the results derived above, to estimate
the radius of our six-manifold. Again, we will concentrate only to
the lowest order in $\alpha'$. The potential for the generic case
is the same as discussed in \radmod\ except, that for our purpose
we need to go to higher orders in $t$. We can write the potential
as
\eqn\potwrit{V(t) = {t^3 \over \alpha'} ~+~ {9 \over 64}~
{\alpha' |f|^4 \over t^3} ~+~ {\cal O}(\alpha'^2),}
which is basically derived from \weneed, by keeping terms to order
$z^2$. Recall also, that we are taking the superpotential as in
\supgenfor. Minimizing $V(t)$ with respect to $t$, gives an
estimate of the radius of the six-dimensional non-K\"ahler
manifold. We can write this explicitly as
\eqn\radofsixman{ t =  {\Big (} {3\alpha'|f|^2
\over 8} {\Big )}^{1/3},}
where $f$ is defined in \pandq. Observe
also that, since $|f|$ can be determined directly from \backgeom,
one can numerically estimate the radius $t$ of our manifold.

Before closing this section let us tie some loose ends. In
deriving the radius for our manifold, we have expanded functions
with respect to the quantity ${1/ p}$, where $p$ is defined in
\pandq. {}From \radofsixman\ we see, that this is possible, if the
flux {\it density} $f$ is large (with a total constant flux).
Therefore, we need the radius $t$ to be fixed but large. The
limit of fixed but large enough radius is also consistent with
the supergravity description. In fact, in the analysis done in
section 3.1 of \beckerD, we imposed
\eqn\hetrequire{ g^{(4)}_{het} \to  0,
\qquad  {\rm and} \qquad V_{het} > > 1,}
in order to have a valid supergravity description. Here
$g^{(4)}_{het}$ is the four dimensional heterotic coupling and
$V_{het}$ is the volume of the six-manifold. However, from
\valueAB\ the reader may wonder, whether such choice is
generically possible, because of the radius stabilization. But
since all K\"ahler moduli are not fixed by our fluxes, we can
constrain the volume of four-cycles in the original ${\cal
M}$-theory picture to be small. Further investigations on this
matter will be presented elsewhere.

There is yet another reason to have a
large but finite radius. The vanishing
of
the gaugino and the gravitino supersymmetry
transformations (in the notations of
equation  4.29 of \beckerD) implies that
\eqn\gaggagino{
\nabla_M({\hat\omega})~\epsilon = 0 =
F^a_{MN}\Gamma^{MN}
\epsilon,}
converting \hetsigma\ to the light-cone RNS (0,1) non-linear
sigma model. This identification is precise {\it only}, when
$d{\cal H} = 0$. However, we do not have this situation and
therefore, we require the corrections to the set of equations
\gaggagino\ to be small. This is possible, if ${t^2 /\alpha'}$
becomes sufficiently large. The above equation \gaggagino\ can
also be viewed as the condition under which \hetsigma\ is
invariant under world-sheet supersymmetry transformations on
$X^i, S^p$. Therefore, when we identify \hetsigma\ with the (0,1)
sigma modeltransformations, we are also inherently assuming, that
the corrections are small. Notice also, that the connection
appearing in \gaggagino\ is ${\hat\omega}$ and therefore, the
fact that the two-form $J_{ij}$ is covariantly constant, is with
respect to this connection. This aspect has been briefly alluded
to in the earlier sections and discussed in much detail in a
series of papers by Hull \HULL. Furthermore,
 having
$d{\cal H} \ne 0$ also means that we have the identity \HULL
\eqn\hulliden{{1\over 2} {\Big [} R_{abcd}(\tilde\omega) -
R_{cdab}(\hat\omega) {\Big ]} = {\cal H}_{[abc, d]},}
implying a non-zero two-loop contribution to the trace
anomalies\foot{Recall, that the Ricci tensor is $R_{ab} \equiv
R^c_{~~acb}({\tilde\omega})$ with respect to the connection
$\tilde\omega$. Being non-zero and finite, this contributes to the
trace anomalies.}. Again, if ${t^2 /\alpha'}$ is large, this
contribution becomes small \HUT. A non-vanishing Riemann tensor in
\hulliden\ can again contribute to the beta function at two-loop
order. This contribution has been calculated in many earlier works
and is shown to be proportional to $R_{abcd}R^{abcd} -
F_{ab}F^{ab}$. This would obviously cancel under the usual
embedding of the spin connection into the gauge connection. For
our case, we can argue, that this is suppressed, if the radius
$t$ is large. More details on this and phenomenological
implications of large but finite radius will be explored in an
upcoming paper \toappear.

\vskip .2in

\newsec{Discussions}

In this paper we studied in detail the geometrical and topological
properties of non-K\"ahler manifolds of the form \backgeom\ and a
large class of generalizations thereof. As we showed, these
manifolds in general have zero Euler characteristics and also
zero first Chern class. For the background studied earlier in
\beckerD, the complete topological properties were determined. The
torsional metric in the presence and absence of gauge fields has
been worked out and was shown in both cases to satisfy the
torsional equations imposed by supersymmetry. In section 3 more
general examples of non-K\"ahler manifolds were found and their
mathematical properties determined. All these examples are
compact and complex. In section 4 we determined the
superpotential for compactifications of the heterotic string on
such non-K\"ahler manifolds. We showed, that many of the moduli
fields of these heterotic compactifications can be stabilized,
once the ${\cal H}$-fluxes are turned on. In particular, we have
computed the potential for the radial modulus and showed, that the
value of this field can be determined.


\subsec{Related Examples of non-K\"ahler Manifolds}

Recently there have been some more examples of these manifolds
discussed in the literature. One particular interesting one is the
Iwasawa manifold first discussed in this context by Strominger
\rstrom\ and more recently discussed in more detail by Cardoso et.
al \carluest. In fact, this example follows directly from the
generic construction that we gave in section 3. The Iwasawa
manifold is a principle torus bundle over a torus constructed
from a set of $3 \times 3$ matrices with complex entries
\eqn\iwasawa{\pmatrix{1&a&b\cr 0&1&c \cr 0&0&1},} and is
therefore, a complex manifold. As discussed in \rstrom\ and
\carluest, if we restrict this to integers $m,n,p$ and induce the
action: \eqn\act{a \to a + m,~~~b \to b + cm + n, ~~~ c \to c +
p,} then, we get a smooth manifold called the Iwasawa manifold.
Therefore, the complete properties of heterotic string
compactifications on the Iwasawa manifold can be determined. The
background geometry is \eqn\iwasawa{\eqalign{&ds^2 = dz d{\bar z}
+ dv d{\bar v} + |du - z dv|^2, ~ ~~ e^{\phi} = {\rm c_o},~~~
\chi = 0, \cr & {\cal H} = - {1\over 4}(du - z dv) \wedge d {\bar
z} \wedge d {\bar v} + {\rm c.c.}, ~~ \{~b_i \} ~=~ (1, 4, 8, 10,
8, 4,1), }} with the anomaly condition $d{\cal H} = - {\rm tr}~ F
\wedge F$, solved with just an abelian gauge field configuration.
The reader can extract more details on this from \carluest. It
will also be interesting to see, whether the Iwasawa manifold can
appear from a four-fold in ${\cal M}$-theory in the same way as
in \sav, \beckerD. It is clear, that the naive identification of
the four-fold as $T^4/ {\cal I}_4 \times T^4$ cannot work because
of two obvious reasons:

\noindent (a) The Euler characteristics of this four-fold is zero
and therefore cannot support fluxes, as the anomaly constraints
will prohibit it \rBB.

\noindent (b) The choice of $T^4/{\cal I}_4$ will tell us, that on
the Type I side we should always have three-forms, that have one
leg along the $z^3$ or ${\bar z}^3$ direction. From the explicit
background constructed in \carluest\ we see, that there are other
components of ${\cal H}$.

Another detailed study of the mathematical aspects of non-K\"ahler
manifolds has appeared recently in \GP. The manifolds considered
there are of the form \eqn\mangp{ ds^2 = e^{2\phi}~g_{CY} + (dx +
\alpha)^2 + (dy + \beta)^2,} where $\phi$ is the warp factor and
$g_{CY}$ is a Calabi-Yau base. These class of examples are also related to
our construction.
The one-forms $\alpha$ and $\beta$
are defined on the Calabi-Yau base. These one-forms can be
identified with (1,1) anti-self-dual forms $\omega_p$ and
$\omega_q$ via $d\alpha = \omega_p$ and $d\beta = \omega_q$,
which give rise to the three-form \eqn\thrforgp{ H_3 = dx \wedge
\omega_p + dy \wedge \omega_q,} in the Type IIB theory, after we
make two T-dualities along the two-cycles of the fiber torus
$T^2$. This three-form is basically the NS-NS three-form of the
Type IIB theory and as such lies in the integer cohomology.

The paper \louisL\ gave examples of new manifolds, that are
neither complex nor K\"ahler in the context of the Type IIA
theory. These manifolds are termed {\it half-flat}, and were
shown to have torsion lying in all the five torsion classes ${\cal
W}_i$. They generically have an $SU(3)$ structure and could be
complex, if the torsion lies in ${\cal W}_3 \oplus {\cal W}_4
\oplus {\cal W}_5$. The vanishing of the Nijenhuis tensor amounts
to having, in this language, vanishing ${\cal W}_1 \oplus {\cal
W}_2$ classes. This classification of torsion classes is well
known in the mathematics literature but in connection to string
theory compactifications it has also been addressed in \carluest.
The previous Type IIA models are mirrors of the Type IIB theory
compactified on Calabi-Yau three-folds with NS-NS fluxes turned
on. However, in \louisL\ not all fields in the Type IIB theory are
given an expectation value. In fact, one has to turn on at least
the RR three-form to completely embed this in string theory. This
has been done recently in \kstt. The authors of \kstt\ gave
explicit examples of mirror manifolds in the Type IIA theory and
showed, that these manifolds have an almost complex structure,
which may or may-not be integrable. When it is integrable, then
the manifold is complex. The model discussed in \kstt\ is the Type
IIB theory on $T^6/\IZ_2$, and therefore making less than six
T-dualities we always remain in either Type IIA or Type IIB
(depending on whether we make an odd or even number of
T-dualities). Hence, we have mirror descriptions in either of
these theories on generic compact non-K\"ahler manifolds with an
almost complex structure $J_{mn}$. For the models presented in
\louisL\ the fundamental two form $J_{mn}$ is not covariantly
constant with respect to the affine connection but {\it is} covariantly
constant with respect to the preferred connection measured by the contorsion
tensor. In the language of ours the contorsion tensor is
precisely ${1\over 2} {\cal H}$ and therefore the connection can
be identified with the $\hat\omega$ discussed in the present
paper.

Further discussions of Type IIB compactifications on
six-dimensional manifolds with fluxes have been recently
discussed in \pktspt.


\subsec{Phenomenological~Applications}

The models considered herein might be rather interesting for
particle phenomenology, as many moduli fields appearing in these
compactifications (including the radial modulus) can be
stabilized and definite predictions for the coupling constants of
the standard model can be made. Furthermore, our
compactifications include a warp factor, which provides one of
the few known mechanisms for solving the gauge hierarchy problem
\kst. Recently it has been speculated that the masses generated by
the fluxes could be even at the phenomenological viable TeV scale
\olgi.

Another important property of our compactification manifolds is,
that they have zero Euler characteristics and a vanishing first
Chern class. One may wonder, if a vanishing Euler number implies a
vanishing number of particle generations in the four-dimensional
theory. It is well known \EdNew, that the net number of
generations minus anti-generations is determined in terms of the
Euler number of the internal manifold as
\eqn\generations{|h^{2,1}-h^{1,1}|={|\chi|\over 2},} for
compactifications on Calabi-Yau three-folds, where $h^{i,j}$
describe the corresponding Hodge numbers of the internal manifold.
However, the above formula is not valid for the models considered
herein, as the spin connection has to be embedded into the gauge
connection to arrive at the above result, even for the case of
K\"ahler compactifications. This is not the case we are interested
in. To determine the number of generations appearing in our
models, we would need to analyze the zero modes of the Dirac
equation for our backgrounds. We shall report  this in a future
publication \toappear.

In this paper we have shown, that the radial modulus for
compactifications of the heterotic string on non-K\"ahler
manifolds receives a potential, which allowed us to estimate the
actual value of the size of the internal manifold. A tantalizing
possibility would be, that a cosmological constant is indeed
induced, after the radial modulus has been stabilized. A priori,
we see here the possibility, that a positive cosmological constant
could be induced, giving us a realization of de Sitter space in
string theory. This is a long standing puzzle, whose solution
could certainly be along these lines. We leave the details of
this fascinating possibility for future work \toappear. It would
be a great triumph of string theory, if the correct relation
between the supersymmetry breaking scale and the cosmological
constant could be predicted in this way \banks.


\subsec{Future Directions}

$\bullet$ There are many interesting directions to pursue in the
future. In the sigma-model section we discussed the fact, that
for our case the simplest embedding of the gauge-connection into
the torsional-spin connection is not allowed. Therefore, there
might be the possibility, that the two-loop beta function is not
vanishing. As discussed earlier and also in \callan\ and \HULL,
these contributions are suppressed, if the size of the
six-manifold is large. In fact, for our case we can have a large
sized manifold, as the size parameter depends on the
flux-density. Furthermore, the background found in \beckerD\
by using U-dualities from a given supergravity background in
${\cal M}$-theory, is a valid solution at least to the {\it
lowest} order in $\alpha'$. Therefore, (to extend the discussion
to all orders in $\alpha'$) two things could happen here:

(a) Apart from the size factor, there could be generic
counter-terms, that could cancel the two loop contributions to the
beta function. At least it has been discussed in some detail in
the literature, that this contribution is cancelled, if we define
the background three-form ${\cal H}$ as in \threeform, because
there exist counter-terms \callan\ and \HULL, that lead to a
vanishing beta function. But this relies on the fact, that we are
embedding the gauge connection into the torsional-spin connection
$\tilde\omega$, which is not what we want to do in the present
case.

(b) The beta function is exactly zero to all orders in $\alpha'$
only for a given size of the six manifold. This is what we might
expect, because the radial modulus is fixed and therefore only,
when the background has the right radial modulus and right complex
structure modulus, the beta function for strings propagating on
this background will be zero. This matter needs further
investigation and more details will be presented elsewhere.

$\bullet$ In this paper we studied $SO(32)$ heterotic strings on
compact non-K\"ahler manifolds or more appropriately $D_4^4$
heterotic strings. It would be interesting to see, how to describe
the $E_8 \times E_8$ heterotic string in this framework. Observe,
that the reason we have gotten the $D_4^4$ heterotic theory is,
because we started with an orientifold model in the Type IIB
theory, that under two T-dualities and an S-duality reproduces the
$D_4^4$ theory \beckerD. To get the $E_8 \times E_8$ heterotic
string we need a similar framework, maybe in ${\cal F}$-theory,
where there is a possibility of having exceptional symmetries
\dasmukhi. But the points, where we can have exceptional
symmetries and orientifold representations are no longer {\it
perturbative} \dasmukhi. Let us elaborate this a little bit. More
details will appear in \toappear.

The models studied in \sav\ and \beckerD\ have an ${\cal
F}$-theory interpretation defined in terms of elliptic curves
\eqn\elipva{y^2 = x^3 + x f(z) + g(z),} where $z$ is the
coordinate on the $CP^1$ base and $f,g$ are polynomials of degree
8 and 12 respectively \vafasen. The modular parameter of the fiber
is given in terms of j-function \vafasen, where $j\propto
{f^3\over \Delta}$ and $\Delta$ (not to be confused with
warp-factor) is the discriminant. The ${\cal F}$-theory
representation, that concretely realizes the model of \beckerD,
has the following choice of $f, g$ and $\Delta$ \eqn\folchoi{f(z)
\sim (z - z_1)^2, ~~~~ g(z) \sim (z - z_1)^3, ~~~~ \Delta(z) \sim
(z - z_1)^6,} near one ``orientifold'' point $z \to z_1$. The
fact that $\Delta$ is proportional to $z^6$ means, that all the
orientifold planes have become dynamical realizing the $D_4$
symmetry. Now, as shown in \dasmukhi, an $E_8$ symmetry is
realized at $z \to z_1$, when $f, g, \Delta$ are \eqn\esing{ f(z)
= 0, ~~~~ g(z) \sim (z - z_1)^5, ~~~~ \Delta(z) \sim (z -
z_1)^{10}.} A couple of immediate points to note here are, that
$K3$ goes to its $Z_6$ orbifold point (with full symmetry of $E_8
\times E_6 \times D_4$) and the fact that we have $g \sim z^5$
will tell us, from Tate's algorithm, that the symmetry is $E_8$.
This immediately tells us, that the pure $E_8 \times E_8$
symmetry is realized, when \eqn\pureeight{g(z) = (z - z_1)^5(z -
z_2)^5(z - z_3)(z - z_4),} with vanishing $f(z)$ as before. This
has a non-perturbative orientifold representation, as was shown in
\dasmukhi. A more concrete way of realizing this construction was
discussed in \zwee. Thus, this could be one possible way of
getting the torsional background for the $E_8 \times E_8$
heterotic theory.

$\bullet$ The non-K\"ahler manifolds discussed in this paper all
have a vanishing Euler characteristics. This is not a problem by
itself, as we discussed above, that a vanishing Euler
characteristics {\it doesn't} imply zero number of generations. It
will be interesting to extend the above analysis to non-zero Euler
characteristics. We would then have to start on the Type IIB side
with a manifold with non-zero Euler characteristics in the {\it
absence} of fluxes. Let us elaborate this a little bit. In the
present paper we have seen, that in the absence of NS-NS and R-R
fluxes in the Type IIB picture, making two T-dualities we end up
in the Type I theory on $K3 \times T^2$. As discussed earlier, by
switching on fluxes, the torus $T^2$ becomes nontrivially fibred
over the base $K3$. However $K3 \times T^2$ has zero Euler
characteristics (because of the $T^2$) and therefore, the
non-K\"ahler manifold of the heterotic compactification also has
zero Euler characteristics. For the generalization of the present
construction to non-zero Euler characteristics, we should start
with a manifold, which looks like $K3 \times Z$, where $Z$ is a
two-dimensional manifold with non-zero Euler characteristics on
the Type IIB side. This is the minimal requirement. Of course, we
can even get a generic six-dimensional manifold $X$, which should
then have the following properties in the {\it absence} of
fluxes: (a) compact and complex with non-zero Euler
characteristics, (b) there exists a four-fold, which is a
non-trivial $T^2$ fibration over $X$, and most importantly (c)
should have an orientifold setup in the Type IIB framework. More
details on this will be addressed in a future publication
\toappear.

$\bullet$ In sub-section 2.5 we discussed the anomaly relation
for the heterotic theory and showed, how the ${\rm tr} ~F \wedge
F$ term can appear in the heterotic Bianchi identity from the Type
IIB side. The allowed gauge bundle is very restricted for the
torsional background, because, as for the K\"ahler
compactifications, we expect the gauge bundle to satisfy
\eqn\gagbund{g^{a \bar b}F_{a \bar b}^n ~ = ~ 0, ~~~~ F_{ab}^n ~
= ~ 0, ~~~~ F_{\bar a \bar b}^n ~ = ~ 0,} which are the
Donaldson-Uhlenbeck-Yau (DUY) equations for gauge fields. It
would be interesting to see, how the DUY constraints can be
derived from a D-term along the lines of \EdNew\ and \becons\ for
the case of compactifications on Calabi-Yau three-folds . In
\beckerD\ it was shown, how in the Type IIB theory this is
realized from the primitivity condition of $G$-fluxes in ${\cal
M}$-theory. Furthermore, because the three-form ${\cal H}$ is
related to the metric by the torsional equations \toreqrel, there
arises another restriction on the allowed gauge bundle
\eqn\anrest{{\rm Tr}~ F \wedge F = 30{\Big [} {\rm tr} ~ R \wedge
R - {i \over \alpha'}~ \del {\bar \del} J {\Big ]}.} Since
${\cal H}$ is globally defined, the constraint \obstr\ follows
from this formula. The relation \anrest\ is non-trivial, because
none of the terms in the right hand side can be zero. Thus, the properties of
the gauge bundle is another important aspect, that needs to be
studied in a future \toappear.

$\bullet$ We discussed in some detail, how many of the moduli for
the heterotic theory, including the radial modulus, can be fixed
for this kind of compactification. In the earlier sections we
argued, that the dilaton modulus does not get fixed in this
process. One might argue, that since the superpotential in Type
IIB theory has an axion-dilaton appearing explicitly in the
formula, the heterotic superpotential, which follows from a set of
U-dualities, should also have a dilaton dependence. This is not
quite so, because after performing two T-dualities and an
S-duality on the Type IIB superpotential one can show, using the
T-duality rules of \mesO, that the dilaton factor {\it does not}
appear in the formula for the heterotic superpotential. Therefore,
it will be interesting to see, how explicitly the dilaton could
appear in the heterotic superpotential. Fixing the dilaton will
immediately guarantee, that we can have a constraint on the size
of the six manifold in the Type IIB framework. This would imply,
that a string theory model in de-Sitter space could be
constructed, along the lines of the first paper of
\renata\foot{We have been informed, that some stable de-Sitter
solutions have recently been found in Type IIB theory with fluxes
\renshamit.}. Furthermore, as discussed in the other papers of
\renata, the Type IIB models with fluxes and controlled moduli
give the supergravity dual of cascading ${\cal N} = 1$ gauge
theories, which are confining in the IR. A better understanding
of the moduli problem in the Type IIB setup can therefore be
used, to understand the dynamics of ${\cal N} = 1$ gauge theories.

$\bullet$ In the examples studied in section 4, we have considered
the case, when the preferred spin connection is not embedded into
the gauge connection and the relation between $A$ and
$\tilde\omega$ is given by \weasfor. This in particular implies,
that we have \eqn\handtilh{d{\cal H} = d\tilde{\cal H} = {\cal
O}(\alpha'),} and therefore in section 4.4 had a simple way to
study the stabilization of the radial modulus. This is not always
the case and, in fact, for a very generic embedding we can have a
situation, where the only possible way to study the three-form is
by iterative technique. However, it is clear by scaling arguments,
that the radius is again fixed for this case (see sec 4.4(c)). It
will therefore be interesting to see, what value of the radius we
get in this generic situation.

\centerline{{\bf Acknowledgements}}

\nobreak It is our pleasure to thank Michael Dine, Lance Dixon,
Edward Goldstein, Shamit Kachru, Joe Polchinski, Sergey
Prokushkin, Savdeep Sethi,
Mohammed M. Sheikh-Jabbari, Andrew Strominger,
Prasanta Tripathy, Sandip Trivedi and Edward Witten for useful
discussions. The work of K.B. was supported by the University of
Utah. The work of M. B. is supported by NSF grant PHY-01-5-23911
and an Alfred Sloan Fellowship. The work of K.D. was supported in
part by a David and Lucile Packard Foundation Fellowship 2000 $-$
13856. K.B. wishes to thank the warm hospitality at the
University of Maryland, where part of this work has been carried
out.

\vfill

\break

\vfill

\eject

\listrefs

\bye